\documentclass{sig-alternate}

\pdfoutput=1

\usepackage{fancyhdr}
\usepackage[normalem]{ulem}
\usepackage[hyphens]{url}
\usepackage{amsmath}
\usepackage{multirow}
\usepackage{subfig}
\usepackage{caption}
\usepackage{float}
\usepackage[hyphens]{url}
\usepackage{amssymb}
\usepackage{mathtools}
\usepackage{xspace}
\usepackage{color}
\usepackage{comment}
\usepackage[flushleft]{threeparttable}
\usepackage[ruled,vlined]{algorithm2e}
\usepackage[noend]{algcompatible}
\usepackage{cite}
\usepackage{epstopdf}
\usepackage{pifont}
\usepackage{listings}
\usepackage{tikz}

\newcommand{\bheading}[1]{{\vspace{2pt}\noindent{\textbf{#1}}\hspace{2pt}}}

\newcommand{\gbytes}{\ensuremath{\mathrm{GB}}\xspace}
\newcommand{\mbytes}{\ensuremath{\mathrm{MB}}\xspace}
\newcommand{\kbytes}{\ensuremath{\mathrm{KB}}\xspace}

\newcommand{\ghertz}{\ensuremath{\mathrm{GHz}}\xspace}

\newcommand{\RNum}[1]{\uppercase\expandafter{\romannumeral #1\relax}}

\newcommand{\etc}{\emph{etc.}\xspace}
\newcommand{\ie}{\emph{i.e.}\xspace}
\newcommand{\eg}{\emph{e.g.}\xspace}

\newcommand{\attackname}{memory DoS attacks\xspace}
\newcommand{\aattackname}{memory DoS attack\xspace}

\newcommand{\protectedVM}{\textsc{Protected VM}\xspace}

\newcommand{\secref}[1]{\mbox{Sec.~\ref{#1}}\xspace}

\newcommand{\cmark}{\ding{52}}
\newcommand{\xmark}{\ding{56}}

\newenvironment{packeditemize}{
\begin{list}{$\bullet$}{
\setlength{\labelwidth}{8pt}
\setlength{\itemsep}{0pt}
\setlength{\leftmargin}{\labelwidth}
\addtolength{\leftmargin}{\labelsep}
\setlength{\parindent}{0pt}
\setlength{\listparindent}{\parindent}
\setlength{\parsep}{0pt}
\setlength{\topsep}{3pt}}}{\end{list}}

\definecolor{dkgreen}{rgb}{0,0.6,0}
\definecolor{gray}{rgb}{0.5,0.5,0.5}
\definecolor{mauve}{rgb}{0.58,0,0.82}

\makeatletter
\def\@copyrightspace{\relax}
\makeatother

\begin{document}
\title{Memory DoS Attacks in Multi-tenant Clouds: \\Severity and Mitigation}

\numberofauthors{3} 
\author{
\alignauthor
Tianwei Zhang\\
       \affaddr{Princeton University}\\
       \email{tianweiz@princeton.edu}
\alignauthor
Yinqian Zhang\\
       \affaddr{The Ohio State University}\\
       \email{yinqian@cse.ohio-state.edu}
\alignauthor
Ruby B. Lee\\
       \affaddr{Princeton University}\\
       \email{rblee@princeton.edu}
}

\maketitle

\begin{abstract}

In cloud computing, network Denial of Service (DoS) attacks are well studied and defenses have been implemented, but severe DoS attacks on a victim's working memory by a single hostile VM are not well understood. Memory DoS attacks are Denial of Service (or Degradation of Service) attacks caused by contention for hardware memory resources on a cloud server. Despite the strong memory isolation techniques for virtual machines (VMs) enforced by the software virtualization layer in cloud servers, the underlying hardware memory layers are still shared by the VMs and can be exploited by a clever attacker in a hostile VM co-located on the same server as the victim VM, denying the victim the working memory he needs. We first show quantitatively the \emph{severity} of contention on different memory resources. We then show that a malicious cloud customer can mount low-cost attacks to cause severe performance degradation for a Hadoop distributed application, and 38$\times$ delay in response time for an E-commerce website in the Amazon EC2 cloud.

Then, we design an effective, new defense against these \attackname, using a statistical metric to detect their existence and \emph{execution throttling} to mitigate the attack damage. We achieve this by a novel re-purposing of existing \textit{hardware performance counters} and \textit{duty cycle modulation} for security, rather than for improving performance or power consumption. We implement a full prototype on the OpenStack cloud system. Our evaluations show that this defense system can effectively defeat \attackname with negligible performance overhead.

\end{abstract}

\section{Introduction}
\label{sec:intro}

Public Infrastructure-as-a-Service (IaaS) clouds provide elastic computing resources on demand to customers at low cost. Anyone with a credit card may host scalable 
applications in these computing environments, and become a {\em tenant} of the 
cloud. To maximize resource utilization, cloud providers schedule virtual machines 
(VMs) leased by different tenants on the same physical machine, sharing the same 
hardware resources.

While software isolation techniques, like VM virtualization, carefully isolate 
memory pages (virtual and physical), most of the underlying hardware memory 
hierarchy is still shared by all VMs running on the same physical machine
in a multi-tenant cloud environment. Malicious VMs can exploit the multi-tenancy 
feature to intentionally cause severe contention on the shared memory resources to 
conduct Denial-of-Service (DoS) attacks against other VMs sharing 
the resources. Moreover, it has been shown that a malicious cloud customer can 
intentionally co-locate his VMs with victim VMs to run on the same physical 
machine \cite{RiTrSh:09, Varadarajan:2015:PVS, Xu:2015:MSC}; 
this co-location attack can serve as a first step for performing \attackname 
against an arbitrary target. 

The severity of memory resource contention has been seriously underestimated. While
it is temping to presume the level of interference caused by resource contention is
modest, and in the worst case, the resulting performance degradation is isolated on
one compute node, we show this is not the case. We present advanced attack techniques 
that, when exploited by malicious VMs, can induce much more intense memory contention 
than normal applications could do, and can degrade the performance of VMs on multiple nodes.

To demonstrate that our attacks work on real applications in real-world 
settings, we applied them to two case studies conducted in a commercial IaaS 
cloud, Amazon Elastic Compute Cloud (EC2). We show that even if the attacker only 
has \textit{one} VM co-located with one of the many VMs of the target multi-node 
application, significant performance degradation can be caused to the entire 
application, rather than just to a single node. In our first case study, we show 
that when the adversary co-locates one VM with one node of a 20-node distributed 
Hadoop application, he may cause up to 3.7$\times$ slowdown of the entire 
distributed application. Our second case study shows that our 
attacks can slow down the response latency of an E-commerce application 
(consisting of load balancers, web servers, database servers and memory caching 
servers) by up to 38 times, and reduce the throughput of the servers down to 13\%.
 
Despite the severity of the attacks, neither current cloud providers nor research 
literature offer any solutions to \attackname. Our communication with cloud 
providers suggests such issues are not currently addressed, in part because the 
attack techniques presented in this paper are non-conventional, and existing 
solutions to network-based DDoS attacks do not help. Research studies have not 
explored defenses against adversarial memory contention either. As will be 
discussed in \secref{sec:defense_past}, existing solutions \cite{DeKo:13, ZhLaMa:14, 
YaBrMa:13, ZhBlFe:10, AhKiHa:12} only aim 
to enhance performance isolation between benign applications. Intentional memory 
abuses that are evident in \attackname are immune to these solutions.

Therefore, a large portion of this paper is devoted to the design and implementation of a novel and effective approach to detect and mitigate all known types of \attackname with low-cost overhead. Our detection strategy provides a generalized method for detecting deviations from the baseline behavior of the victim VM due to \attackname. We collect the baseline behaviors of the monitored VM at runtime, by creating a \emph{pseudo isolated period}, without completely pausing co-tenant VMs. This provides periodic (re)establishment of baseline behaviors that adapt to changes in program phases and workload characteristics. Once \attackname are detected, we show how malicious VMs can be identified and their attacks mitigated, using a novel form of selective execution throttling.

We implemented a prototype of our defense solution on the opensource OpenStack cloud 
software, and extensively evaluated its effectiveness and efficiency.  Our
evaluation shows that we can accurately detect
\attackname and promptly and effectively mitigate the attacks. The performance
overhead of persistent performance monitoring is lower than 5\%, which is low
enough to be used in production public clouds.  Because our solution does not 
require modifications of CPU hardware, hypervisor or guest operating systems, it
minimally impacts the existing cloud implementations. Therefore, we envision our
solution can be rapidly deployed in public clouds as a new security service to
customers who require higher security assurances (like in Security-on-Demand clouds \cite{JaSzPe:13, ZhLe:15}).

In summary, the contributions of this paper are: 

\begin{packeditemize}

\item  A set of attack techniques to perform \attackname. Measurement of the 
severity of the resulting Degradation of Service (DoS) to the victim VM.

\item Demonstration of the severity of \attackname in public clouds (Amazon 
EC2) against Hadoop applications and E-commerce websites. 

\item A novel, generalizable, attack detection method to detect abnormal 
probability distribution deviations at runtime, that adapts to program phase 
changes and different workload inputs.

\item A novel method for detecting the attack VM using selective execution throttling.

\item A new, rapidly deployable, defense for all \attackname with accurate detection and low-overhead, using existing hardware processor features.

\end{packeditemize}

We first discuss our threat model 
and background of memory resources in \secref{sec:overview}. Techniques to 
perform \attackname are presented in \secref{sec:separate_attack}. We 
show the power of these attacks in two case studies conducted in Amazon EC2 
in \secref{sec:ec2}. Our new defense techniques are described and evaluated in
\secref{sec:discuss}. We summarize related work in \secref{sec:related} and
conclude in \secref{sec:conclusion}.

\section{Background}
\label{sec:overview}

\subsection{Threat Model and Assumptions}
\label{sec:threat_model}

We consider security threats from malicious tenants of public IaaS clouds.  We 
assume the adversary has the ability to launch at least one VM on the cloud 
servers on which the victim VMs are running. Techniques required to do so have 
been studied~\cite{RiTrSh:09, Varadarajan:2015:PVS, Xu:2015:MSC}, and are orthogonal to
our work. The adversary can run any program inside his own VM. We do not assume that the 
adversary can send network packets to the victim directly, thus resource freeing 
attacks~\cite{VaKoFa:12} or network-based DoS attacks~\cite{Li:10} are not 
applicable. We do not consider attacks from the cloud providers, or any attacks 
requiring direct control of privileged software.

\begin{figure}[t]
\centerline{\mbox{\includegraphics[width=0.8\linewidth]{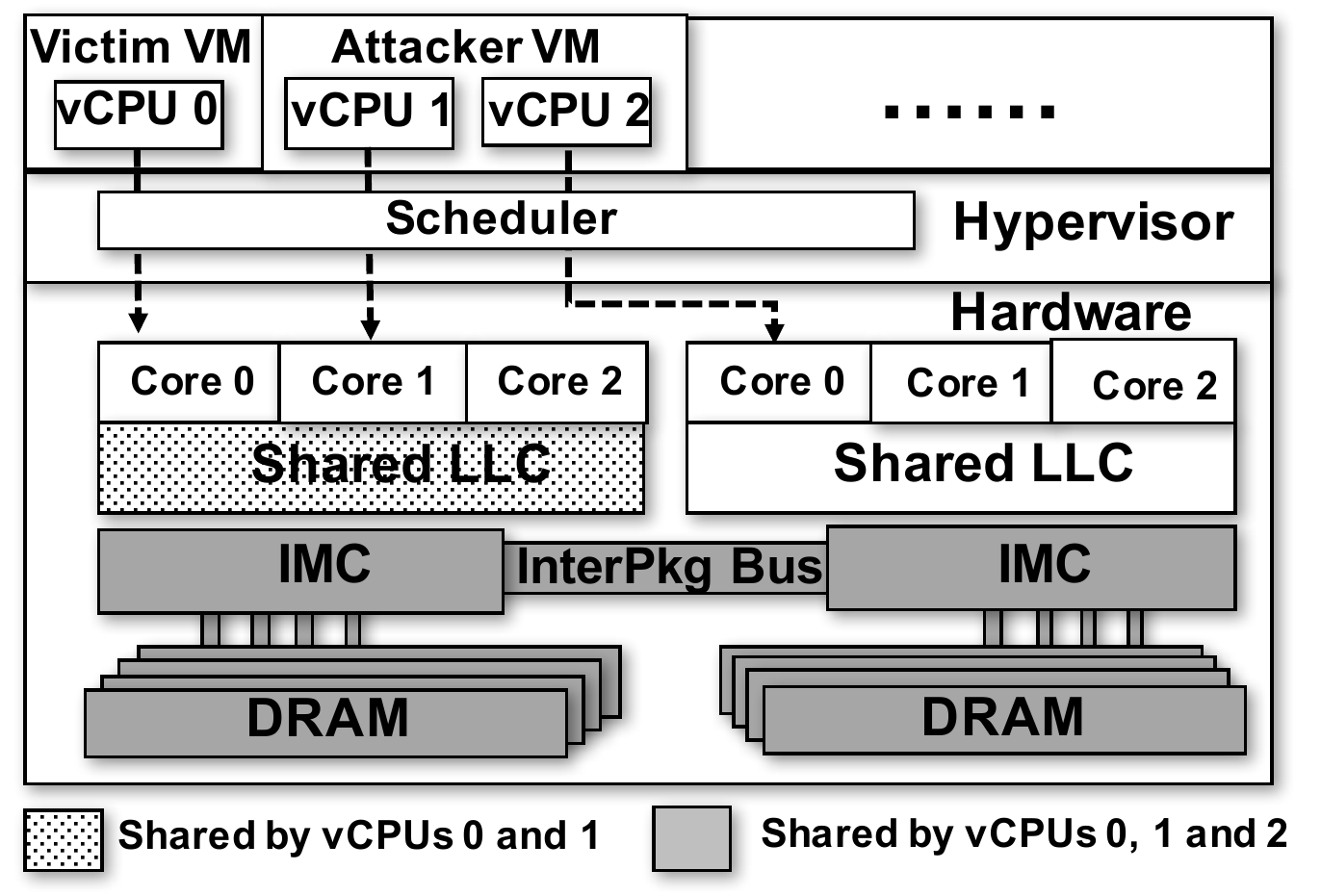}}}
\caption{An attacker VM (with 2 vCPUs) and a victim VM share multiple layers of memory resources.}
\label{fig:sharing}
\end{figure}

We assume the software and hardware isolation mechanisms function correctly as 
designed. A hypervisor virtualizes and manages the hardware resources (see Figure 
\ref{fig:sharing}) so that each VM thinks it has the entire computer. A server can 
have multiple processor packages, where all processor cores in a package share a 
Last Level Cache (LLC), while L1 and L2 caches are private to a processor core and not 
shared by different cores. All processor packages share the Integrated Memory 
Controller (IMC), the inter-package bus and the main memory storage (DRAM 
chips). Each VM is designated a disjoint set of virtual CPUs (vCPU), which can be 
scheduled to operate on any physical cores based on the hypervisor's scheduling 
algorithms. A program running on a vCPU may use all the hardware resources 
available to the physical core it runs on. Hence, different VMs may simultaneously 
share the same hardware caches, buses, memory channels and DRAM bank buffers. We 
assume the cloud provider may schedule VMs from different customers on the same 
server (as co-tenants), but likely on different physical cores. As is the case 
today, software-based VM isolation by the hypervisor only isolates accesses to 
virtual and physical memory pages, but not to the underlying hardware memory 
resources shared by the physical cores.

\subsection{Hardware Memory Resources}
\label{sec:overview:resource}

\begin{figure*}[t]
\centerline{\mbox{\includegraphics[width=0.9\linewidth]{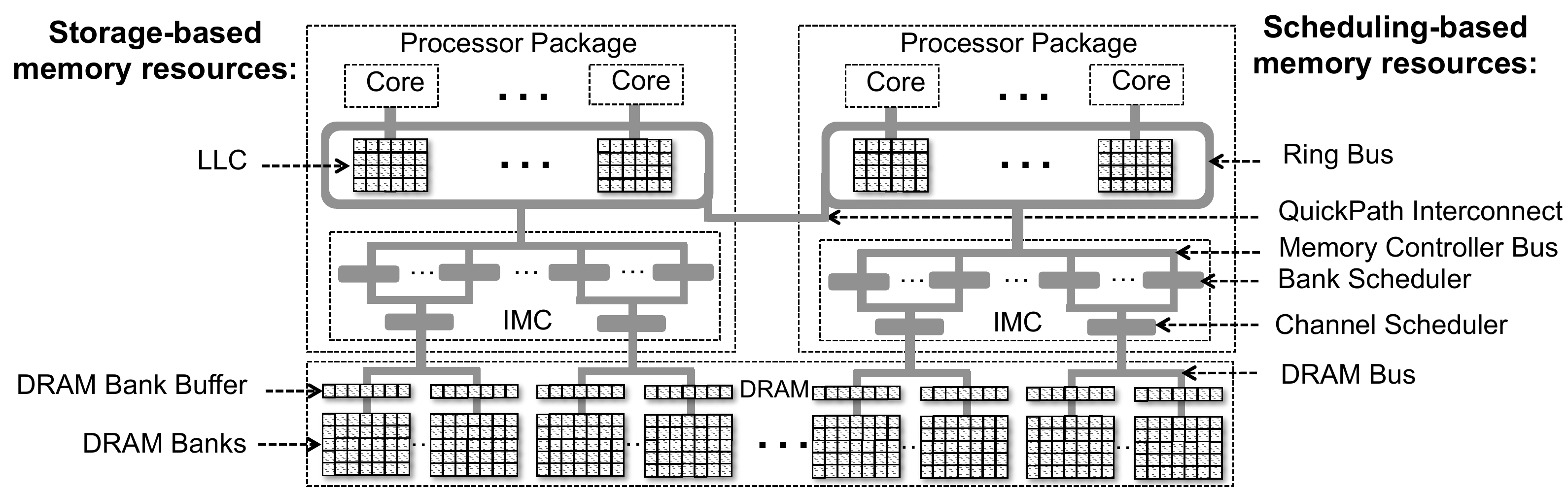}} }
\caption{Shared storage-based and scheduling-based hardware memory resources in multi-core cloud servers.}
\label{fig:memory_system}
\end{figure*}

Figure \ref{fig:memory_system} shows the hardware memory resources in modern computers. Using Intel processors as examples, modern X86-64 processors usually consist of multiple processor packages, each of which consists of several physical processor cores. Each physical core can execute one or two hardware threads in parallel with the support of Hyper-Threading Technology. A hierarchical memory subsystem, from the top to the bottom, is composed of different levels of \emph{storage-based} components (e.g., caches, the DRAMs). These memory components are inter-connected by a variety of \emph{scheduling-based} components (e.g., memory buses and controllers), with various schedulers arbitrating their communications. Memory resources shared by different cores are described below:

\bheading{Last Level Caches (LLC).}
An LLC is shared by all cores in one package (older processors may have one package supported by multiple LLCs). Intel LLCs usually adopt an inclusive cache policy: every cache line maintained in the upper-level caches (i.e., core-private Level 1 and Level 2 caches in each core - not shown in Figure \ref{fig:memory_system}) also has a copy in the LLC. In other words, when a cache line in the LLC is evicted, so are the copies in the upper-level caches. A subsequent access to the memory block mapped to this cache line will result in an LLC miss, which will lead to the much slower main memory access. On recent Intel processors (since Nehalem), LLCs are split into multiple slices, each of which is associated with one physical core, although every core may use the entire LLC. Intel employs static hash mapping algorithms to translate the physical address of a cache line to one of the LLC slices that contains the cache line. These mappings are unique for each processor model and are not released to the public. So it is harder for attackers to generate LLC contention using the method from \cite{WoLe:07}.

\bheading{Memory Buses.}
Intel uses a ring bus topology to interconnect components in the processor package, 
\eg, processor cores, LLC slices, Integrated Memory Controllers (IMCs), QuickPath 
Interconnect (QPI) agents, etc. The high-speed QPI provides point-to-point 
interconnections between different processor packages, and between each processor 
package and I/O devices. The memory controller bus connects the LLC slices to the 
bank schedulers in the IMC, and the DRAM bus connects the IMC's schedulers to the 
DRAM banks. Current memory bus designs with high bandwidth make it very difficult 
for attackers to saturate the memory buses. Also, elimination of bus locking
operations for normal atomic operations make bus locking attacks via normal atomic 
operations (e.g., \cite{WoLe:07}) less effective. However, some exotic atomic bus
locking operations still exist.

\bheading{DRAM banks.}
Each DRAM package consists of several banks, each of which can be thought of as a two dimensional data array with multiple rows and columns. Each bank has a bank buffer to hold the most recently used row to speed up DRAM accesses. A memory access to a DRAM bank may either be served in the bank buffer, which is a buffer-hit (fast), or in the bank itself, which is a buffer-miss (slow).

\bheading{Integrated Memory Controllers (IMC).}
Each processor package contains one or multiple IMCs. The communications between an IMC and the portion of DRAM it controls are supported by multiple memory channels, each of which serves a set of DRAM banks. When the processor wants to access the data in the DRAM, it first calculates the bank that stores the data based on the physical address, then it sends the memory request to the IMC that controls the bank. The processor can request data from the IMC in the local package, as well as in a different package via QPI. The IMCs implement a bank priority queue for each bank they serve, to buffer the memory requests to this bank. A bank scheduler is used to schedule requests in the bank priority queue, typically using a First-Ready-First-Come-First-Serve algorithm that gives high scheduling priority to the request that leads to a buffer-hit in the DRAM bank buffer, and then to the request that arrived earliest. Once requests are scheduled by the bank scheduler, a channel scheduler will further schedule them, among requests from other bank schedulers, to multiplex the requests onto a shared memory channel. The channel scheduler usually adopts a First-Come-First-Serve algorithm, which favors the earlier requests. Modern DRAM and IMCs can handle a large amount of requests concurrently, so the it is less effective to flood the DRAM and IMCs to generate severe contention, as shown in \cite{MoMu:07}.

\section{Memory DoS Attacks}
\label{sec:separate_attack}

\subsection{Fundamental Attack Strategies}
\label{sec:attack_strategies}
We have classified all memory resources into either storage-based or scheduling-based resources. This helps us formulate the following two fundamental attack strategies for \attackname:

\begin{packeditemize}

\item \textbf{Storage-based contention attack.} The fundamental attack strategy to cause contention on storage-based resources is to \emph{reduce the probability that the victim's data is found in an upper-level memory resource (faster), thus forcing it to fetch the data from a lower-level resource (slower).}

\item \textbf{Scheduling-based contention attack.} The fundamental attack strategy on a scheduling-based resource is to \emph{decrease the probability that the victim's requests are selected by the scheduler, e.g., by locking the scheduling-based resources temporarily, tricking the scheduler to improve the priority of the attacker's requests, or overwhelming the scheduler by submitting a huge amount of requests simultaneously}.

\end{packeditemize}

We systematically show how \attackname can be constructed on different layers of memory resources (LLC in \secref{sec:uncore_cache_contention}, bus in \secref{sec:bus_contention}, memory controller and DRAM in \secref{sec:IMC_contention}). For each memory component, we first study the basic techniques the attacker can use to generate resource contention and affect the victim's performance. We measure the effectiveness of the attack techniques on the victim VM with different vCPU locations and program features. Then we propose some practical attacks and evaluate their impacts on real-world benchmarks.

\bheading{Testbed configuration.}
To demonstrate the severity of different types of \attackname, we use a server configuration, representative of many cloud servers, configured as shown in Table \ref{table:testbed}. We use Intel processors, since they are the most common in cloud servers, but the attack methods we propose are general, and applicable to other processors and platforms as well.

\begin{table}[ht]
\centering
\caption{Testbed Configuration}
\label{table:testbed}
\resizebox{\linewidth}{!}{
\begin{threeparttable}
\begin{tabular}{|l|l|}
  \hline
  Server & Dell PowerEdge R720 \\
  \hline
  Processor Packages & Two 2.9\ghertz Intel Xeon E5-2667 (Sandy Bridge)\\
  \hline
  Cores per Package & 6 physical cores, or 12 hardware threads with Hyper-Threading\\
  \hline
  Core-private & L1 I and L1 D: each 32\kbytes, 8-way set-associative;\\
  Level 1 and Level 2 caches &L2 cache: 256\kbytes, 8-way set-associative\\
  \hline
  Last Level Cache (LLC)& 15\mbytes, 20-way set-associative, shared by cores in package, divided\\
   &into 6 slices of 2.5\mbytes each; one slice per core\\
  \hline
  Physical memory & Eight 8\gbytes DRAMs, divided into 8 channels, and 1024 banks\\
  \hline
  \hline
  Hypervisor & Xen version 4.1.0 \\
  \hline
  VM's OS & Ubuntu 12.04 Linux, with 3.13 kernel \\
  \hline
  \end{tabular}
  \end{threeparttable}}
\end{table}

In each of the following experiments, we launched two VMs, one as the attacker 
and the other as the victim. By default, each VM was assigned a single vCPU. We 
select a mix set of benchmarks for the victim: (1) We use a modified stream 
program \cite{stream_benchmark, MoMu:07} as a micro benchmark to explore the effectiveness 
of the attacks on victims with different features. This program 
allocates two array buffers with the same size, one as the source and the other 
as the destination. It copies data from the source to the destination in loops 
repeatedly, either in a sequential manner (resulting a program with \textit{high 
memory locality}) or in a random manner (\textit{low memory locality}). We chose 
this benchmark because it is memory-intensive and allows us to alter the size of 
memory footprints and the locality of memory resources. (2) To fully evaluate the 
attack effects on real-world applications, we choose 8 macro benchmarks (6 from 
SPEC2006~\cite{SPEC2006} and 2 from PARSEC~\cite{Bi:11}) and cryptographic 
applications based on OpenSSL as the victim program. Each experiment was repeated 
10 times, and the mean values and standard deviations are reported.

\subsection{Cache Contention (Storage Resources)}
\label{sec:uncore_cache_contention}

Of the storage-based contention attacks, we found that the LLC contention 
results in the most severe performance degradation. The root vulnerability is that 
an LLC is shared by all cores of the same CPU package, without access control or 
quota enforcement.  Therefore a program in one VM can evict LLC cache lines 
belonging to another VM. Moreover, inclusive LLCs (\eg, most modern Intel LLCs) 
will propagate these cache line evictions to core-private L1 and L2 caches, 
further aggravating the interference between programs (or VMs) in CPU caches. 

\subsubsection{Contention Study}

\bheading{Cache cleansing.}
To cause LLC contention, the adversary can allocate a memory buffer to cover the 
entire LLC. By accessing one memory address per memory block in the buffer, the 
adversary can cleanse the entire cache and evict all of the victim's data from the 
LLC to the DRAM.

The optimal buffer used by the attacker should {\em exactly map} to the LLC, which 
means it can fill up each cache set in each LLC slice without {\em self-conflicts} 
(\ie, evicting earlier lines loaded from this buffer). 
For example, for a LLC with $n^{s}$ slices, $n^{c}$ sets in each slice, and $n^w$-way 
set-associativity, the attacker would like $n^{s}\times n^{c}\times n^{w}$ memory blocks to 
cover all cache lines of all sets in all slices. 
There are two challenges that make this 
task difficult for the attacker: the host physical addresses of the buffer to 
index the cache slice are unknown to the attacker, and the mapping from physical 
memory addresses to LLC slices is not publicly known. 

{\em Mapping LLC cache slices:} 
To overcome these challenges, the attacker can first allocate a 1\gbytes Hugepage which is guaranteed to have continuous host physical addresses; thus he need not worry about virtual to physical page translations which he does not know. Then for each LLC cache set $\mathrm{S^i}$ in all slices, the attacker sets up an empty group $\mathbb{G}^i$, and starts the following loop: (i) select block $\mathrm{A^k}$ from the Hugepage, which is mapped to set $\mathrm{S^i}$ by the same index bits in the memory address; (ii) add $\mathrm{A^k}$ to $\mathbb{G}^i$; (iii) access all the blocks in $\mathbb{G}^i$; and (iv) measure the average access latency per block. A longer latency indicates block $\mathrm{A^k}$ causes self-conflict with other blocks in $\mathbb{G}^i$, so it is removed from $\mathbb{G}^i$. The above loop is repeated until there are $n^{s}\times n^{w}$ blocks in $\mathbb{G}^i$, which can exactly fill up set $\mathrm{S^i}$ in all slices. Next the attacker needs to distinguish which blocks in $\mathbb{G}^i$ belong to each slice: He first selects a new block $\mathrm{A^n}$ mapped to set $\mathrm{S^i}$ from the Hugepage, and adds it to $\mathbb{G}^i$. This should cause a self-conflict. Then he executes the following loop: (i) select one block $\mathrm{A^m}$ from $\mathbb{G}^i$; (ii) remove it from $\mathbb{G}^i$; (iii) access all the blocks in $\mathbb{G}^i$; and (iv) measure the average access latency. A short latency indicates block $\mathrm{A^m}$ can eliminate the self-conflict caused by $\mathrm{A^n}$, so it belongs to the same slice as $\mathrm{A^n}$. The attacker keeps doing this until he discovers $n^w$ blocks that belong to the same slice as $\mathrm{A^n}$. These blocks form the group that can fill up set $\mathrm{S^i}$ in one slice. The above procedure is repeated till the blocks in $\mathbb{G}^i$ are divided into $n^s$ slices for set $\mathrm{S^i}$. After conducting the above process for each cache set, the attacker obtains a memory buffer with non-consecutive blocks that map exactly to the LLC. Such self-conflict elimination is also useful in improving side-channel attacks \cite{YaLiGe:15}. 

\begin{figure}[ht]
\centerline{\mbox{\includegraphics[width=\linewidth]{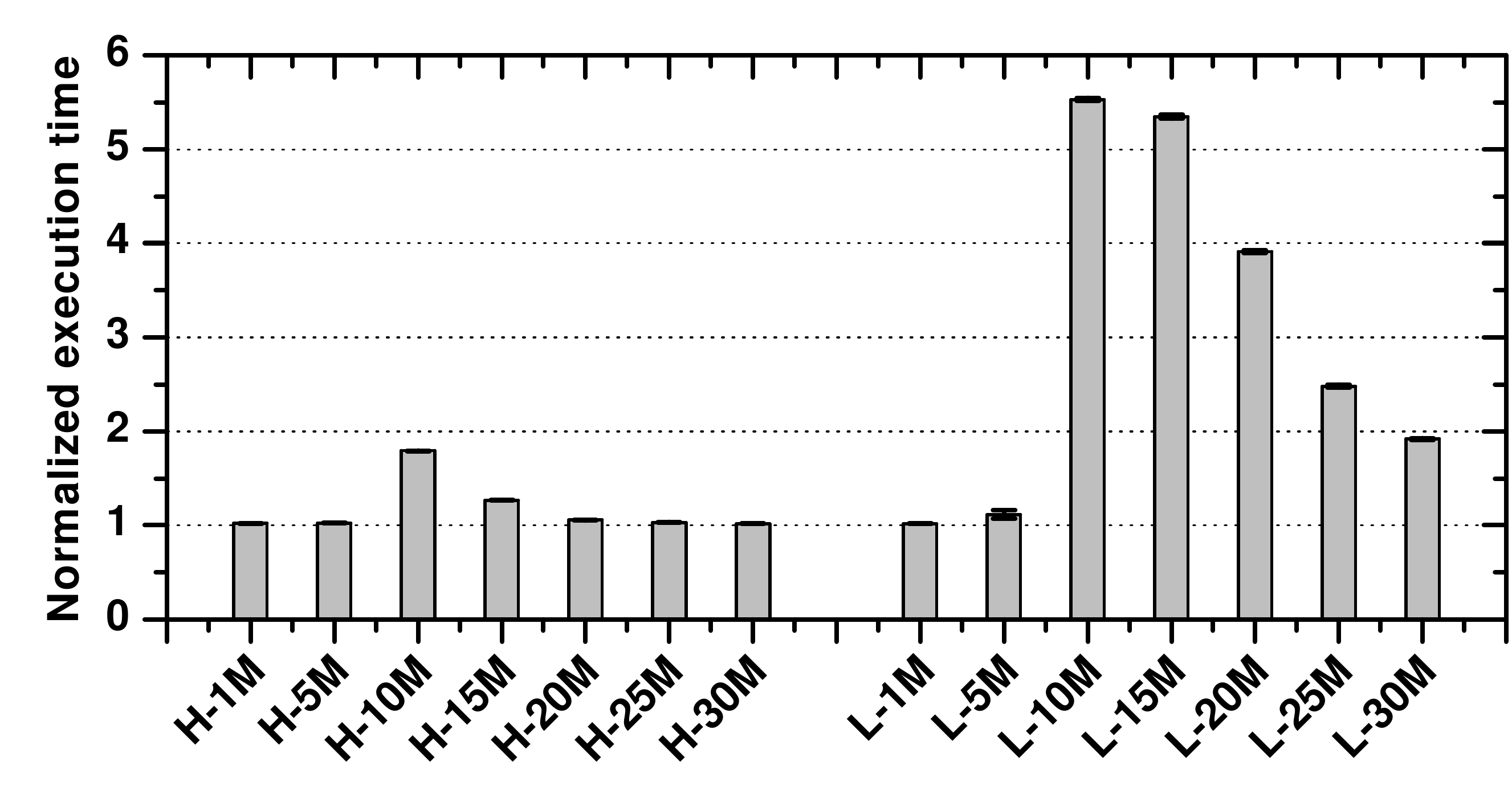}} }
\caption{Performance slowdown due to LLC cleansing contention. We use ``H-\emph{x}" or ``L-\emph{x}" to denote the victim program has high or low memory locality and has a buffer size of \emph{x}.}
\label{fig:llc_testing_perf}
\end{figure}

To test the effectiveness of cache cleansing, we arranged the attacker VM and 
victim VM on the same processor package, thus sharing the LLC and all memory 
resources in lower layers.  The adversary first identified the memory buffer that 
maps to the LLC. Then he cleansed the whole LLC repeatedly. The resulting 
performance degradation of the victim application is shown in Figure \ref{fig:llc_testing_perf}. The victim suffered from the most significant performance 
degradation when the victim's buffer size is around 10\mbytes (1.8$\times$ 
slowdown for the high locality program, and 5.5$\times$ slowdown for the low 
locality program).  When the buffer size is smaller than 5\mbytes (data are 
stored mainly in upper-level caches), or the size is larger than 25\mbytes (data 
are stored mainly in the DRAM), the impact of cache contention on LLC is 
negligible. 

The results can be explained as follows: the maximum performance degradation 
can be achieved on victims with memory footprint smaller than, but
close to, the LLC size, which is 15\mbytes, because
the victim suffers the least from self conflicts in 
LLC and the most from the attacker's LLC cleansing. Moreover, as a low locality 
program accesses its data in a random order, hardware prefetching is less effective in enhancing the program's access speed. So the program accesses the cache at a relatively lower rate. Its data will be evicted out of the LLC by the attacker with higher probability.
That is why the LLC cleansing has a larger impact on low locality programs
than on high locality programs.

\bheading{Takeaways.} 
LLC contention is (more) effective when (1) the attacker and victim VMs share
the same LLC, (2) the victim program's memory footprint is about the
size of LLC, and (3) the victim program has lower memory locality.

\subsubsection{Practical Attack Evaluation}
We improve this attack by increasing the cleansing speed, and the accuracy 
of evicting (thus contending with) the victim's data. 

\bheading{Multi-threaded LLC cleansing.}
To speed up the LLC cleansing, the adversary may split the cleansing task into $n$ 
threads, with each running on a separate vCPU and cleansing only a non-overlapping 
$1/n$ of the LLC simultaneously. This effectively increases the cleansing speed by 
$n$ times.  

\begin{figure}[ht]
\centerline{\mbox{\includegraphics[width=\linewidth]{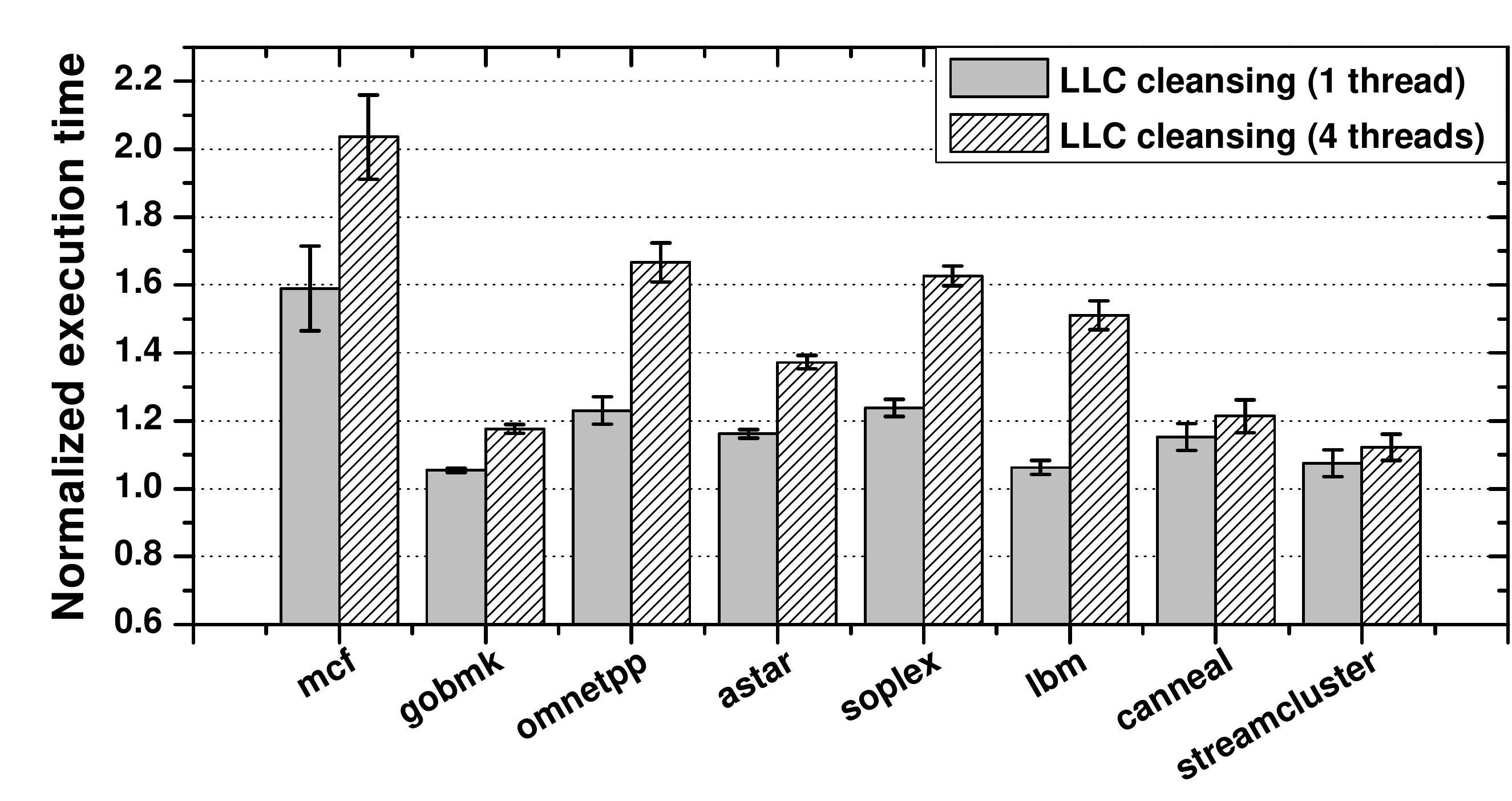}} }
\caption{Performance slowdown due to multi-threaded LLC cleansing attack}
\label{fig:llc_mem_eval}
\end{figure}

In our experiment, the attacker VM and the victim VM were arranged to share the
LLC. The attacker VM was assigned 4 vCPUs.  It first prepared the memory buffer
that exactly mapped to the LLC. Then he cleansed the LLC with (1) one vCPU; (2)
4 vCPUs (each cleansing 1/4 of the LLC). Figure \ref{fig:llc_mem_eval} shows
that the attack can cause $1.05\sim1.6\times$ slowdown to the victim VM when using 
one thread, and $1.12\sim2.03\times$ slowdown when using four threads.

\bheading{Adaptive LLC cleansing.}
The basic LLC cache cleansing technique does not work when the victim's program 
has a memory footprint ($<$1\mbytes) that is much smaller than an LLC (\eg, 
15\mbytes), since it takes a long time to finish one complete LLC cleansing, where 
most of the memory accesses do not induce contention with the victim. To achieve 
finer-grained attacks, we developed a cache probing technique to pinpoint the 
cache sets in the LLC that map to the victim's memory footprint, and cleanse 
only these selected sets. 

The attacker first allocates a memory buffer covering the entire LLC in his own
VM. Then he conducts cache probing in two steps: (1) In the \textsc{Discover
Stage}, while the victim program runs, for each cache set, the attacker accesses
some cache lines belonging to this set and figures out the maximum number of
cache lines which can be accessed without causing cache conflicts. If this number 
is smaller than the set associativity, this cache set will be selected to conduct
adaptive cleansing attacks, because the victim has frequently occupied some cache 
lines in this set; (2) In the \textsc{Attack Stage}, the attacker keeps accessing 
these selected cache sets to cleanse the victim's data. Algorithm \ref{alg:algorithm2} shows the steps to perform the adaptive LLC
cleansing. 

\begin{algorithm}[t]
\scriptsize
\SetAlgoLined
 \KwIn{}
 \Indp 
   cache\_set\{\}: all the sets in the LLC \\
         cache\_buffer\{\}: cover the entire LLC \\
         cache\_assoc\_num: the associativity of LLC \\
\Indm
\Begin{

  /* \textsc{Discover Stage} */\\
  victim\_set=$\emptyset$ \\
  \For {\em each set \emph{i} in cache\_set\{\}}{
    Find out \emph{j}, \emph{s.t.}, accessing \emph{j} cache lines in set \emph{i} from cache\_buffer\{\} has no cache conflict (low accessing time), but accessing \emph{j}+1 cache lines in set \emph{i} from cache\_buffer\{\} has cache conflict (high accessing time) \\
    \If{\em \emph{j}$<$cache\_assoc\_num}{
      add \emph{i} to victim\_set\{\} \\
    }
  }
  \BlankLine
  \BlankLine
  /* \textsc{Attack Stage}: */ \\
  \While{\em attack is not finished}{
    \For{\em each set \emph{i} in victim\_set\{\}}{
      access cache\_assoc\_num cache lines in set \emph{i} from cache\_buffer\{\}
    }
  }
}
 \caption{Adaptive LLC cleansing}
\label{alg:algorithm2}
\end{algorithm}

\begin{figure}[ht]
\centerline{\mbox{\includegraphics[width=\linewidth]{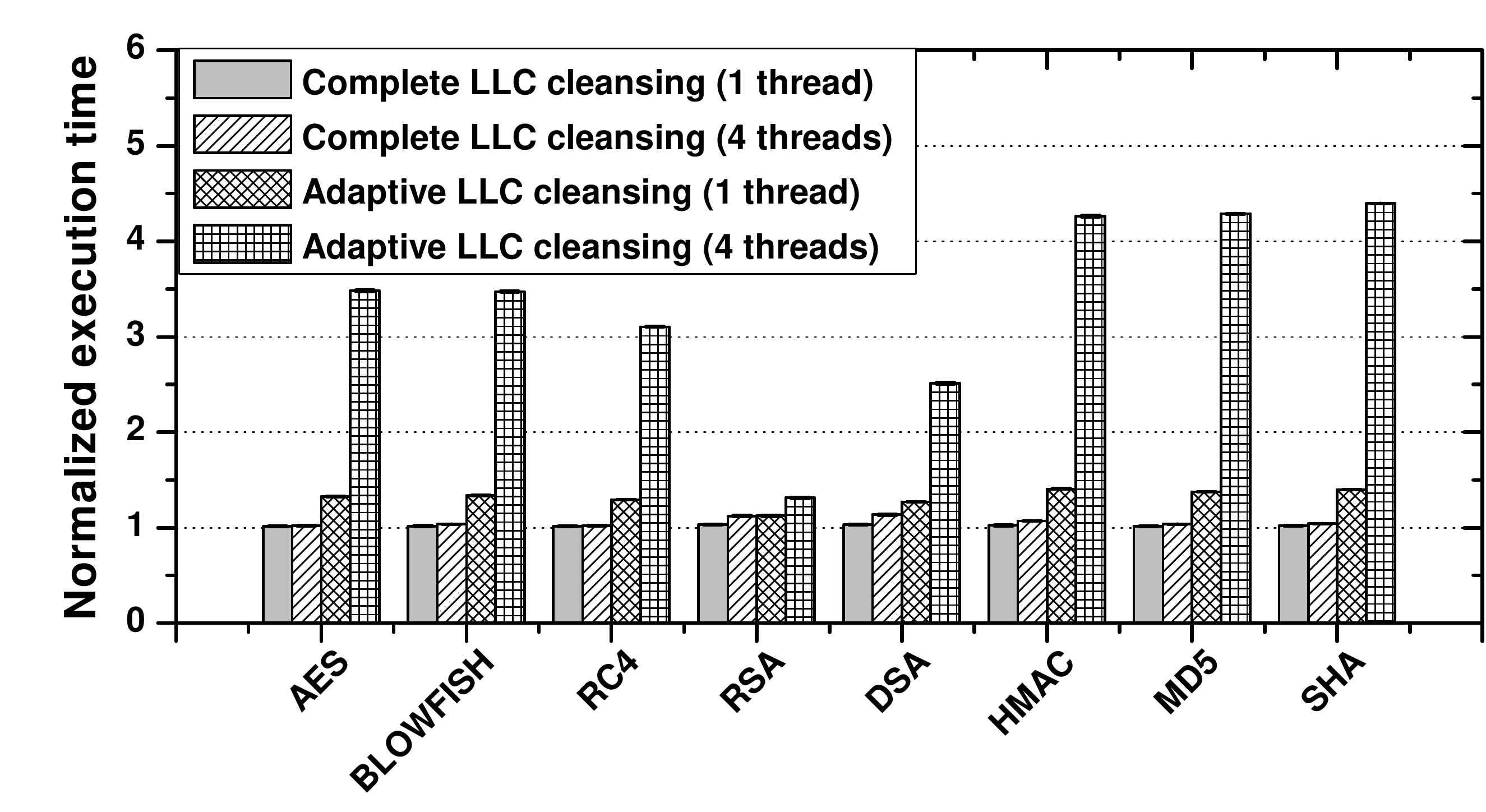}} }
\caption{Performance slowdown due to adaptive LLC cleansing attacks}
\label{fig:adaptive_llc}
\end{figure}

Figure \ref{fig:adaptive_llc} shows the results of the attacker's multi-threaded 
adaptive cleansing attacks against victim applications with cryptographic operations. 
While the basic cleansing did not have any effect, the adaptive attacks can achieve 
around 1.12 to 1.4 times runtime slowdown with 1 vCPU, and up to 4.4$\times$ slowdown 
with 4 vCPUs.

\subsection{Bus Contention (Scheduling Resources)}
\label{sec:bus_contention}

The availability of internal memory buses can be compromised by overwhelming or temporarily locking down the buses. We study the effects of these techniques. 

\subsubsection{Contention Study}

\bheading{Bus saturation.}
One intuitive approach for an adversary is to create numerous memory requests to 
saturate the buses \cite{WoLe:07}. However, the bus bandwidth in modern 
processors may be too high for a single VM to saturate.

\begin{figure}[t]
     \centering
     \subfloat[][Same package]{
     \includegraphics[width=0.48\linewidth]{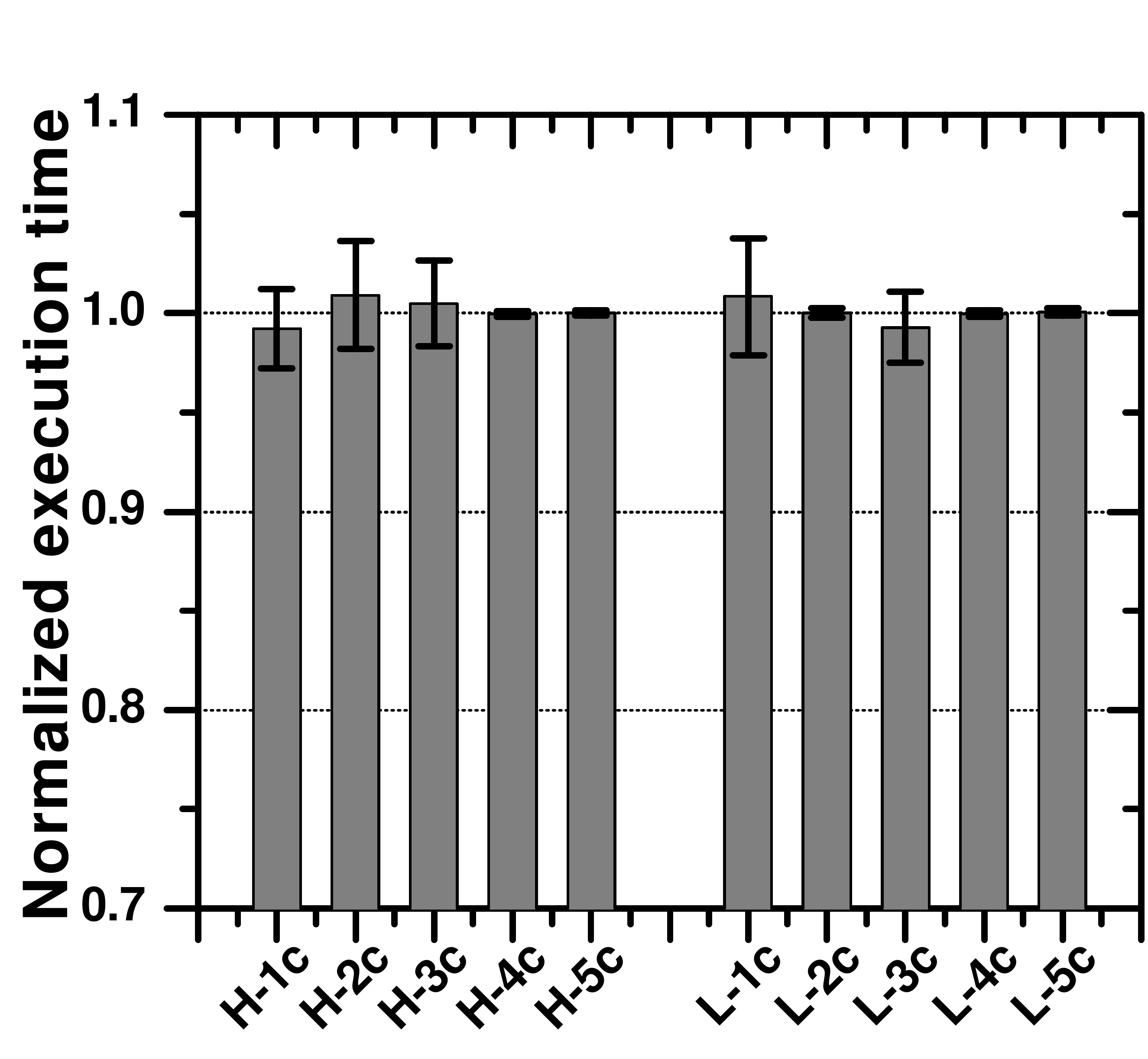}
     \label{fig:bus_saturate_perf1}}
     \subfloat[][Different packages]{
     \includegraphics[width=0.48\linewidth]{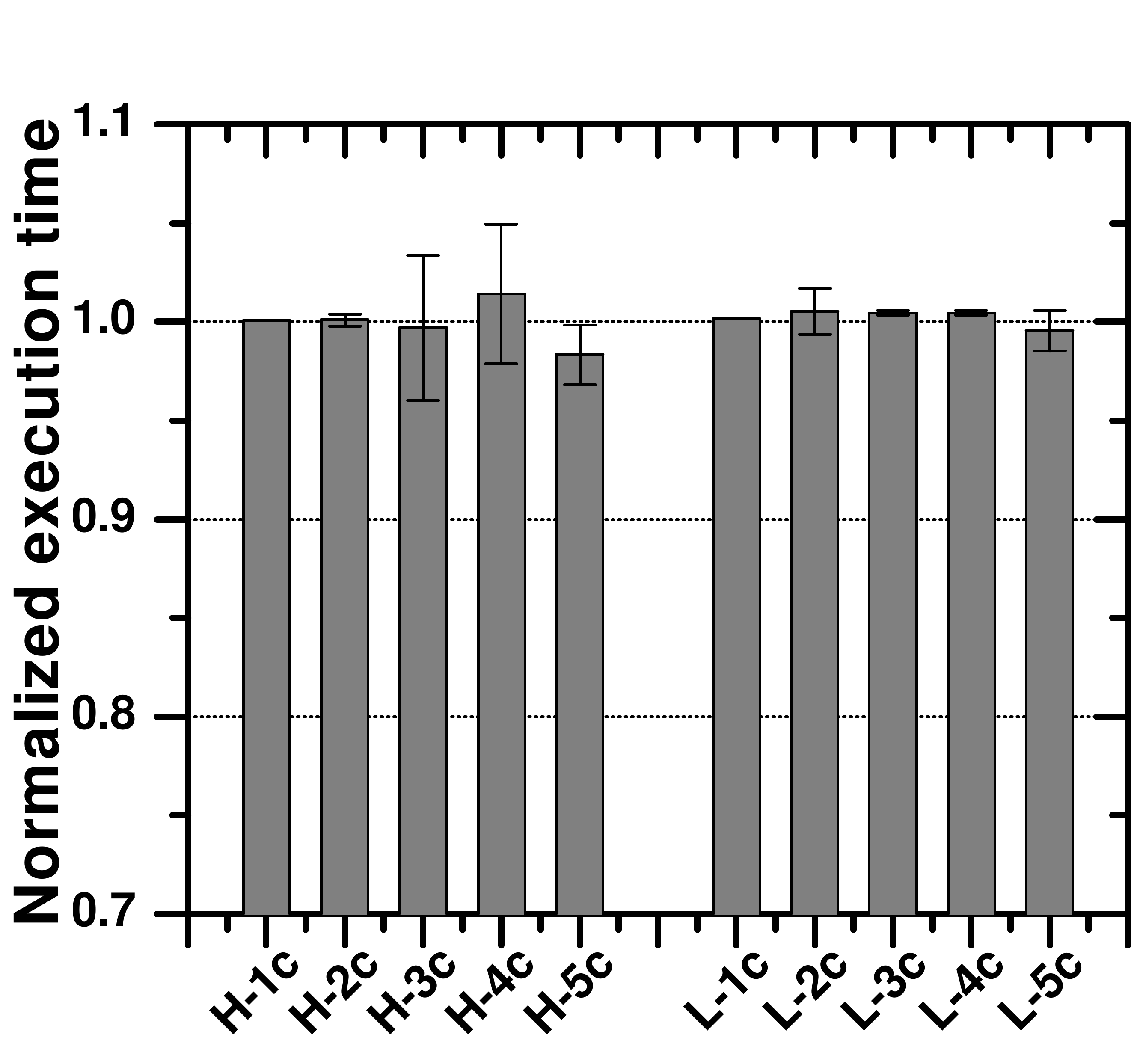}
     \label{fig:bus_saturate_perf2}}
    \caption[The LOF caption]{Performance slowdown due to bus saturation contention. We use
``H-\emph{x}c" or ``L-\emph{x}c" to denote the configuration that the
victim program has high or low locality, and both of the attacker and victim use \emph{x} cores to contend for the bus.}
    \label{fig:bus_saturate_perf}
\end{figure}

To examine the effectiveness of \emph{bus saturation contention}, we conducted two sets of 
experiments. In the first set of
experiments, the victim VM and the attacker VM were located in the
same processor package but on different physical cores (Figure \ref{fig:bus_saturate_perf1}). They accessed
different parts of the LLC, without touching the DRAM. Therefore the
attacker VM causes contention in the ring bus that connects LLC
slices without causing contention in the LLC itself. In the second
set of experiments, the victim VM and the attacker VM were pinned on
different processor packages (Figure~\ref{fig:bus_saturate_perf2}). 
They accessed different memory channels,
without inducing contention in the memory controller and DRAM
modules. Therefore the attacker and victim VMs only contend in buses that
connect LLCs and IMCs, as well as the QPI buses.  The attacker and victim were
assigned increasing number of vCPUs to cause more bus traffic. 
Results in Figure \ref{fig:bus_saturate_perf} show that these buses 
were hardly saturated and the impact on the victim's performance was negligible
in all cases.

\bheading{Bus locking.}
To deny the victim from being scheduled by a scheduling resource, the adversary 
can temporarily lock down the internal memory buses. 
Intel processors provide locked atomic operations for managing shared data 
structures between multi-processors \cite{intel_manual}. Before Intel Pentium (P5) 
processors, the locked atomic operations always generate LOCK signals on the 
internal buses to achieve operation atomicity. So other memory accesses are blocked
until the locked atomic operation is completed. For processor families after P6, 
the bus lock is transformed into a cache lock: the cache line is locked instead
of the bus and the cache coherency mechanism is used to ensure operation atomicity.
This causes much smaller scheduling lockdown times.

However, we have found two exotic atomic operations the adversary can still use to 
lock the internal memory buses: (1) \emph{Locked atomic accesses to unaligned 
memory blocks}: the processor has to fetch two adjacent cache lines to complete 
this unaligned memory access. To guarantee the atomicity of accessing the two 
adjacent cache lines, the processors will flush in-flight memory accesses issued 
before, and block memory accesses to the bus, until the unaligned memory access is 
finished. (2) \emph{Locked atomic accesses to uncacheable memory blocks}: when 
uncached memory pages are accessed in atomic operations, the cache coherency 
mechanism does not work. Hence, the memory bus must be locked to guarantee atomicity. 
Listings \ref{lst:unaligned} and \ref{lst:uncached} show the codes for issuing unaligned 
and uncached atomic operations. The two programs keep conducting the addition 
operation of a constant (\texttt{x}) and a memory block (\texttt{block\_addr}) 
(line 5 -- 11): in line 7, the \texttt{lock} prefix indicates this operation is atomic. 
The instruction \texttt{xaddl} indicates this is an addition operation. The first operand
is the register \texttt{eax}, which stores \texttt{x} (line 9). The second operand is 
the first parameter of line 9 (data denoted by the address \texttt{block\_addr}). The 
results will be loaded to the register \texttt{eax} (line 8). 
In Listing \ref{lst:unaligned}, we set this memory block as unaligned (line 
4). In Listing \ref{lst:uncached}, we added a new system call to set the page 
table entries of the memory buffer as cache disabled (line 2).

\begin{figure*}[t]
  \centering
  \begin{minipage}{.46\textwidth}
\begin{lstlisting}[caption={Attack using unaligned atomic operations}, label={lst:unaligned}]
  char *buffer = mmap(0, BUFFER_SIZE, PROT_READ|PROT_WRITE, MAP_PRIVATE|MAP_ANONYMOUS, -1, 0);

  int x = 0x0;
  int *block_addr = (int *)(buffer+CACHE_LINE_SIZE-1);
  while (1) {
    __asm__(
      "lock; xaddl %%eax, %1\n\t"
      :"=a"(x)
      :"m"(*block_addr), "a"(x)
      :"memory"); 
  }
\end{lstlisting}  
\end{minipage}
  \begin{minipage}{.46\textwidth}
\begin{lstlisting}[caption={Attack using uncached atomic operations}, label={lst:uncached}]
  char *buffer = mmap(0, BUFFER_SIZE, PROT_READ|PROT_WRITE, MAP_PRIVATE|MAP_ANONYMOUS, -1, 0);
  syscall(__NR_UnCached, (unsigned long)buffer);
  int x = 0x0;
  int *block_addr = (int *)buffer;
  while (1) {
    __asm__(
    "lock; xaddl %%eax, %1\n\t"
    :"=a"(x)
    :"m"(*block_addr), "a"(x)
    :"memory"); 
  }
\end{lstlisting}
  \end{minipage}
  \vspace{-15pt}
\end{figure*}

\begin{figure*}[t]
     \centering
     \subfloat[][Same package]{
     \includegraphics[width=0.48\linewidth]{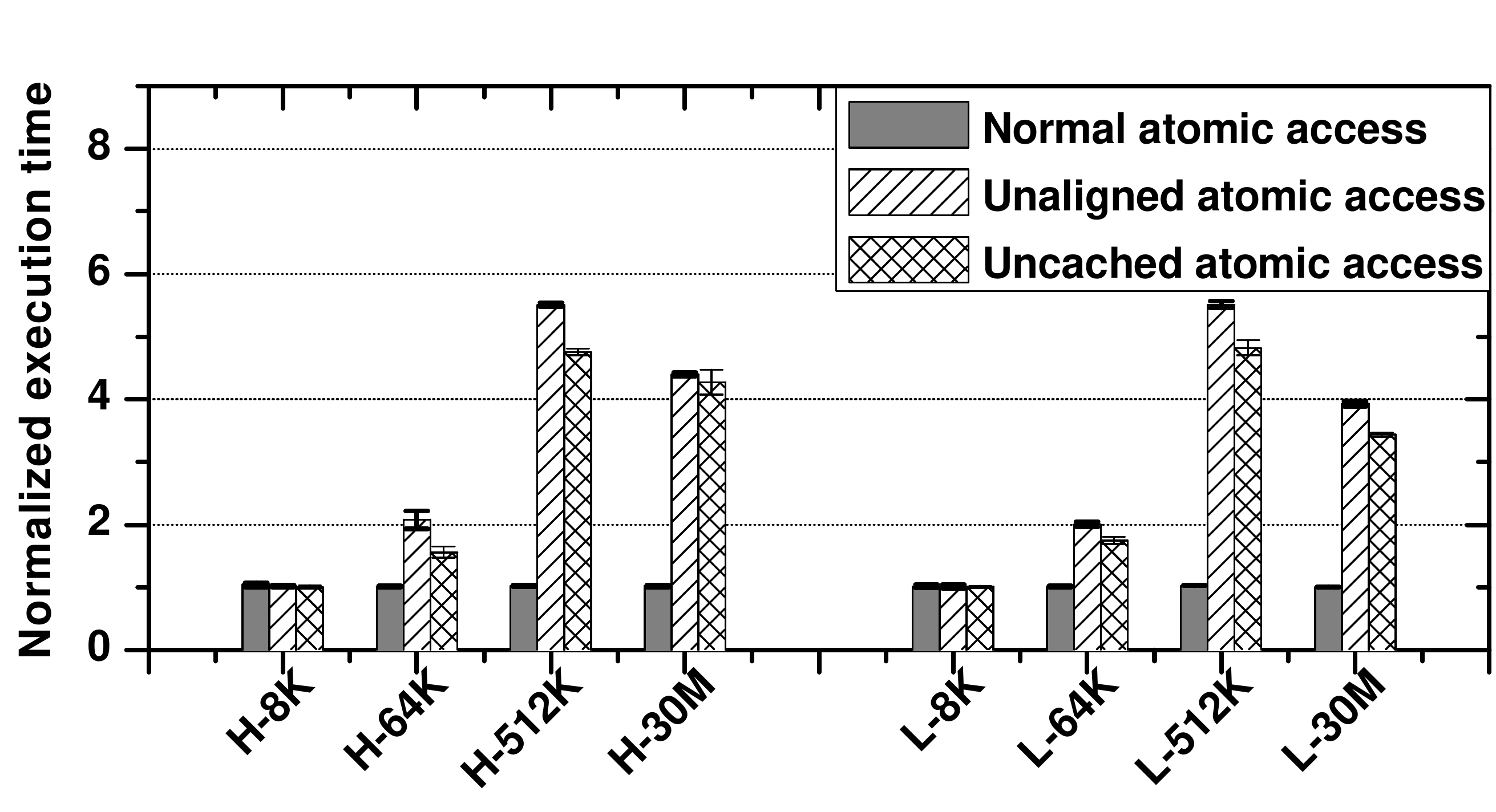}
     \label{fig:bus_testing_perf}}
     \subfloat[][Different packages]{
     \includegraphics[width=0.48\linewidth]{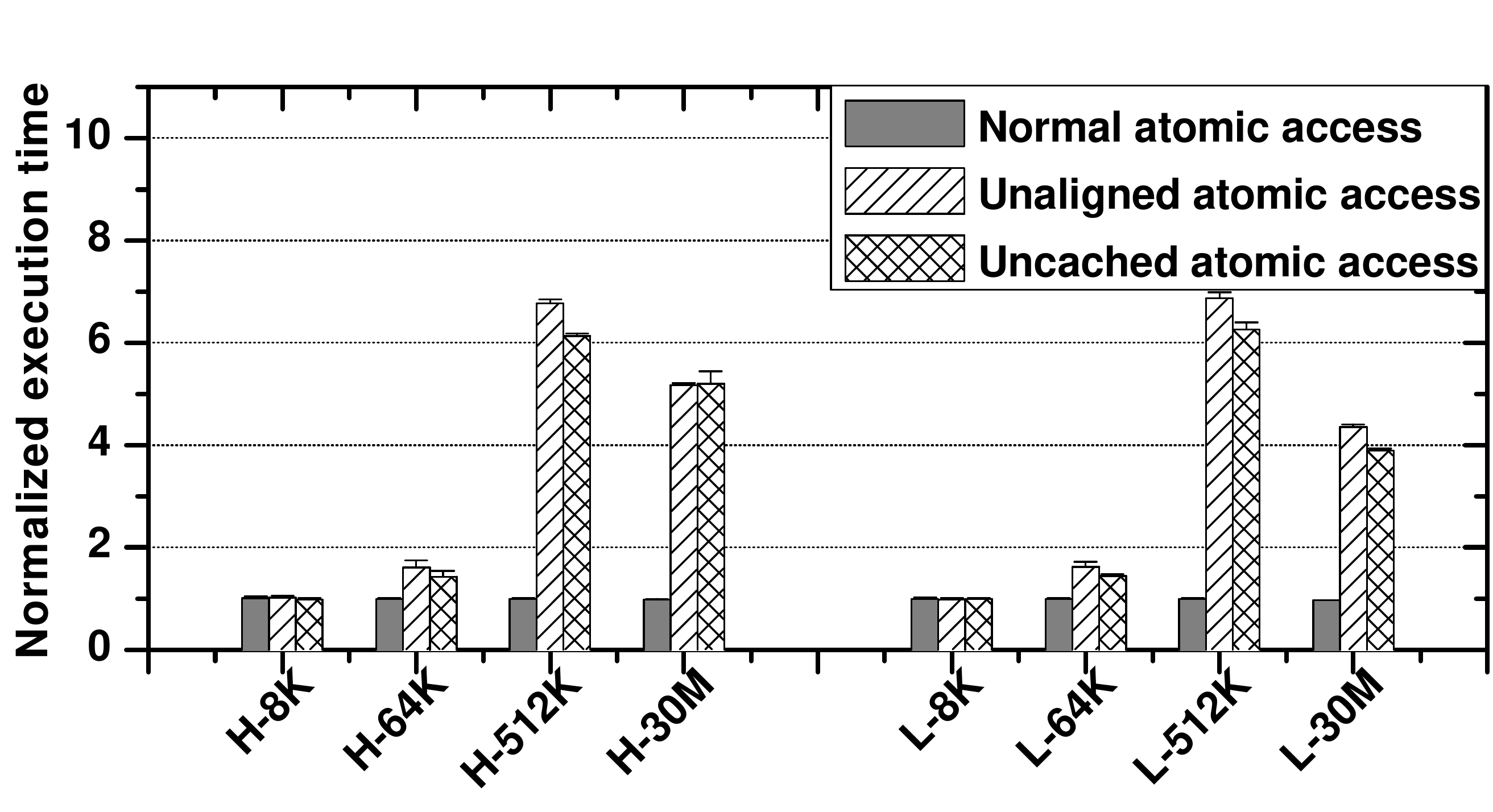}
     \label{fig:bus_testing_perf1}}
    \caption[The LOF caption]{Performance slowdown due to bus locking contention. We use ``H-\emph{x}" or ``L-\emph{x}" to denote the victim program has high or low memory locality and has a buffer size of \emph{x}.}
    \label{fig:bus_testing11}
\end{figure*}

To evaluate the effects of \emph{bus locking contention}, we 
chose the footprint size of the victim program as (1) 8\kbytes, with which 
the L1 cache was under-utilized, (2) 64\kbytes, with which the 
L1 cache was over-utilized but the L2 cache was under-utilized, 
(3) 512\kbytes, with which the L2 cache was over-utilized but the LLC
was under-utilized, and (4) 30\mbytes, with which the LLC was over-utilized.
The attacker VM kept issuing unaligned atomic or uncached atomic memory accesses 
to lock the memory buses. For comparison, we 
also run another group of experiments, where the attacker kept issuing normal locked memory 
accesses. We considered two scenarios: (1) the attacker and victim
shared the same processor package, but run on different 
cores; (2) they were scheduled on different processor packages. The
normalized execution time of the victim program is shown in Figure
\ref{fig:bus_testing11}.

We observe that the victim's performance was significantly affected when the its buffer 
size was larger than the L2 caches. This is because the attacker who kept requesting atomic,
unaligned memory accesses was only able to lock the buses within its physical cores, the
ring buses around the LLCs in each package, the QPI, and the buses from each
package to the DRAM. So when the victim's buffer size was smaller than
LLC, it fetched data from the private caches in its own core without
being affected by the attacker.
However, when the victim's buffer size was larger than the L2 caches, its access to the LLC
would be delayed by the bus locking operations, and the performance is degraded
(up to $6\times$ slowdown for high locality victim programs and
$7\times$ slowdown for low locality victim programs).

\bheading{Takeaways.} We explored two approaches to bus contention.
Saturating internal buses is unlikely to cause noticeable performance
degradation. Bus locking shows promise when the victim program makes
heavy use of the shared LLC or lower layer memory resources, whenever 
the victim VM and attacker VM are on the same processor package or 
different packages.

\subsubsection{Practical Attack Evaluation}
\begin{figure}[ht]
\centerline{\mbox{\includegraphics[width=\linewidth]{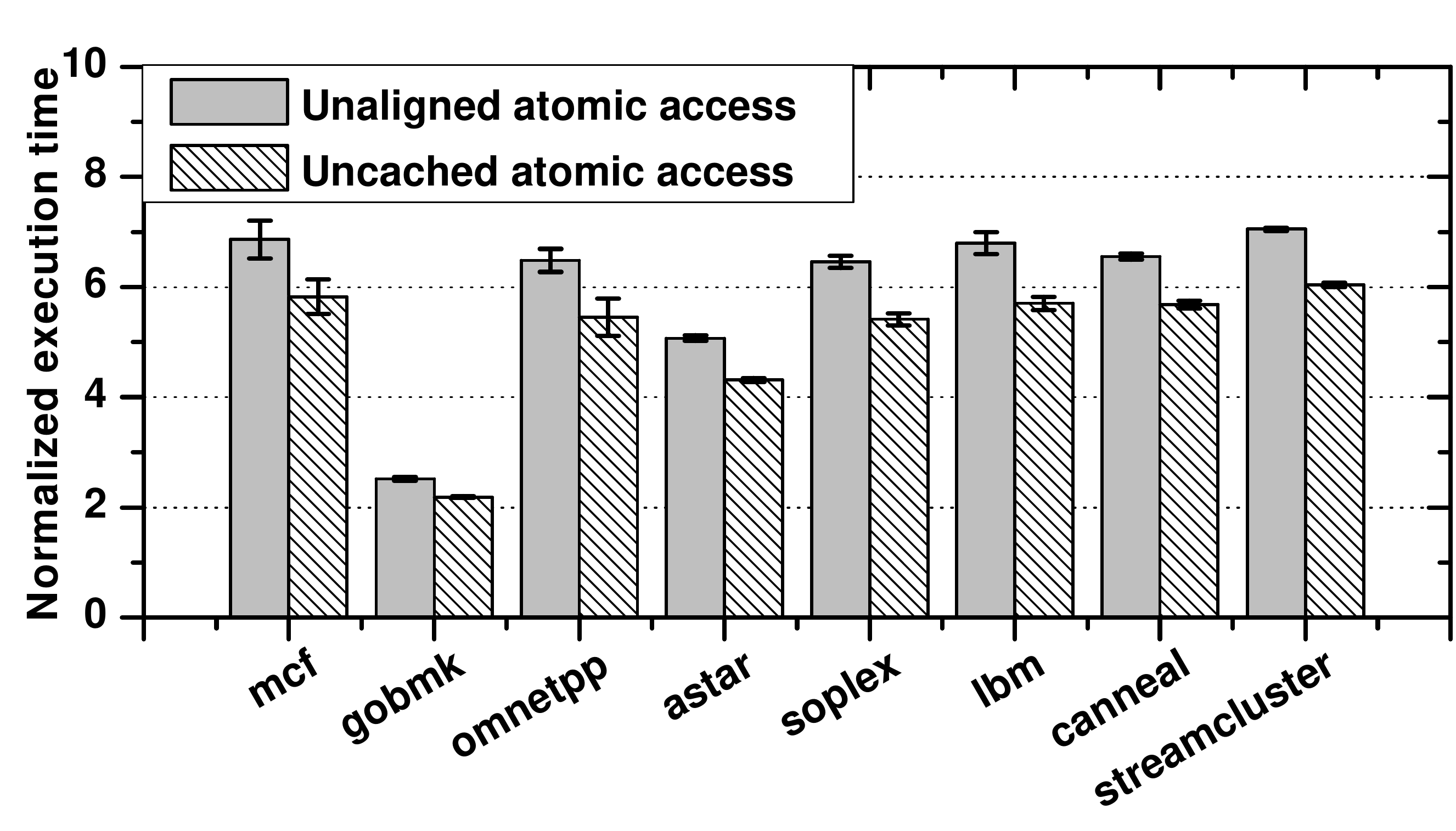}} }
\caption{Performance slowdown due to bus locking attacks.}
\label{fig:bus_lock}
\end{figure}

To evaluate the effectiveness of \emph{atomic locking attacks} on real-world applications, 
we scheduled the attacker VM and victim VM on different processor packages. The attacker 
VM kept generating atomic locking signals by (1) requesting unaligned atomic memory 
accesses, or (2) requesting uncached atomic memory accesses. The normalized execution time 
of the victim program is shown in Figure \ref{fig:bus_lock}. We observe that the 
victim's performance can be degraded as much as 7 times when the attacker conducted
exotic atomic operations.

\subsection{Memory Contention (Combined Resources)}
\label{sec:IMC_contention}

An IMC uses the bank scheduler and channel scheduler to select the memory requests for each DRAM access. Therefore an adversary may contend on these two schedulers by frequently issuing memory requests that result in bank buffer hits to boost his priority in the scheduler. Moreover, each memory bank is equipped with only one bank buffer to hold the recently used bank row, so the adversary can easily induce storage-based contention on bank buffers by frequently occupying them with his own data.

\subsubsection{Contention Study}

\bheading{Memory flooding.}
Since channel and bank schedulers use First-Come-First-Serve policies, an attacker can send a large amount of memory requests to flood the target memory channels or DRAM banks. These requests will contend on the scheduling-based resources with the victim's memory requests. In addition, the attacker can issue requests in sequential order to achieve high row-hit locality and thus high priority in the bank scheduler, to further increase the effect of flooding. Furthermore, when the adversary keeps flooding the IMCs, these memory requests can also evict the victim's data out of the DRAM bank buffers. The victim's bank 
buffer hit rate is decreased and its performance is further degraded.

\begin{figure}[h]
\centerline{\mbox{\includegraphics[width=\linewidth]{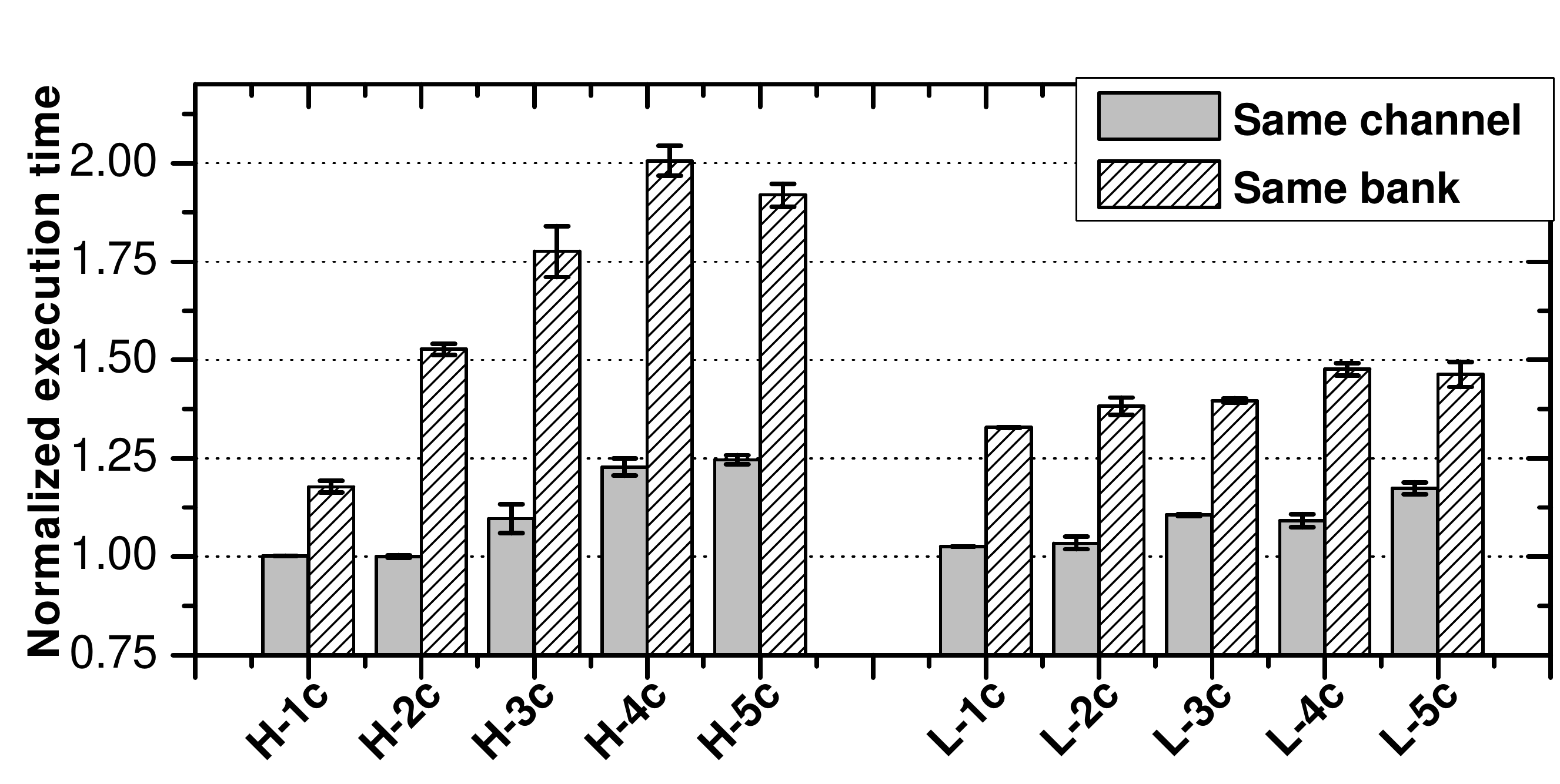}} }
\caption{Performance degradation due to memory channel and bank
contention. We use ``H-\emph{x}c" or ``L-\emph{x}c" to denote the configuration that the
victim program has high or low locality, and both of the attacker and victim use \emph{x} cores to contend for the bus.}
\label{fig:dram_testing_perf}
\end{figure}

To demonstrate the effects of DRAM contention, we configure one attacker VM to operate 
a memory flooding program, which kept accessing memory 
blocks in the same DRAM bank directly without going through 
caches (\ie, uncached accesses). The victim VM did exactly the same
with either high or low memory locality. We conducted
two sets of experiments: (1) The two VMs access the same bank in the same
channel (Same bank in Figure~\ref{fig:dram_testing_perf}); (2) the two VMs access two different banks
in the same channel (Same channel in Figure~\ref{fig:dram_testing_perf}). 
To alter the memory request rate issued by the two VMs, we also changed 
the number of vCPUs in the attacker and victim VMs. The normalized execution 
time of the victim program is shown in Figure \ref{fig:dram_testing_perf}. 

Three types of contention were observed in these experiments. First,
channel scheduling contention was observed when the attacker and the
victim access different banks in the same channel. It was enhanced
with increased number of attacker and victim vCPUs, thus increasing the
memory request rate (around $1.2\times$ slowdown for ``H-5c" and
``L-5c"). Second, bank scheduling contention was also observed when
the attacker and victim accessed the same DRAM bank. When the memory
request rate was increased, the victim's performance was further
degraded by an additional 70\% and 25\% for ``H-5c"
and ``L-5c", respectively. Third, contention in DRAM bank buffers
was observed when we compare the results of ``Same bank'' in Figure~\ref{fig:dram_testing_perf}
between high locality and low
locality victim programs ---low locality victims already suffer from
row-misses and the additional performance degradation in high locality
victims is due to bank buffer contention ($1.9\times$ slowdown for
H\_5c verses $1.45\times$ slowdown for
L\_5c).

\begin{figure}[h]
\centerline{\mbox{\includegraphics[width=\linewidth]{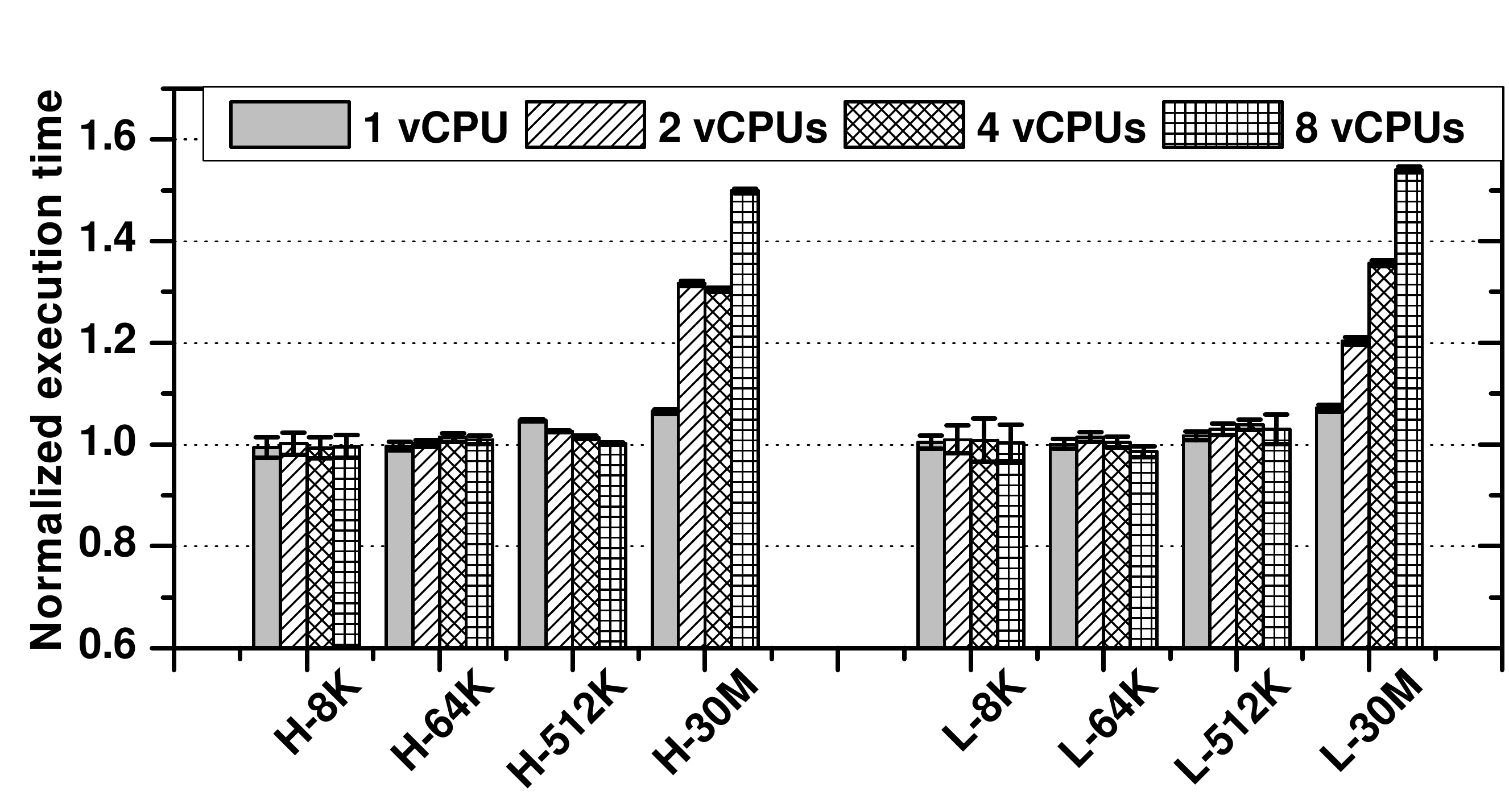}} }
\caption{Performance degradation due to memory flooding contention. We use ``H-\emph{x}" or ``L-\emph{x}" to denote the victim program has high or low memory locality and has a buffer size of \emph{x}.}
\label{fig:dram_testing_perf1}
\end{figure}

We consider the overall effect of memory flooding contention. 
In this experiment, the victim VM runs a high locality or low 
locality stream benchmark on its only vCPU. The attacker VM allocates 
a memory buffer with the size $20\times$ that of the LLC and runs a 
stream program which keeps accessing memory blocks sequentially 
in this buffer to generate contention in every channel and every bank. 
To increase bus traffic, the attacker employed multiple vCPUs to perform 
the attack simultaneously. The performance degradation, as we can 
see in Figure \ref{fig:dram_testing_perf1}, was significant when the victim's
memory accesses footprint was mostly in the DRAM, and more vCPUs of the attacker VM
were used in the attack.
The attacker can use 8 vCPUs to induce about $1.5\times$ slowdown to 
the victim with the buffer size larger than LLC.

\bheading{Takeaways.}
Contention can be induced in channel schedulers, bank schedulers and
bank buffers between different programs from different processor packages.
This contention is especially significant when the victim program's memory
footprint is larger than the LLC.

\subsubsection{Practical Attack Evaluation}
We evaluate two advanced memory flooding attacks.

\begin{figure}[t]
\centerline{\mbox{\includegraphics[width=\linewidth]{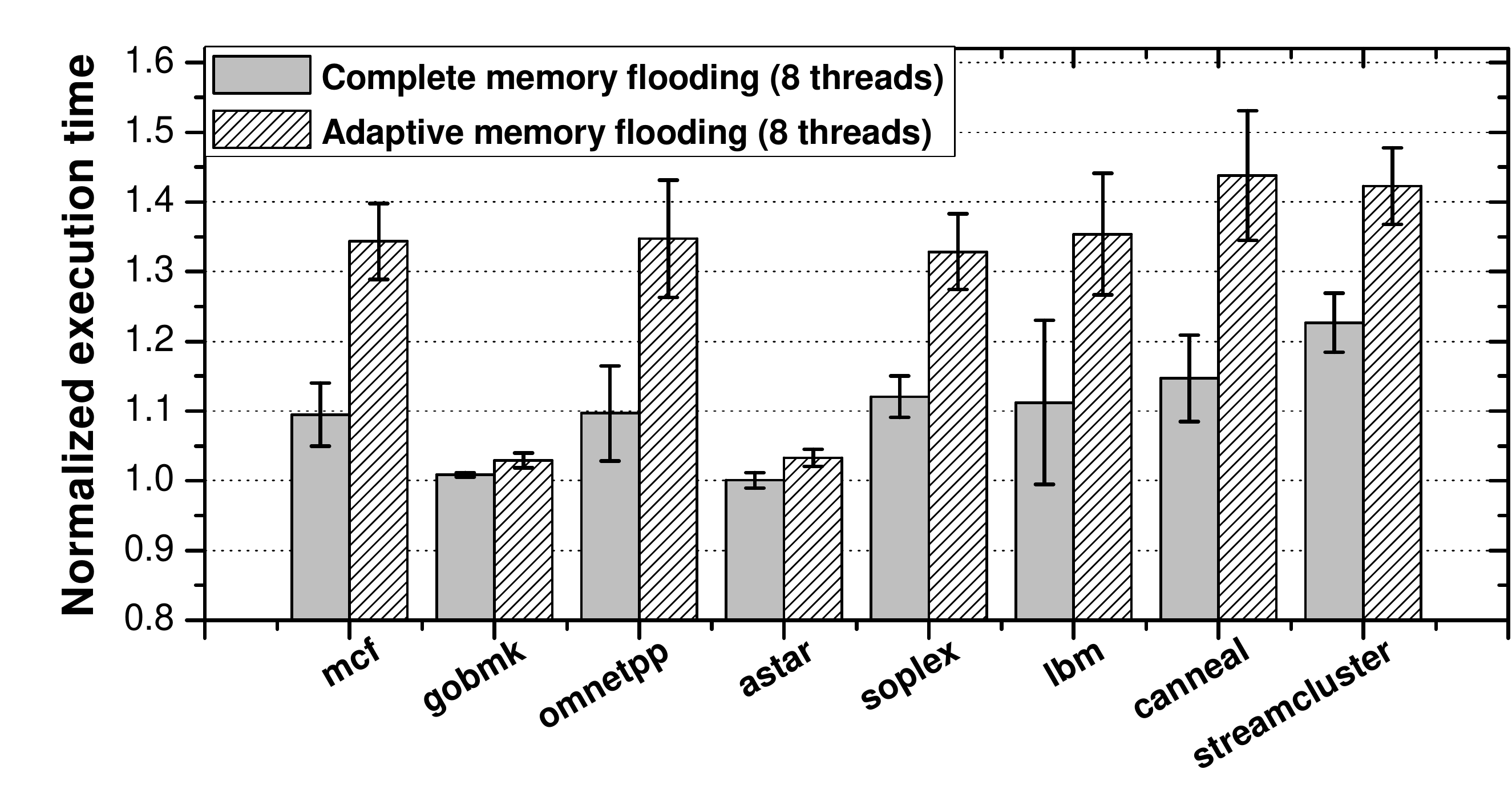}}}
\caption{Performance overhead due to multi-threaded and adaptive memory flooding attacks.}
\label{fig:adaptive_mem} 
\end{figure}

\bheading{Multi-threaded memory flooding.}
The attacker can use more threads to increase the memory flooding speed, as we demonstrated in the previous section. We evaluated this attack on real-world applications. The attacker and victim VMs are located in two different processor packages, so they only share the IMCs and DRAM. The attacker VM issues frequent, highly localized memory requests to flood every DRAM bank and every channel. To increase bus traffic, the attacker employed 8 vCPUs to perform the attack simultaneously. Figure \ref{fig:adaptive_mem}  shows that the victim experiences up to a $1.22\times$ runtime slowdown when the attacker uses 8 vCPUs to generate contention (Complete Memory Flooding bars).

\bheading{Adaptive memory flooding.}
For a software program with smaller memory footprint, only a few memory channels will be involved in its memory accesses. We developed a \emph{novel} approach with which an adversary may identify memory channels that are more frequently used by a victim program. To achieve this, the attacker needs to reverse engineer the unrevealed algorithms that map the physical memory addresses to memory banks and channels, in order to accurately direct the flows of the memory request flood.

{\em Mapping DRAM banks and channels:} The attacker can leverage methods due to Liu et al. \cite{LiCuXi:12} to identify the bits in physical memory addresses that index the DRAM banks. The attacker first allocates a 1\gbytes Hugepage with continuous physical addresses, which avoids the unknown translations from guest virtual addresses to machine physical addresses. Then he selects two memory blocks from the Hugepage whose physical addresses differ in only one bit. He then flushes these two blocks out of caches and accesses them from the DRAM alternatively. A low latency indicates these two memory blocks are served in two banks as there is no contention on bank buffers. In this way, the attacker is able to identify all the bank bits. Next, the attacker needs to identify the channel bits among the bank bits. We design a \textit{novel} algorithm which is shown in Algorithm \ref{alg:channel_index} to achieve this goal. The attacker selects two groups of memory blocks from the Hugepage, whose bank indexes differ in only one bit. The attacker then allocates two threads to access the two groups simultaneously. If the different bank index bit is also a channel index bit, then the two groups will be in two different channels, and a shorter access time will be observed since there is no channel contention. 

\begin{algorithm}
\scriptsize
\SetAlgoLined
 \KwIn{}
 \Indp bank\_bit\{\} // bank index bits\\
          memory\_buffer\{\} // a memory buffer\\
\Indm \KwOut{}
\Indp channel\_bit\{\} \\
\Indm
\Begin{
  channel\_bit\{\}=$\emptyset$ \\
  \For{\emph{each bit} i $\in$ \emph{bank\_bit}\{\}} {
    buffer\_A\{\}=memory\_buffer\{\} \\
    buffer\_B\{\}=memory\_buffer\{\} \\
    \For{\emph{each memory block} d\_a $\in$ \emph{buffer\_A}\{\}}{
            \emph{m\_a} = physical address of \emph{d\_a}\\
              \If{\emph{(}m\_a\emph{'s bit} i\emph{)} $\neq$ 0}{
          delete \emph{d\_a} from buffer\_A\{\} \\ 
          \textbf{break} \\
      }
      \For{\emph{each bit} j $\in$ \emph{bank\_bit}\{\} \textbf{\emph{and}} i $\neq$ j} {   
        \If{\emph{(}m\_a\emph{'s bit} j\emph{)} $\neq$ 0}{
          delete \emph{d\_a} from buffer\_A\{\} \\ 
          \textbf{break} \\
        }
      }
    }
    \For{\emph{each memory block} d\_b $\in$ \emph{buffer\_B}\{\} }{
            \emph{m\_b} = physical address of \emph{d\_b}\\
              \If{\emph{(}m\_b\emph{'s bit} i\emph{)} $\neq$ 1}{
          delete \emph{d\_b} from buffer\_B\{\} \\ 
          \textbf{break} \\
      }
      \For{\emph{each bit} j $\in$ \emph{bank\_bit}\{\} \textbf{\emph{and}} i $\neq$ j} {   
        \If{\emph{(}m\_b\emph{'s bit} j\emph{)} $\neq$ 0}{
          delete \emph{d\_b} from buffer\_B\{\} \\ 
          \textbf{break} \\
        }
      }
    }
    thread\_A: // access \emph{buffer\_A} in an infinite loop\\
    \While{\textbf{\emph{(true)}}}{
      \For{\emph{each memory block} d\_a $\in$ \emph{buffer\_A}\{\}}{
        access \emph{d\_a} (uncached)\\
      }
    }
    thread\_B:// access \emph{buffer\_B} N times and measure time\\
    \For{i=0 \emph{to} N-1}{
      \For{\emph{each memory block} d\_b $\in$ \emph{buffer\_B}\{\}}{
        access \emph{d\_b} (uncached)\\
      }
    }
    total\_time = thread\_B's execution time;\\
    \If{\emph{total\_time}$<$Threshold} {
      add \emph{i} to channel\_bit\{\}
    } 
  }
    \KwRet{\emph{channel\_bit}\{\}}
}
 \caption{Discovering channel index bits}
 \label{alg:channel_index}
\end{algorithm}

Then the attacker performs two stages: in the \textsc{Discover Stage}, the attacker keeps accessing each memory channel for a number of times and measures his own memory access time to infer contention from the victim program. By identifying the channels with a longer access time, the attacker can detect which channels are heavily used by the victim.  Note that the attacker only needs to discover the channels used by the victim, but does not need to know the exact value of channel index bits for a given channel. In the \textsc{Attack Stage}, the attacker floods these selected memory channels. Algorithm \ref{alg:avail:algorithm3} shows the two steps to conduct adaptive memory flooding attacks. 

\begin{algorithm}[t]
\scriptsize
\SetAlgoLined
 \KwIn{}
 \Indp 
   memory\_channel\{\}: all the channels in the memory \\
         memory\_buffer\{\} \\
\Indm
\Begin{
  /* \textsc{Discover Stage} */ \\
  victim\_channel=$\emptyset$ \\
  \For {\em each channel \emph{i} in memory\_channel\{\}}{
    access the addresses belonging to channel \emph{i} from memory\_buffer\{\}, and measure the total time (repeat for a number of times) \\
    \If{\em total time is high}{
      add \emph{i} to victim\_channel\{\} \\
    }
  }
  \BlankLine
  \BlankLine
  /* \textsc{Attack Stage} */ \\
  \While{\em attack is not finished}{
    \For{\em each channel \emph{i} in victim\_channel\{\}}
    {
      access the addresses belonging to channel \emph{i} from memory\_buffer\{\}
    }
  }
}
 \caption{Adaptive memory flooding}
  \label{alg:avail:algorithm3}
\end{algorithm}

Figure~\ref{fig:adaptive_mem} shows the results when the attacker VM uses 8 vCPUs to generate contention in selected memory channels which are heavily used by the victim. These adaptive memory flooding attacks cause $3\%\sim44\%$ slowdown while indiscriminately flooding the entire memory causes only $0.07\sim22\%$ slowdown.

\section{Case Studies in Amazon EC2}
\label{sec:ec2}

We now evaluate our \attackname in a real cloud environment, Amazon EC2. We 
provide two case studies: \attackname against distributed applications, and 
against E-Commerce websites. 

\bheading{Legal and ethical considerations.}
As our attacks only involve memory accesses within the attacker VM's own address 
space, the experiments we conducted in this section conformed with EC2 customer 
agreement. Nevertheless, we put forth our best efforts in 
reducing the duration of the attacks to minimally impact other users in the cloud.

\bheading{VM configurations.}
We chose the same configuration for the attacker and victim VMs: t2.medium 
instances with 2 vCPUs, 4\gbytes memory and 8\gbytes disk. Each VM ran Ubuntu 
Server 14.04 LTS with Linux kernel version 3.13.0-48-generic, in full 
virtualization mode. All VMs were launched in the us-east-1c region. Information 
exposed through \texttt{lscpu} indicated that these VMs were running on 2.5\ghertz
Intel Xeon E5-2670 processors, with a 32\kbytes L1D and L1I cache, a 256\kbytes L2 
cache, and a shared 25\mbytes LLC. 

For all the experiments in this section, the attacker employs exotic atomic 
locking (\secref{sec:bus_contention}) and LLC cleansing attacks 
(\secref{sec:uncore_cache_contention}), where each of the 2 
attacker vCPUs was used to keep locking the memory and cleansing the LLC. 
Memory contention attacks (Section \ref{sec:IMC_contention}) are not
used since they cause much lower performance degradation (availability loss)
to the victim.

\bheading{VM co-location in EC2.}
The \attackname require the attacker and victim VMs to co-locate on the same 
machine. Past work \cite{RiTrSh:09, Varadarajan:2015:PVS, 
Xu:2015:MSC} have proven the feasibility of such co-location attacks in public
clouds. While cloud providers adopt new technologies (e.g., Virtual Private Cloud \cite{amazon_vpc}) to 
mitigate prior attacks in \cite{RiTrSh:09}, new ways are discovered to test and detect 
co-location in \cite{Varadarajan:2015:PVS, Xu:2015:MSC}. Specifically, Varadarajan et al. 
\cite{Varadarajan:2015:PVS} achieved co-location in Amazon EC2, Google Compute Engine and 
Microsoft Azure with low-cost (less than \$8) in the order of minutes. They verified
co-location with various VM configurations, launch delay between attacker and victim, 
launch time of day, datacenter location, \etc. Xu et al . \cite{Xu:2015:MSC} used similar 
ideas to achieve co-location in EC2 Virtual Private Cloud. We also applied these techniques 
to achieve co-location in Amazon EC2. In our experiments, we simultaneously launched a large 
number of attacker VMs in the same region as the victim VM. A machine outside EC2 under our 
control sent requests to static web pages hosted in the target victim VM. Each time we 
select one attacker VM to conduct \attackname and measure the victim VM's response latency. 
Delayed HTTP responses from the victim VM indicates that this attacker was sharing the 
machine with the victim.

\subsection{Attacking Distributed Applications}
\label{sec:distribute}

We evaluate \attackname on a multi-node distributed application deployed in a 
cluster of VMs, where each VM is deployed as one node. We show how much performance 
degradation an adversary can induce to the victim cluster with minimal cost, using a single co-located attacker VM. 

\bheading{Experiment settings.}
We used Hadoop as the victim system. Hadoop consists of two layers: MapReduce for 
data processing, and Hadoop Distributed File System (HDFS) for data storage. A 
Hadoop cluster includes a single master node and multiple slave nodes. The master 
node acts as both the Job Tracker for scheduling map or reduce jobs and the 
NameNode for hosting HDFS indexes. Each slave node acts as both the Task Tracker 
for conducting the map or reduce operations and the DataNode for storing data 
blocks in HDFS.  We deployed the Hadoop system with different numbers of VMs (5, 
10, 15 or 20), where one VM was selected as the master node and the rest were the 
slave nodes. 

\begin{figure*}[t]
     \centering
     \subfloat[][MRBench]{
     \includegraphics[width=0.24\linewidth]{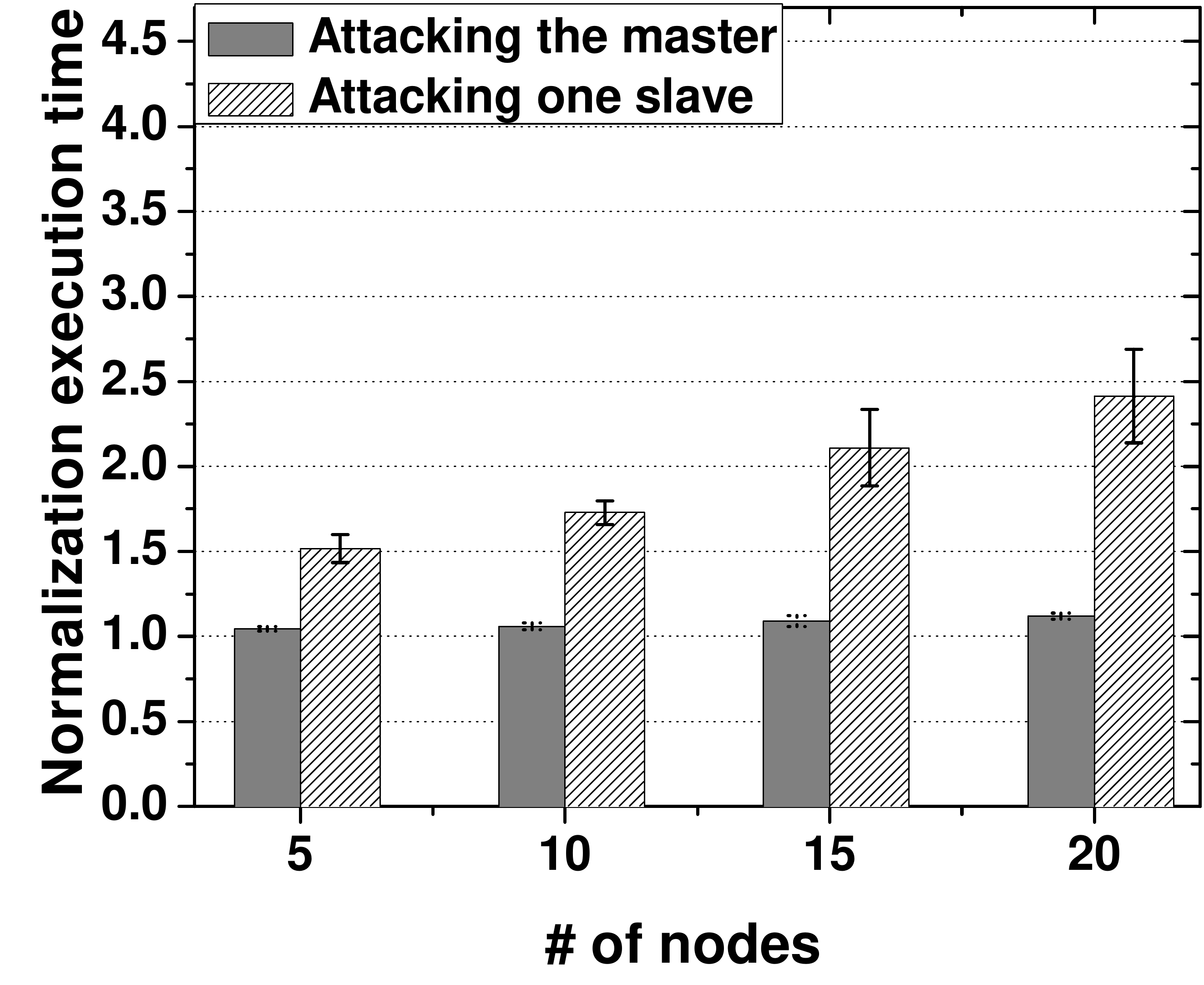}
     \label{fig:hadoop_mrbench}}
     \subfloat[][TestDFSIO]{
     \includegraphics[width=0.24\linewidth]{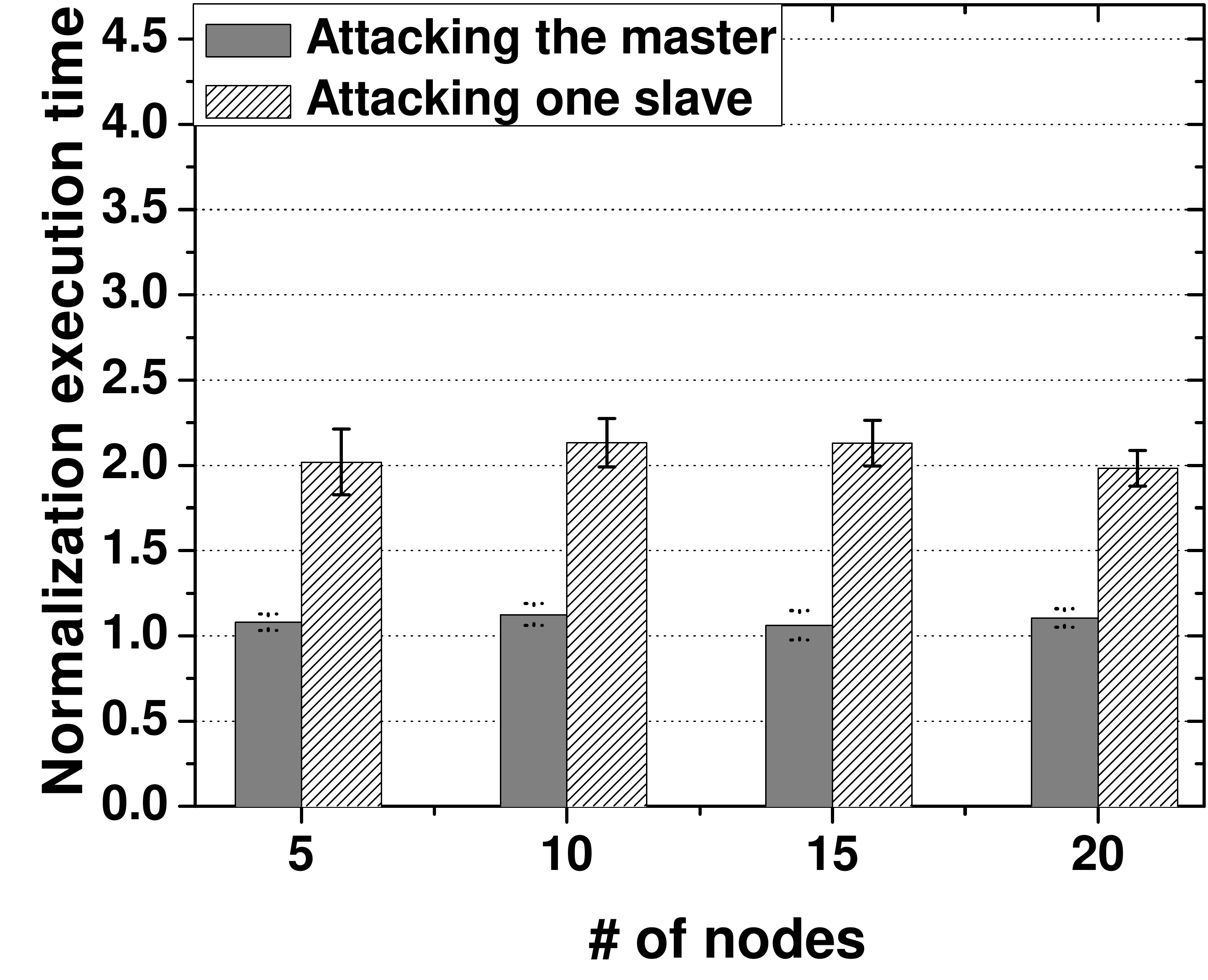}
     \label{fig:hadoop_testdfsio}}
     \subfloat[][NNBench]{
     \includegraphics[width=0.24\linewidth]{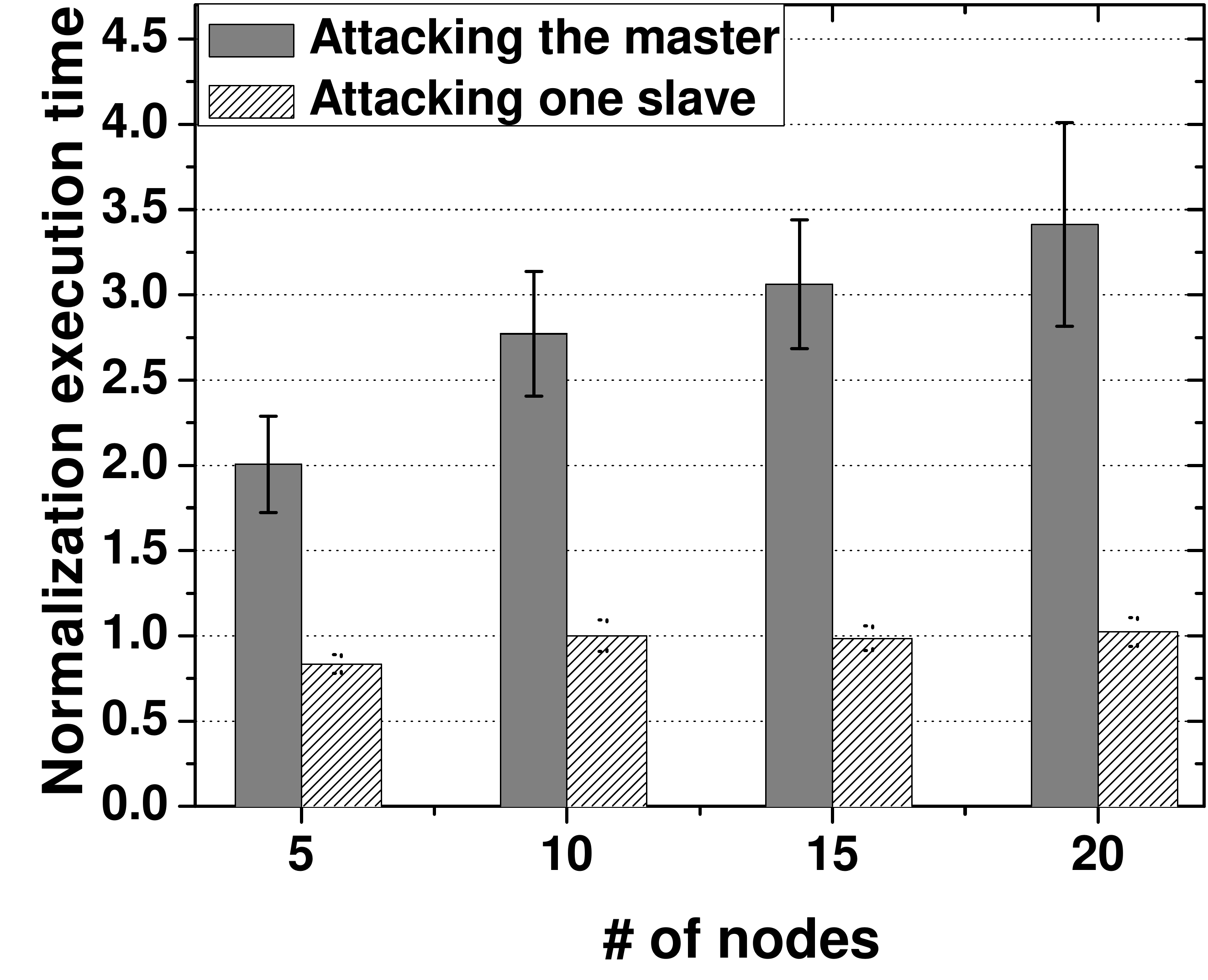}
     \label{fig:hadoop_nnbench}}
     \subfloat[][TeraSort]{
     \includegraphics[width=0.24\linewidth]{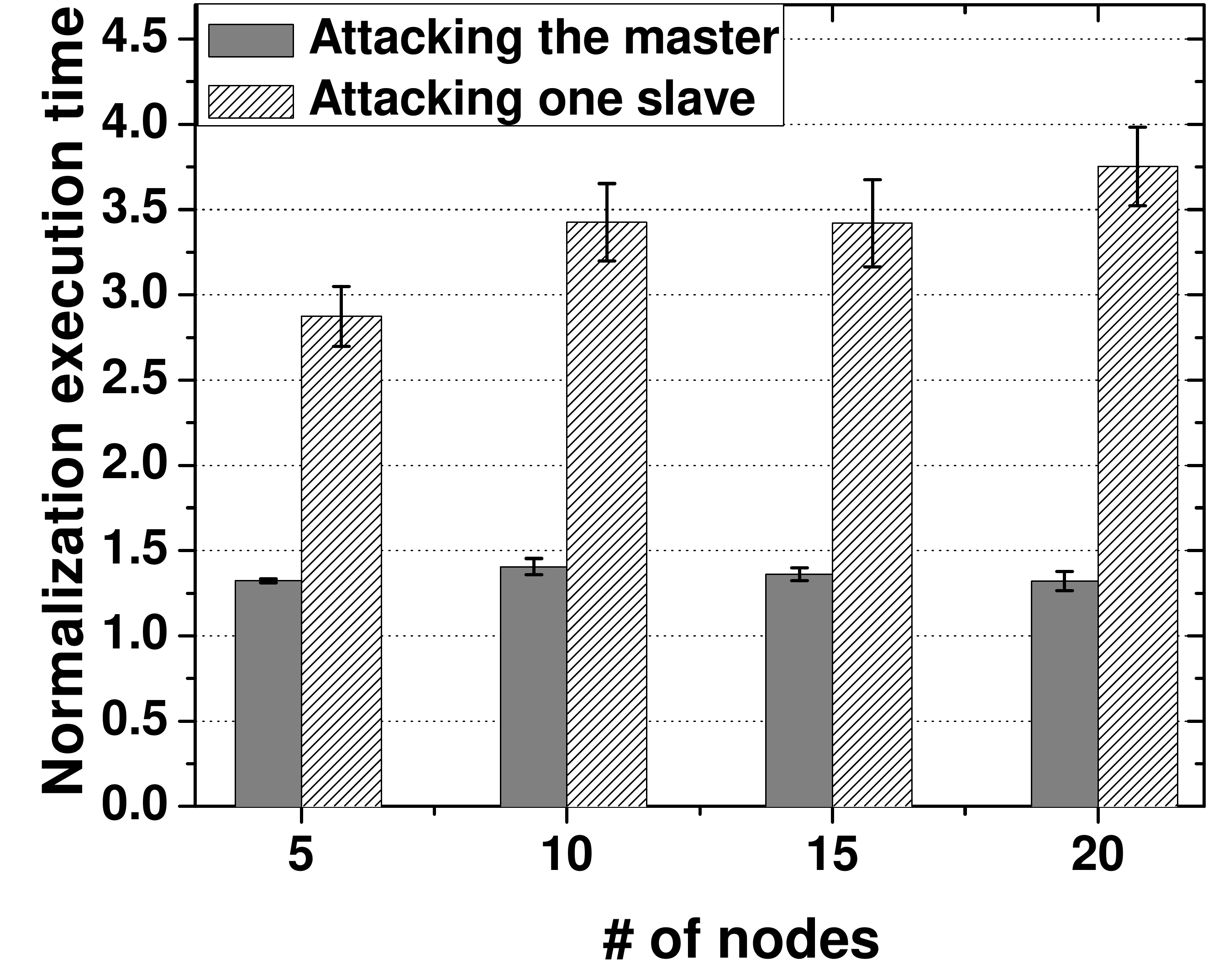}
     \label{fig:hadoop_terasort}}    
    \caption[The LOF caption]{Performance slowdown of the Hadoop applications due to \attackname.}
    \label{fig:hadoop_perf}
\end{figure*}

The attacker only used \textit{one} VM to attack the cluster. He either co-located 
the malicious VM with the master node or one of the slave nodes. We ran four 
different Hadoop benchmarks to test how much performance degradation the single 
attacker VM can cause to the Hadoop cluster. Each experiment was repeated 5 times.
Figure \ref{fig:hadoop_perf} shows the mean values of normalized execution time 
and one standard deviation.

\bheading{MRBench:} This benchmark tests the performance of the  
MapReduce layer of the Hadoop system: it runs a small MapReduce job of text 
processing for a number of times. We set the number of mappers and reducers as the 
number of slave nodes for each experiment.
%\footnote{Configurations of the applications were different as node number changes, and thus increased performance degradation with more nodes does not reflect a trend.}
Figure \ref{fig:hadoop_mrbench} shows that attacking a slave node is 
more effective since the slave node is busy with the map and reduce tasks. In a 
large Hadoop cluster with 20 nodes, attacking just one slave node introduces 
$2.5\times$ slowdown to the entire distributed system. 

\bheading{TestDFSIO:} We use TestDFSIO to evaluate HDFS performance. This 
benchmark writes and reads files stored in HDFS. We configured it 
to operate on $n$ files with the size of 500\mbytes, where $n$ is the number of 
slave nodes in the Hadoop cluster. Figure \ref{fig:hadoop_testdfsio} shows that 
attacking the slave node is effective: the adversary can achieve about 
2$\times$ slowdown.

\bheading{NNBench:} This program is also used to benchmark HDFS in Hadoop. It 
generates HDFS-related management requests on the master node of HDFS. We 
configured it to operate on $200n$ small files, where $n$ is the number of slave 
nodes in the Hadoop cluster. Since the master node is heavily used for serving the 
HDFS requests, attacking the master node can introduce up to 3.4$\times$ slowdown 
to the whole Hadoop system, as shown in Figure \ref{fig:hadoop_nnbench}.

\bheading{Terasort:} We use this benchmark to test the overall performance of both 
MapReduce and HDFS layers in the Hadoop cluster. TeraSort generates a large set of 
data and uses map/reduce operations to sort the data. For each experiment, we set 
the number of mappers and reducers to $n$, and the size of data to be sorted to 
$100n$ \mbytes, where $n$ is the number of slave nodes in the Hadoop cluster. 
Figure \ref{fig:hadoop_terasort} shows that attacking the slave node is very 
effective: it can bring $2.8\sim3.7$ $\times$ slowdown to the entire Hadoop system.

\bheading{Summary.} 
The adversary can deny working memory availability to the victim VM and thus degrade an important 
distributed system's performance with minimal costs: it can use just one VM to 
interfere with one of 20 nodes in the large cluster. The slowdown of a single 
victim node can cause up to 3.7$\times$ slowdown to the whole system. 

\subsection{Attacking E-Commerce Websites}
\label{sec:ecommerce}

A web application consists of load balancers, web servers, database servers and 
memory caching servers. Memory DoS attacks can disturb an E-commerce 
web application by attacking various components.

\bheading{Experiment settings.}
We chose a popular open source E-commerce web application, Magento~\cite{magento}, 
as the target of the attack. The victim application consists of five VMs: a load 
balancer based on Pound for balancing network requests; two Apache web servers to 
process and deliver web requests; a MySQL database server to store customer and 
merchandise information; and a Memcached server to speed up database transactions. 
The five VMs were hosted on different cloud servers in EC2. The adversary is able 
to co-locate his VMs with one or multiple VMs that host the victim application. We 
measure the application's latency and throughput to evaluate the effectiveness of 
the attack.

\begin{figure}[ht]
     \centering
     \subfloat[][Latency]{
     \includegraphics[width=0.48\linewidth]{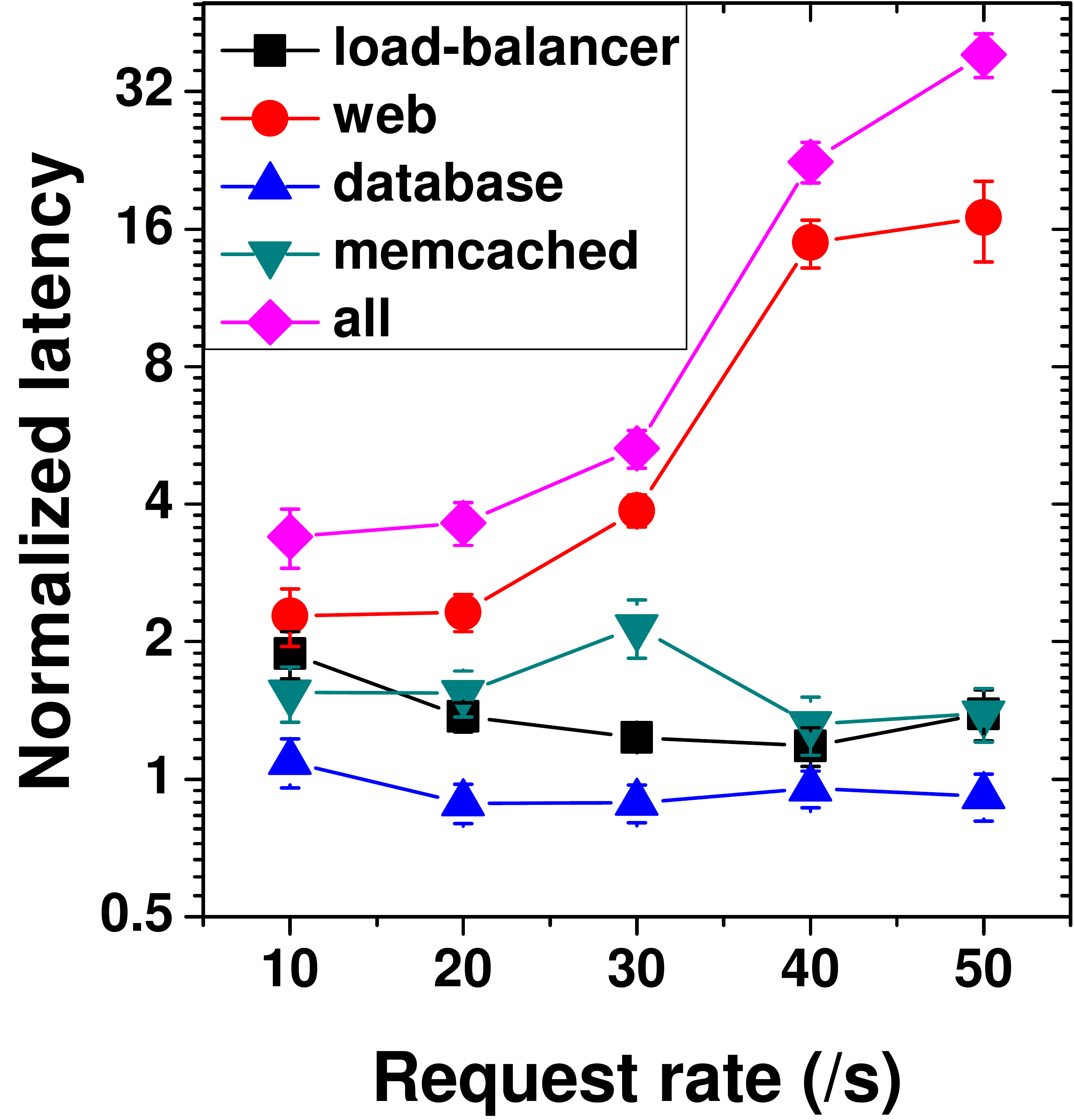}
     \label{fig:web_latency_2}}
     \subfloat[][Throughput]{
     \includegraphics[width=0.48\linewidth]{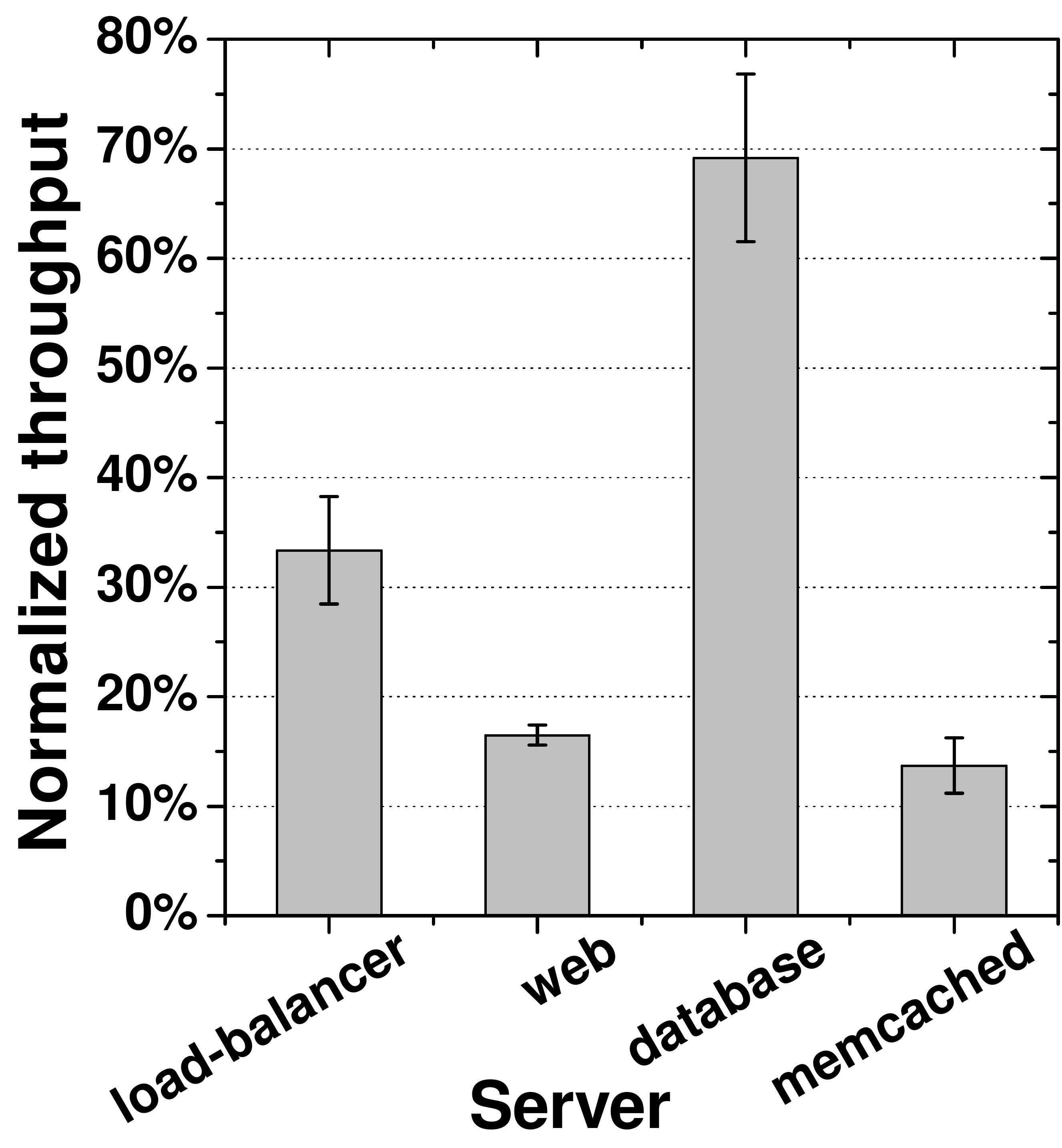}
     \label{fig:web_throughput}}
    \caption{Latency and throughput of the Magento application due to \attackname.}
    \label{fig:web_perf}
\end{figure}

\bheading{Latency.} 
We launched a client on a local machine outside of EC2. The client employed httperf
\cite{httperf} to send HTTP requests to the load balancer with different rates 
(connections per second) and we measured the average response time. We evaluated the 
attack from one or all co-located VMs. Each experiment was repeated 10 times and 
the mean and standard deviation of the latency are reported in Figure 
\ref{fig:web_latency_2}. This shows that memory contention on database, load 
balancer or memcached servers do not have much impact on the overall performance 
of the web application, with only up to 2$\times$ degradation. This is probably 
because these servers were not heavily used in these cases. Memory DoS attacks on 
web servers were the most effective (17$\times$ degradation). When the adversary 
can co-locate with all victim servers and each attacker VM induces contention with 
the victim, the web server's HTTP response time was delayed by $38\times$, for a 
request rate of 50 connections per second.

\bheading{Server throughput.} 
Figure \ref{fig:web_throughput} shows the results of another experiment, where we 
measured the throughput of each victim VM individually, under \attackname. We used 
ApacheBench \cite{ab} to evaluate the load balancer and web servers, SysBench 
\cite{sysbench} to evaluate the database server and memtier\_benchmark 
\cite{memtier} to evaluate the memcached server. This shows \attackname on these 
servers were effective: the throughput can be reduced to only 13\% $\sim$ 70\% 
under malicious contention by the attacker.

\bheading{Summary.} 
The adversary can compromise the quality of E-commerce service and cause financial 
loss in two ways: (1) long response latency will affect customers' satisfaction 
and make them leave this E-commerce website \cite{PoCaGa:11}; (2) it can cause 
throughput degradation, reducing the number of transactions completed in a unit 
time. The cost for these attacks is relatively cheap: the adversary only needs a 
few VMs to perform the attacks, with each t2.medium instance costing \$0.052 per 
hour.

\section{Defense against Memory DoS Attacks}
\label{sec:discuss}

We propose a novel, general-purpose approach to detecting and mitigating 
\attackname in the cloud. Unlike some past work, our defense does not require prior
profiling of the memory resource usage of the applications. 
Our defense can be provided by the cloud providers as a 
new security service to customers. We denote as \protectedVM{s} those VMs for 
which the cloud customers require protection. To detect \attackname, lightweight 
statistical tests are performed frequently to monitor performance changes of the 
\protectedVM{s} (\secref{sec:detection}). To mitigate the attacks, \emph{execution 
throttling} is used to reduce the impact of the attacks (\secref{sec:mitigation}). 
A novelty of our approach is the 
combined use of two existing hardware features: \textit{event counting} using 
hardware performance counters controllable via the Performance Monitoring Unit 
(PMU) and \textit{duty cycle modulation} controllable through the 
\texttt{IA32\_CLOCK\_MODULATION} Model Specific Register (MSR).

\subsection{Detection Method}
\label{sec:detection}
\begin{figure}[t]
\centerline{\mbox{\includegraphics[width=\linewidth]{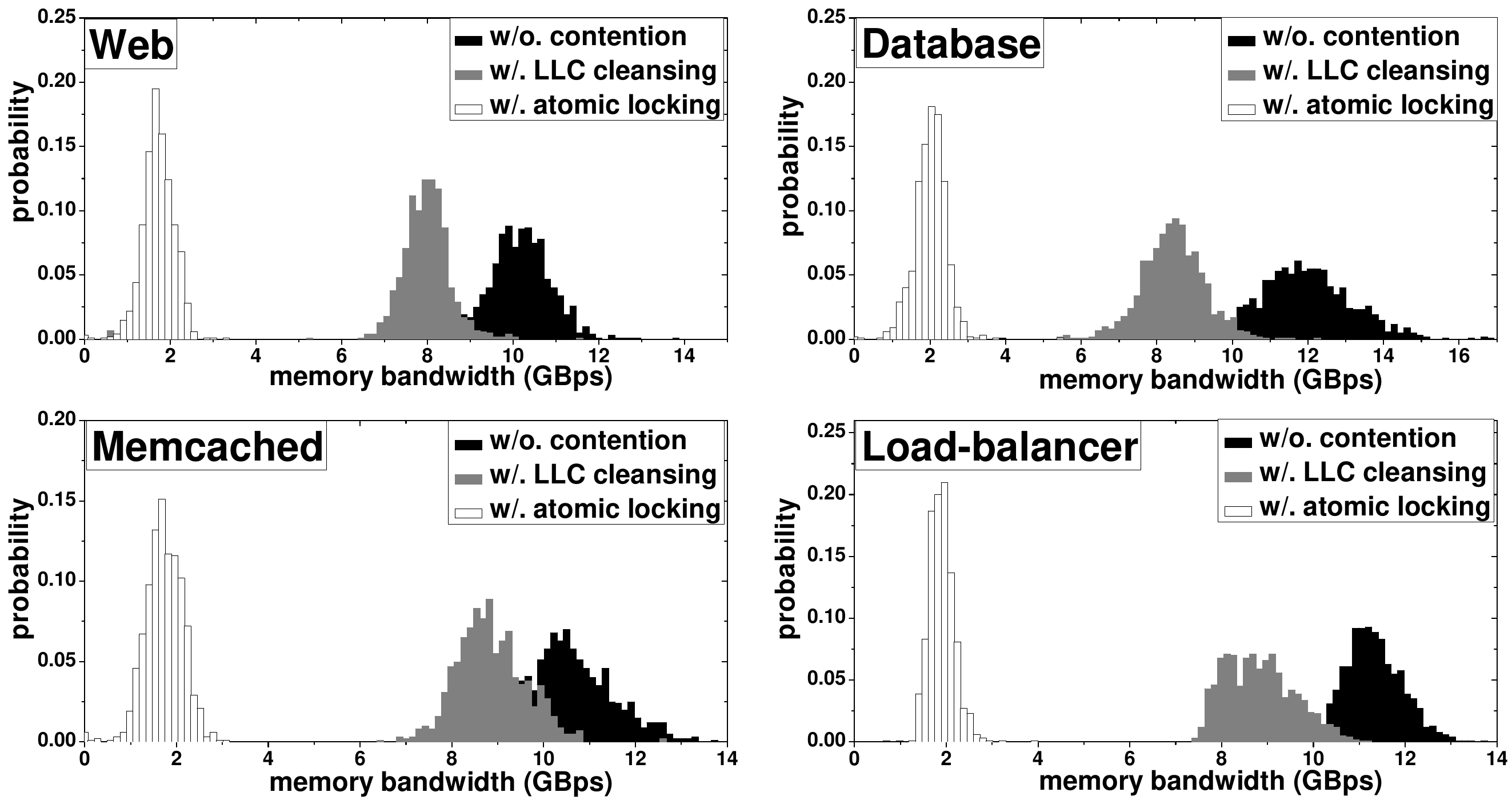}}}
\caption{Probability distributions of the \protectedVM's memory bandwidth.}
\label{fig:prob_dist}
\end{figure}

The key insight in detecting \attackname is that {\em such attacks are caused by
abnormal resource contention between \protectedVM{s} and attacker VMs, and such
resource contention can significantly alter the memory usage of the \protectedVM,
which can be observed by the cloud provider.} We postulate that the
statistics of accesses to memory resources, by a phase of a software program, 
follow certain probability distributions. When a \aattackname happens, these 
probability distributions will change. Figure \ref{fig:prob_dist} shows the 
probability distributions of the \protectedVM's memory access statistics, without 
attacks (black), and with two kinds of attacks (gray and shaded), when it runs one 
of four applications introduced in \secref{sec:ecommerce}, \ie, the Apache web server, 
Mysql database, Memcached and Pound load-balancer. 
When an attacker is present, the probability distribution of the \protectedVM's 
memory access statistics (in this case, memory bandwidth in GigaBytes per second) 
changes significantly.

In practice, only samples drawn from the underlying probability distribution are
observable. Therefore, the provider's task is to collect two sets of samples:
[$X^{\textsc{R}}_1$, $X^{\textsc{R}}_2$, ..., $X^{\textsc{R}}_{n^{\textsc{R}}}$]
are reference samples collected from the probability distribution when we are sure that there 
are no attacks;
[$X^{\textsc{M}}_1$, $X^{\textsc{M}}_2$, ..., $X^{\textsc{M}}_{n^{\textsc{M}}}$]
are monitored samples collected from the \protectedVM at runtime,
when attacks may occur. If these two 
sets of samples are not drawn from the same distribution, we can conclude that the 
performance of the \protectedVM is hindered by its neighboring VMs. When the 
distance between the two distributions is large, we may conclude the \protectedVM 
is under some  \attackname.

We propose to use the two-sample Kolmogorov-Smirnov (KS) tests~\cite{Ma:51}, as a 
metric for whether two samples belong to the same probability distribution. The 
KS statistic is defined in Equation \ref{eq:kstest}, where $F_n(x)$ is the 
empirical distribution function of the samples $[X_1, X_2, ..., X_n]$, and $sup$ 
is the supremum function (\ie, returning the maximum value). Superscripts 
$\textsc{M}$ and $\textsc{R}$ denote the monitored samples and reference samples, 
respectively. $n^M$ and $n^R$ are the number of monitored samples and reference 
samples.

\begin{equation}
\begin{aligned}
\label{eq:kstest}
 D_{n^{\textsc{M}},\; n^{\textsc{R}}} = \sup\limits_x \mid F_{n^{\textsc{M}}}^{\textsc{M}}(x) - F_{n^{\textsc{R}}}^{\textsc{R}}(x) \mid
\end{aligned}
\end{equation}

\begin{equation}
\begin{aligned}
\label{eq:ksvalue}
 D^{\alpha}_{n^{\textsc{M}},\; n^{\textsc{R}}} = \sqrt{\frac{n^{\textsc{M}}+n^{\textsc{R}}}{n^{\textsc{M}}\times n^{\textsc{R}}}}\sqrt{-0.5\times ln(\frac{\alpha}{2})}
 \end{aligned}
\end{equation}

\bheading{Null hypothesis for KS test.} 
We establish the null hypothesis that currently monitored samples are drawn from
the same distribution as the reference samples. Benign performance contention with 
non-attacking, co-tenant VMs will not alter the probability distribution of
the \protectedVM's monitored samples significantly, so the KS statistic is small 
and the null hypothesis is held. Equation \ref{eq:ksvalue} introduces $\alpha$:  
We can reject the null hypothesis with confidence level $1-\alpha$ if the KS 
statistic, $D_{n^{\textsc{M}},\; n^{\textsc{R}}}$, is greater than predetermined 
critical values $D^{\alpha}_{n^{\textsc{M}},\; n^{\textsc{R}}}$. Then, the cloud 
provider can assume, with confidence level $1-\alpha$, that a \aattackname exists, 
and trigger a mitigation strategy.

While monitored samples, $X^{\textsc{M}}_i$, are simply collected at runtime, 
reference samples, $X^{\textsc{R}}_i$, ideally should be collected when the 
\protectedVM is not affected by other co-located VMs. The technical challenge here 
is that if these samples are collected offline, we need to assume the memory 
access statistics of the VM never change during its life time, which is 
unrealistic. If samples are collected at runtime, all the co-locating VMs need to 
be paused during sample collection, which, if performed frequently, can cause 
significant performance overhead to benign, co-located VMs.

\begin{figure*}[t]
     \centering
     \subfloat[][Monitoring the \protectedVM.]{
     \includegraphics[width=0.5\linewidth]{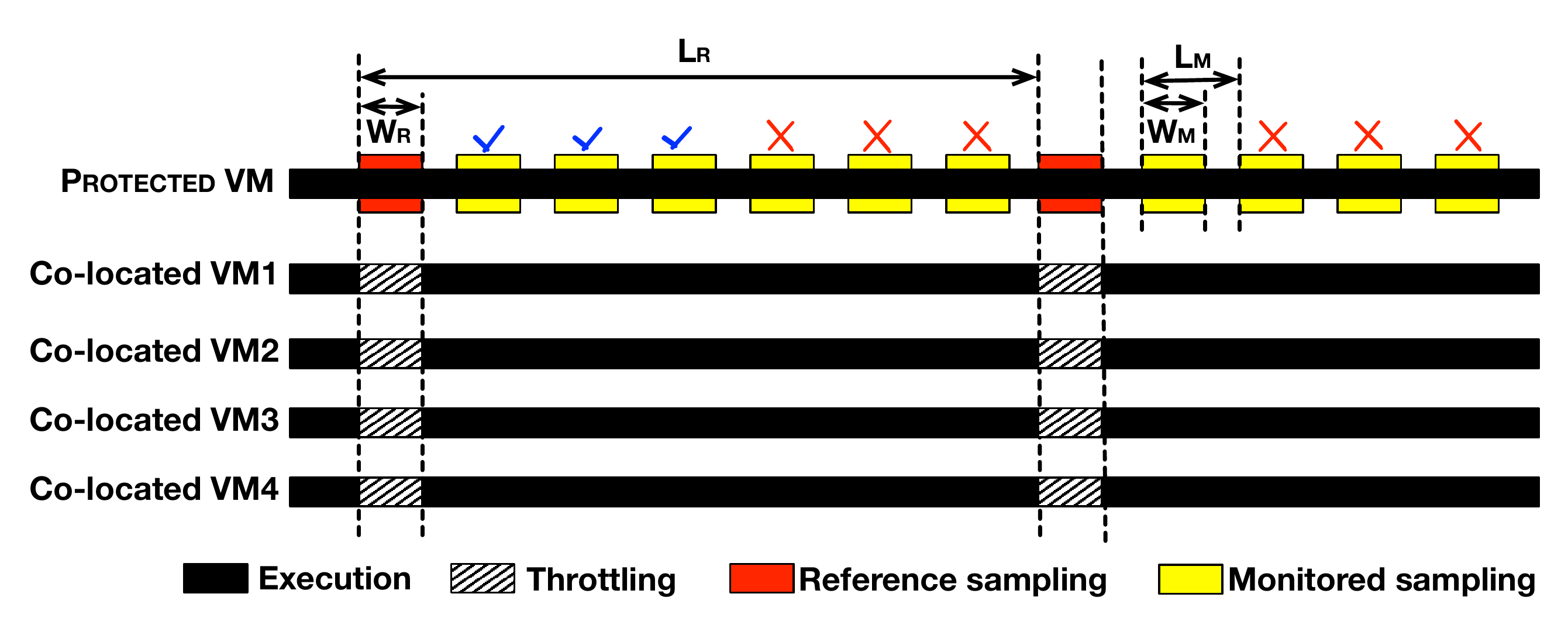}
     \label{fig:throttle_example}}
     \subfloat[][Identifying co-located VM3 as the attacker VM.]{
     \includegraphics[width=0.5\linewidth]{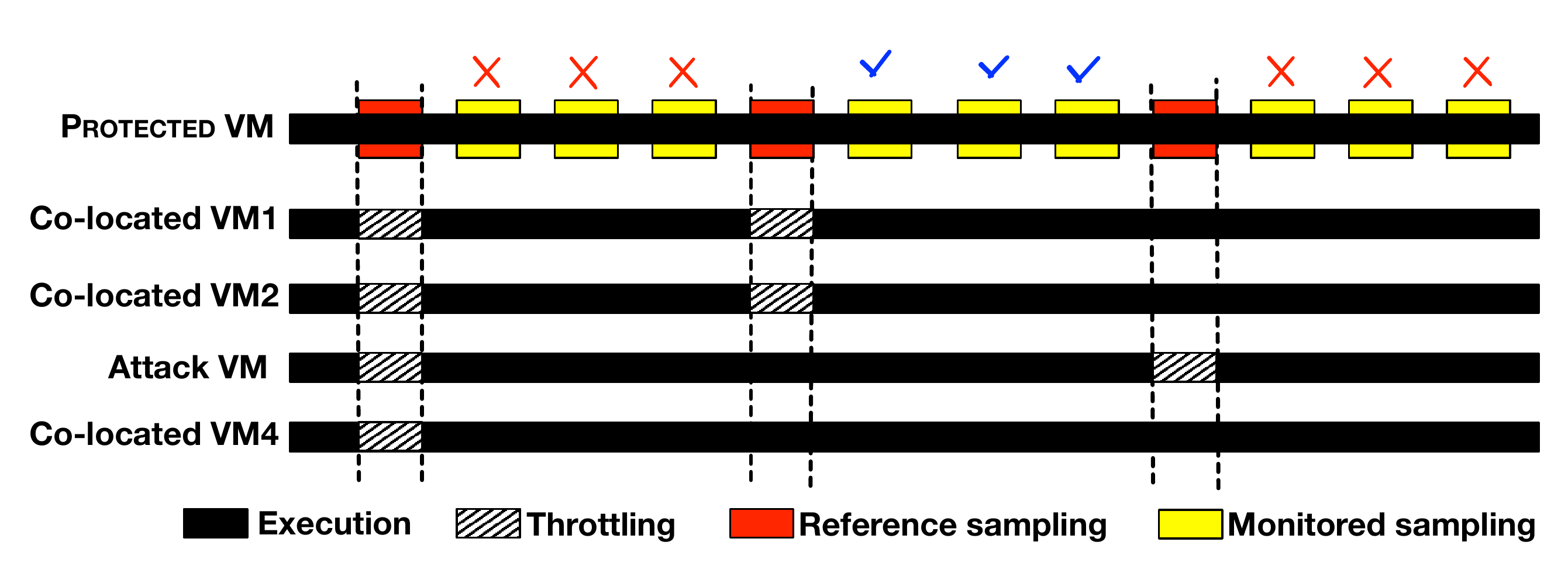}
     \label{fig:attack_iden}} \\
    \caption{Illustration of monitoring the \protectedVM (a) and identifying the attack VM (b). The blue ``\cmark" means the null hypothesis is accepted; while the red ``\xmark" means the null hypothesis is rejected.}
    \label{fig:throttle_ill}
\end{figure*}

\bheading{Pseudo Isolated Reference Sampling.} 
To address this technical challenge, we use \textit{execution throttling} to 
collect the reference samples at runtime. The basic idea is to throttle down the 
execution speed of other VMs, but maintain the \protectedVM's speed during the 
reference sampling stage. This can reduce the co-located VMs' interference without 
pausing them.

\emph{Execution throttling} is based on a feature provided in Intel Processors 
called \textit{duty cycle modulation} \cite{intel_manual}, which is designed to 
regulate each core's execution speed and power consumption. The processor allows 
software to assign ``duty cycles'' to each CPU core: the core will be active 
during these duty cycles, and inactive during the non-duty cycles. For example, 
the duty cycle of a core can be set from 16/16 (no throttling), 15/16, 14/16, ..., 
down to 1/16 (maximum throttling). Each core uses a model specific register (MSR), 
\texttt{IA32\_CLOCK\_MODULATION}, to control the duty cycle ratio: bit 4 of this MSR 
denotes if the duty cycle modulation is enabled for this core; bits 0-3 represent 
the number of 1/16 of the total CPU cycles set as duty cycles. 

In execution throttling, the execution speed of other VMs will be throttled down 
and very little contention is induced to the \protectedVM. As such, reference 
samples collected during the execution throttling stage are drawn from a quasi 
contention-free distribution. 

Figure \ref{fig:throttle_example} illustrates the high-level strategy for
monitoring \protectedVM{s}. The reference samples are collected during the 
reference sampling periods ($W_R$), where other VMs' execution speeds are 
throttled down. The monitored samples are collected during the monitored sampling 
periods ($W_M$), where co-located VMs run normally, without execution throttling. 
KS tests are performed right after each monitored sample is collected, and
probability distribution divergence is estimated by comparing with the most recent 
reference samples. Monitored samples are collected periodically at a time interval 
of $L_M$, and reference samples are collected periodically at a time interval of 
$L_R$. We can also randomize the intervals $L_M$ and $L_R$ for each period to 
prevent the attacker from reverse-engineering the detection scheme and scheduling the attack phases to avoid detection.

If the KS test results reject the null hypothesis, it may be because the 
\protectedVM is in a different execution phase with different memory access 
statistics, or it may be due to \attackname. To rule out the first possibility, 
double checking automatically occurs since reference samples are re-collected and 
updated after a time interval of $L_R$. If deviation of the probability 
distribution still exists, attacks can be confirmed.

\subsection{Mitigation Method}
\label{sec:mitigation}

The cloud provider has several methods to mitigate the attack. One is VM 
migration, which can be achieved either by reassigning the vCPUs of a VM to a
different CPU package, when the memory resource being contended is in the same
package (\eg, LLC), or by migrating the entire VM to another server, when the 
memory resource contended is shared system-wide (\eg, memory bus). However, such 
VM migration can not completely eliminate the attacker VM's impact on other VMs.
 
An alternative approach is to identify the attacker VM, and then employ 
\emph{execution throttling} to reduce the execution speed of the malicious VM, 
while meanwhile the cloud provider conducts further investigation and/or notifies 
the customer of the suspected attacker VM of observed resource abuse activities. 
%Sending emails to notify customers before shutting down VMs is already a common practice in Amazon EC2. 

\bheading{Identifying the attacker VM.} 
Once \attackname are detected, to mitigate the threat, the cloud provider needs
to identify which of the co-located VMs is conducting the attack. Here we propose 
a novel approach to identify malicious VMs based on \textit{selective execution 
throttling in a binary search manner}: First, half of the co-located VMs keep 
normal execution speed while the rest of VMs are throttled down during reference 
sampling periods (Figure \ref{fig:attack_iden}, 2nd Reference Sampling period). If in this case, reference samples and monitored samples are drawn 
from the same distribution, then there are malicious VMs among the ones not 
throttled down during the reference sampling period. Then, we select half of the 
remaining VMs to be throttled while all the other VMs are in normal speed, to
collect the next reference samples. In Figure \ref{fig:attack_iden}, this is the 3rd Reference 
Sampling period, where only VM3 is throttled. Since the subsequent monitored 
samples have a different distribution compared to this Reference Sample, VM3 is 
identified as the attack VM. Note that if there are multiple attacker VMs on the 
server, we can use the above procedure to find one VM each time and repeat it 
until all the attacker VMs are found. By organizing this search for the attacker 
VM or VMs as a binary search, the time taken to identify the source of memory 
contention is O (log $n$), where $n$ is the number of co-tenant VMs on the 
\protectedVM's server. Algorithm \ref{alg:detect_attacker} shows how to detect 
attacker VMs using this \emph{selective execution throttling}.

\begin{algorithm}[ht]
\scriptsize
\SetAlgoLined

\SetAlgoLined\DontPrintSemicolon
\KwIn{}
 \Indp VM[1,...,n] $\kern 1.6pc $ /* \emph{set of co-tenant VMs} */\\
  \BlankLine
  \BlankLine
\Indm
\SetKwFunction{proc}{IdentifyAttacker}
\SetKwProg{myalg}{function}{}{end}

\myalg{\proc{\emph{sub\_VM}}}{
  /* \emph{sub\_VM: set of VMs to identify} */\\
  \eIf{\emph{sub\_VM.\texttt{length}() = 1}}{
    \KwRet sub\_VM[0]
  }{
    imin = 0 \\
    imax = sub\_VM.\texttt{length}()-1 \\
    imid = $\lceil$(imin+imax)/2$\rceil$ \\
    \texttt{ThrottleDown}(sub\_VM[0,...,imid-1]) \\
    reference\_sample = \texttt{DataCollect}() \\
    \texttt{ThrottleUp}(sub\_VM[0,...,imid-1]) \\
    monitor\_sample = \texttt{DataCollect}() \\
    result = \texttt{KSTest}(reference\_sample, monitor\_sample) \\
    \eIf{\emph{result = Reject}}{
      \KwRet \proc{\emph{sub\_VM[0,...,imid-1]}}
    }{
      \KwRet \proc{\emph{sub\_VM[imid,...,imax]}}
    }
  }
}{}
  \BlankLine
  \BlankLine

\Begin{
      vm = \proc{\emph{VM}} \\
      \texttt{ThrottleDown}([vm])
}

\caption{Identifying and mitigating the attacker VMs that cause severe resource contention.}
 \label{alg:detect_attacker}
\end{algorithm}

\subsection{Implementation}
\label{sec:implementation}

\begin{figure}[t]
\centerline{\mbox{\includegraphics[width=\linewidth]{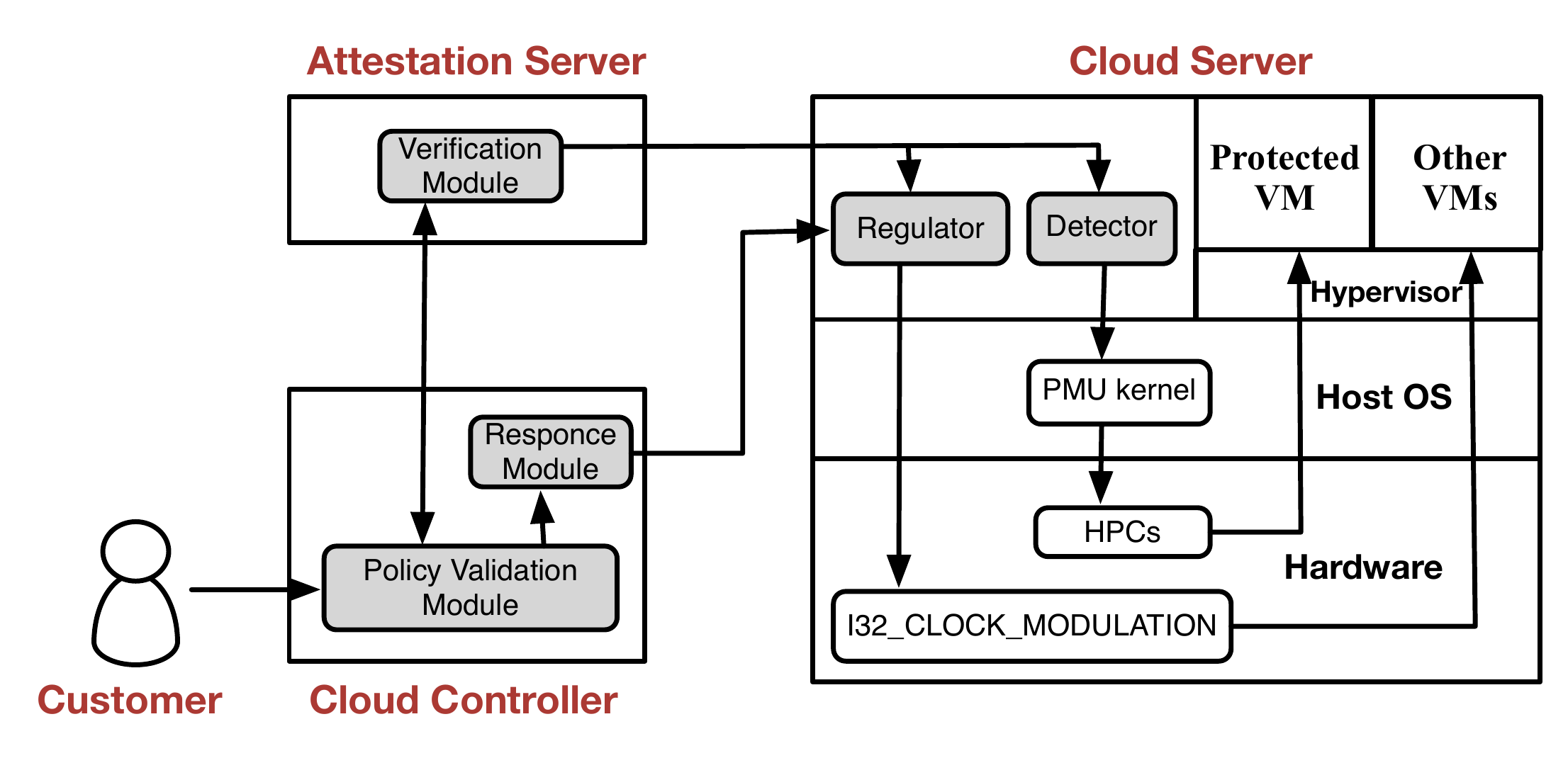}}}
\caption{Architecture overview.}
\label{fig:arch_overview}
\end{figure}

We implement a prototype system of our proposed defense on the OpenStack platform. 
Figure \ref{fig:arch_overview} shows the defense architecture overview.
We adopt the \emph{CloudMonatt} architecture from \cite{ZhLe:15}. 
Specifically, the system includes three types of servers. The Cloud Controller is the cloud manager that manages the VMs. It has a \texttt{Policy Validation Module} to receive and analyze customers' requests. It also has a \texttt{Response Module}, which can throttle down the attacker VMs' execution speed to mitigate \attackname. The Attestation Server is a centralized server for monitoring and detection of \attackname. It has a \texttt{Verification Module} that receives \protectedVM's performance probability distribution, detects \attackname and identifies malicious VMs.

On each of the cloud servers, we use the KVM hypervisor which is the default setup for OpenStack. Other virtualization platforms, such as Xen and HyperV, can also be used. Two software modules are installed on the host OS. A \texttt{Detector} measures the memory access characteristics of the \protectedVM using Performance Monitoring Units (PMU), which are commonly available in most modern processors. A PMU provides a set of Hardware Performance Counters to count hardware-related events. In our implementation, we use the linux kernel API \texttt{perf\_event} to measure the memory access statistics for the number of \textit{LLC accesses} per sampling period. A \texttt{Regulator} is in charge of controlling VMs' execution speed. It uses the \texttt{wrmsr} instruction to modify the \texttt{IA32\_CLOCK\_MODULATION} MSR to control the duty cycle ratio.

In our implementation, the parameters involved in reference and monitored sampling 
are as follows: $W_R = W_M = 1s$, $L_M = 2s$,  $L_R=30s$. These values were 
selected to strike a balance between the performance overhead due to execution
throttling and detection accuracy. In each sampling period, $n = 100$ samples 
are collected, with each collected during a period of 10ms. We choose 10ms 
because it is short enough to provide accurate measurements, and long enough to 
return stable results. In the KS tests, the confidence level, $1-\alpha$, is set 
as 0.999, and the threshold to reject the null hypothesis is $D^{\alpha}=0.276$ 
(given $\alpha=0.001$). If 4 consecutive KS statistics larger than 0.276 are 
observed (the choice of 4 is elaborated in \secref{sec:evaluation}), it is assured 
that the \protectedVM's memory access statistics have been changed. Then to 
confirm that such changes are due to \attackname, reference samples will be 
refreshed and the malicious VM will be identified.

\subsection{Evaluation}
\label{sec:evaluation}

Our lab testbed comprised three servers. A Dell R210II Server (equipped with 
one quad-core, 3.30\ghertz, Intel Xeon E3-1230v2 processor with 8\mbytes LLC) was 
configured as the Cloud Controller as well as the Attestation Server. Two Dell 
PowerEdge R720 Servers (one has two six-core, 2.90\ghertz Intel Xeon E5-2667 processors with 
15\mbytes LLC, the other has one eight-core, 2.90\ghertz Intel Xeon E5-2690 
processor with 20\mbytes LLC) were deployed to function as VM hosting servers. 

\begin{figure*}[t]
     \centering
     \subfloat[][LLC cleansing attack]{
     \includegraphics[width=0.48\linewidth]{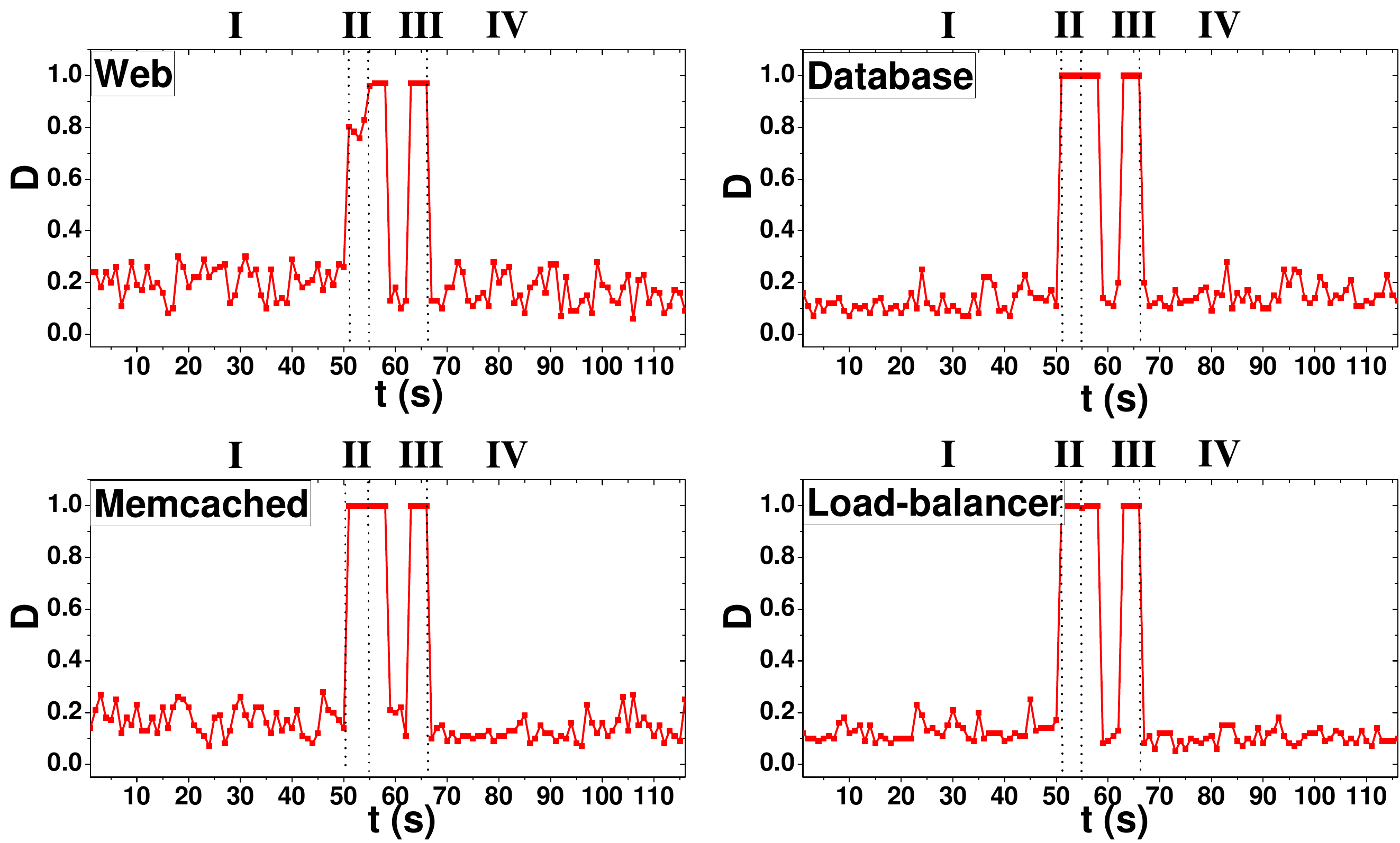}
     \label{fig:llc_detect_vm}} \hspace*{1.5em}
     \subfloat[][Atomic locking attack]{
     \includegraphics[width=0.48\linewidth]{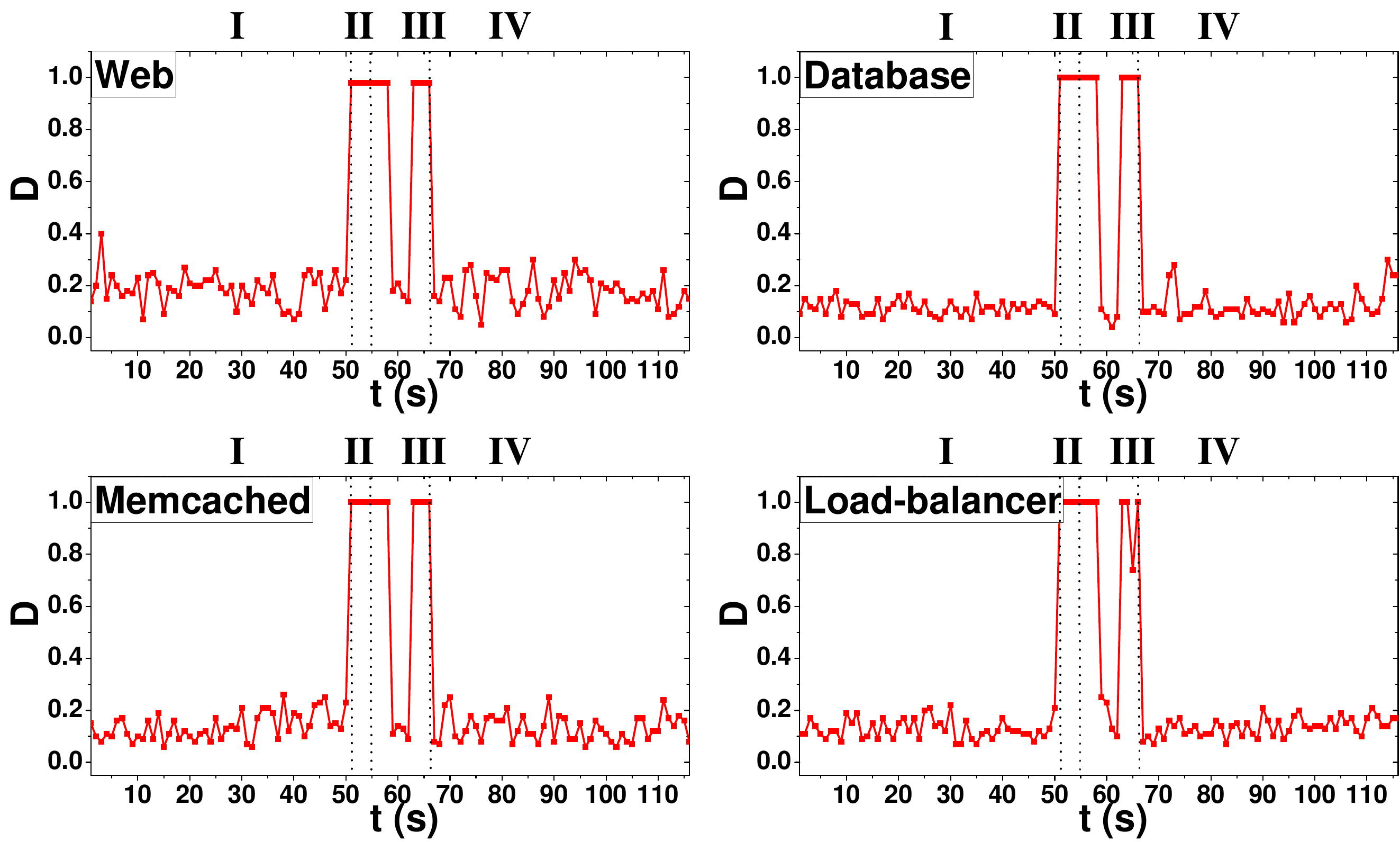}
     \label{fig:bus_detect_vm}} \\
    \caption{KS statistics of the \protectedVM for detecting and mitigating \attackname.}
    \label{fig:detect_vm}
\end{figure*}

\bheading{Detection accuracy.}
We deployed a \protectedVM sharing a cloud server with 8 other VMs. Among these 8 
VMs, one VM was an attacker VM conducting a multi-threaded LLC cleansing attack 
with 4 threads (\secref{sec:uncore_cache_contention}), or an atomic locking attack
(\secref{sec:bus_contention}). The remaining 7 VMs were benign VMs running common 
linux utilities. The \protectedVM runs one of the web, database, 
memcached or load-balancer applications in the Magento application (\secref{sec:ecommerce}). The 
experiments consisted of four stages; the KS statistics of each of the four 
workloads during the four stages under the two types of attacks are shown in 
Figure~\ref{fig:detect_vm}.

In stage \RNum{1}, the \protectedVM runs while the attacker is idle. The KS
statistic in this stage is relatively low. So we accept the null hypothesis that
the memory accesses of the reference and monitored samples follow the same
probability distribution. In stage \RNum{2}, the attacker VM conducts the LLC
cleansing or atomic locking attacks. We observe the KS statistic is much higher
than 0.276. The null hypothesis is rejected, signaling detection of potential
\attackname. In stage \RNum{3}, the cloud provider runs \textit{three} rounds of
reference resampling to pinpoint the malicious VM. Resource contention
mitigation is performed in stage~\RNum{4}: the cloud provider throttles down the
attacker VM's execution speed. After this stage, the KS
statistic falls back to normal which suggests that the attacks are mitigated.

\begin{figure}[ht]
     \centering
     \subfloat[][True positive]{
     \includegraphics[width=0.48\linewidth]{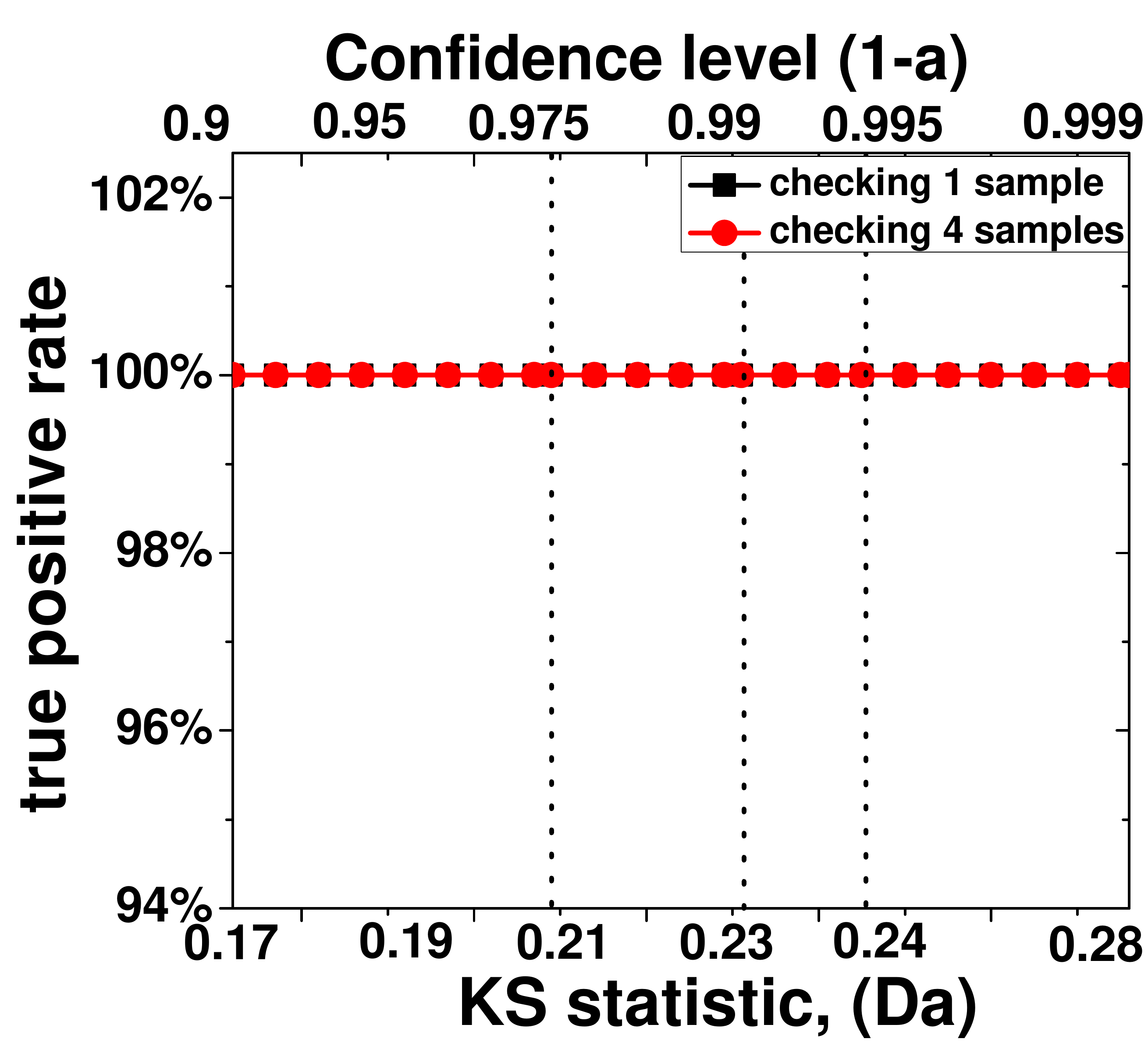}
     \label{fig:detect_tp}}
     \subfloat[][False positive]{
     \includegraphics[width=0.48\linewidth]{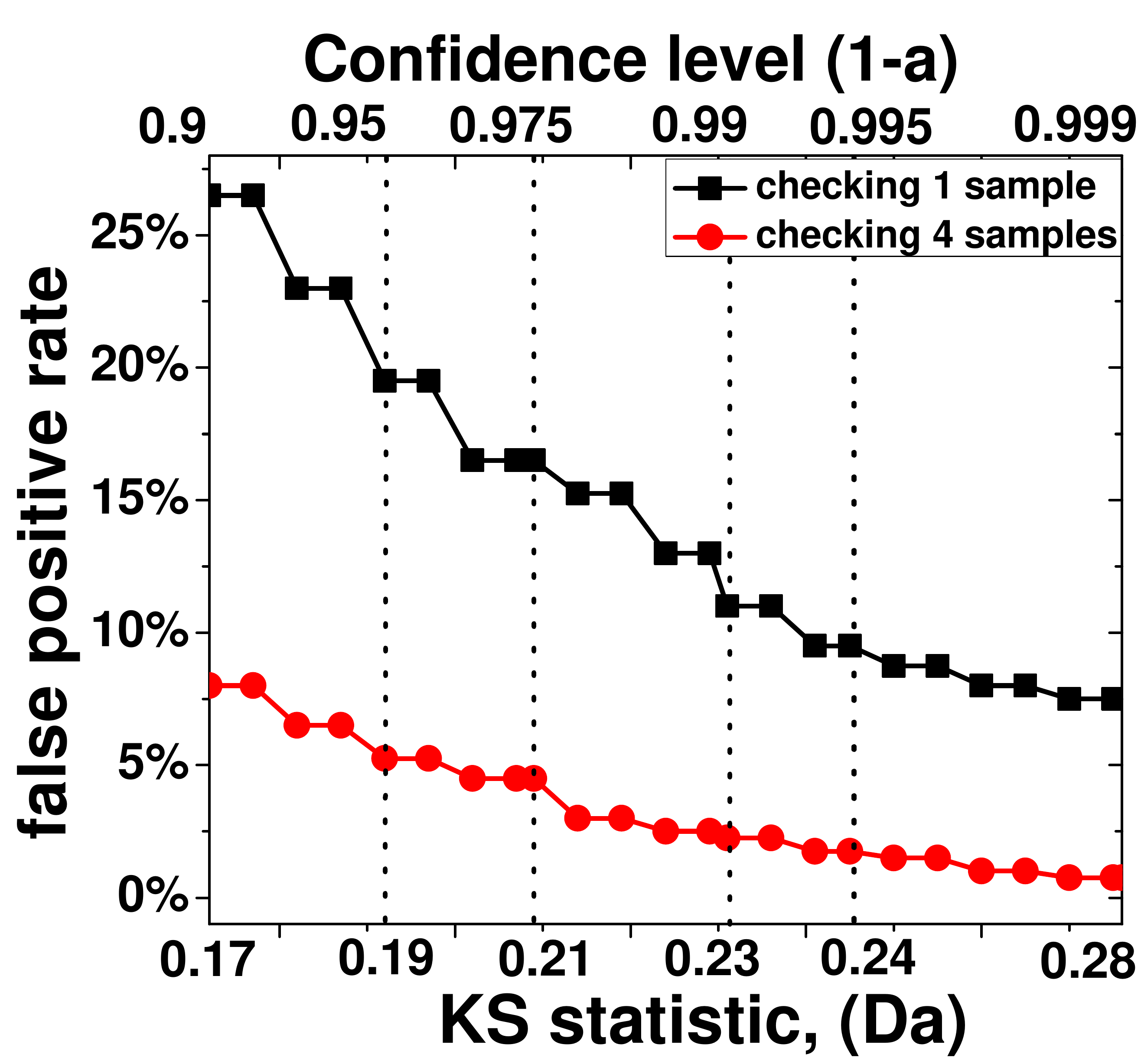}
     \label{fig:detect_fp}}
    \caption{Detection accuracy.}
    \label{fig:detect_accuracy}
\end{figure}

We also evaluated the false positive rates and false negative rates of two 
different criteria for identifying a memory access anomaly: {\em 1} abnormal KS
statistic (larger than the critical value $D^{\alpha}$) or {\em 4} consecutive 
abnormal KS statistics. Figure \ref{fig:detect_tp} shows the true positive rate of 
LLC cleansing and atomic locking attack detection, at different confidence levels
$1-\alpha$.  We observe that the true positive rate is always one (thus zero false 
negatives), regardless of the detection criteria (1 vs 4 abnormal KS tests).  
Figure \ref{fig:detect_fp} shows the false positive rate, which can be caused by 
background noise due to other VMs' executions. This figure shows that using 4 
consecutive abnormal KS statistics significantly reduces the false positive rate.

\begin{figure*}[t]
     \centering
     \subfloat[][Throttling LLC cleansing attacks]{
     \includegraphics[width=0.48\linewidth]{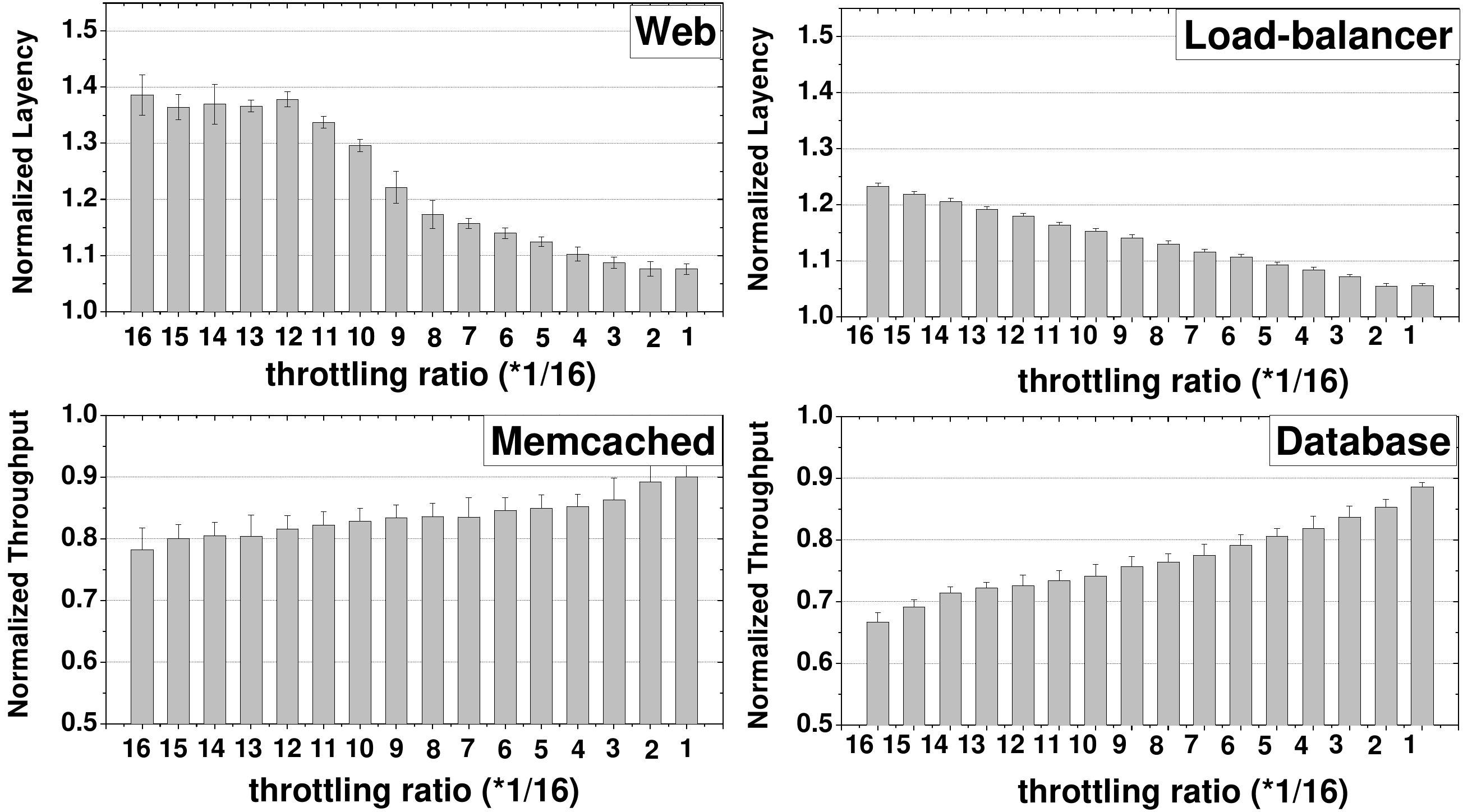}
     \label{fig:throttle_llc}} \hspace*{1.5em}
     \subfloat[][Throttling atomic locking attacks]{
     \includegraphics[width=0.48\linewidth]{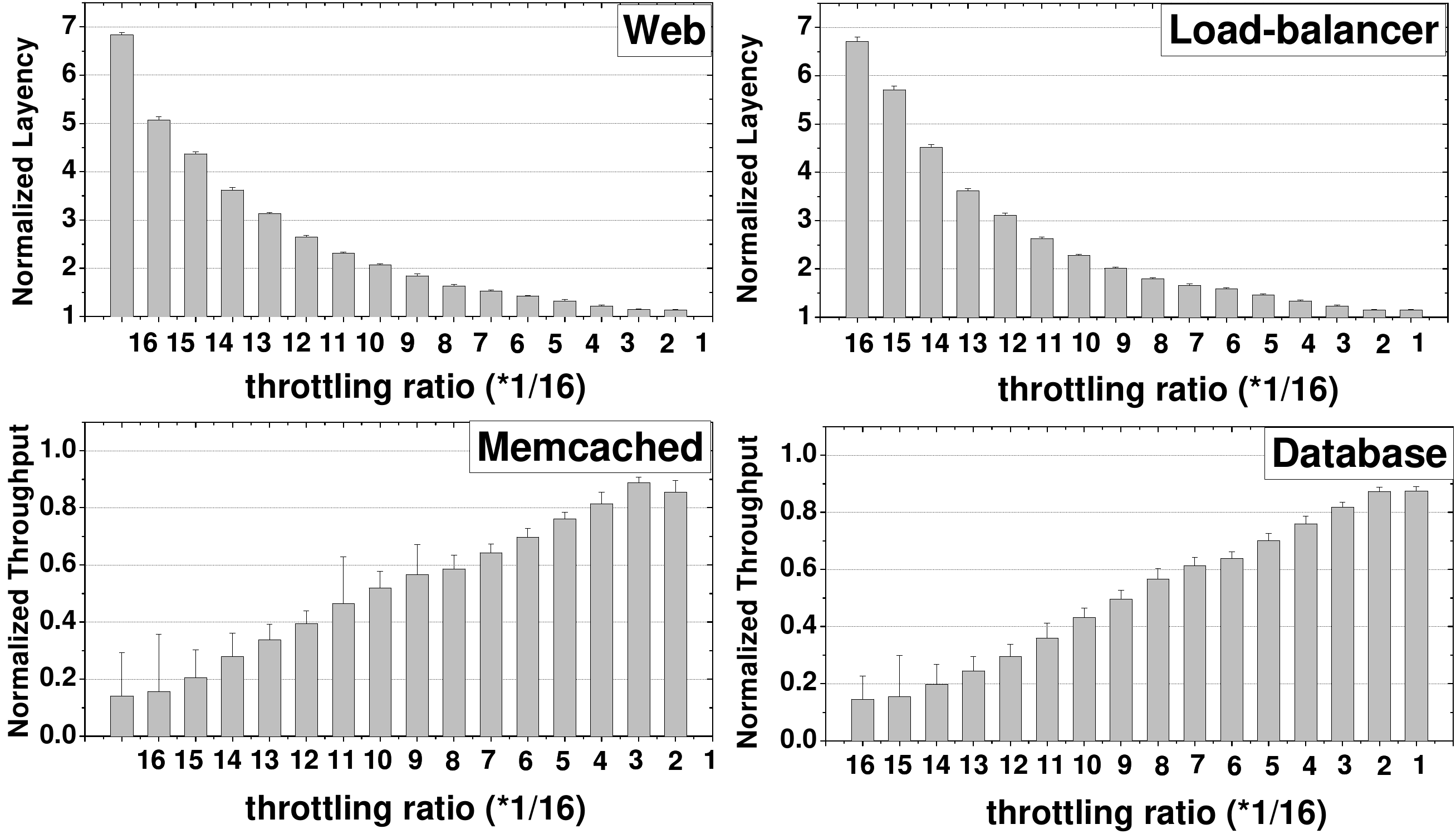}
     \label{fig:throttle_bus}} \\
    \caption{Normalized performance of the \protectedVM with throttling of \attackname.}
    \label{fig:throttle}
\end{figure*}

\bheading{Effectiveness of mitigation.}
We evaluated the effectiveness of execution throttling
based mitigation. The \protectedVM runs the cloud benchmarks from the Magento 
application while the attacker VM runs LLC cleansing or atomic locking attacks. We 
chose different duty cycle ratios for the attacker VM. Figures 
\ref{fig:throttle_llc} and \ref{fig:throttle_bus} show the normalized
performance of the \protectedVM with different throttling ratios, under LLC 
cleansing and atomic locking attacks, respectively. The x-axis shows the duty 
cycle (\emph{x} $\times$ 1/16) given to the co-located VMs, going from no 
throttling on the left to maximum throttling on the right of each figure. The 
y-axis shows the \protectedVM's response latency (for web and load-balancer) or throughput 
(for memcached and database) normalized to the ones without attack. A high latency or a 
small throughput indicates that the performance of the \protectedVM is highly 
taffected by the attacker VM. We can see that a 
smaller throttling ratio can effectively reduce the attacker's impact on the 
victim's performance. When the ratio is set as 1/16, the victim's performance degradation 
caused by the attacker is kept within 12\% (compared to $23\%\sim50\%$ 
degradation with no throttling) for LLC cleansing attacks. It is within 14\% for 
atomic locking attacks (compared to $7\times$ degradation with no throttling).

\begin{figure}[ht]
     \centering
     \subfloat[][LLC cleansing attack]{
     \includegraphics[width=0.48\linewidth]{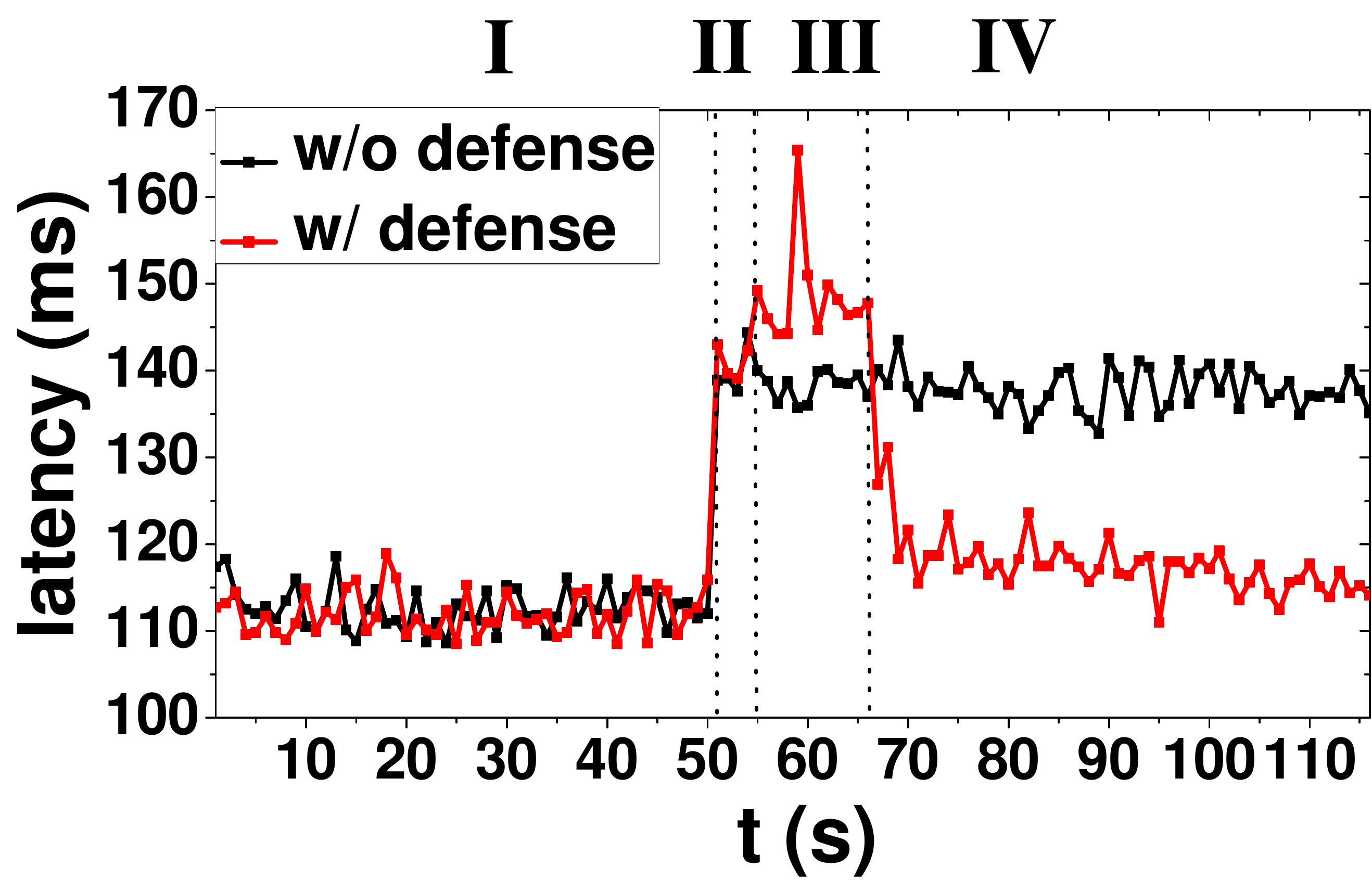}
     \label{fig:llc_detect_perf}}
     \subfloat[][Atomic locking attack]{
     \includegraphics[width=0.48\linewidth]{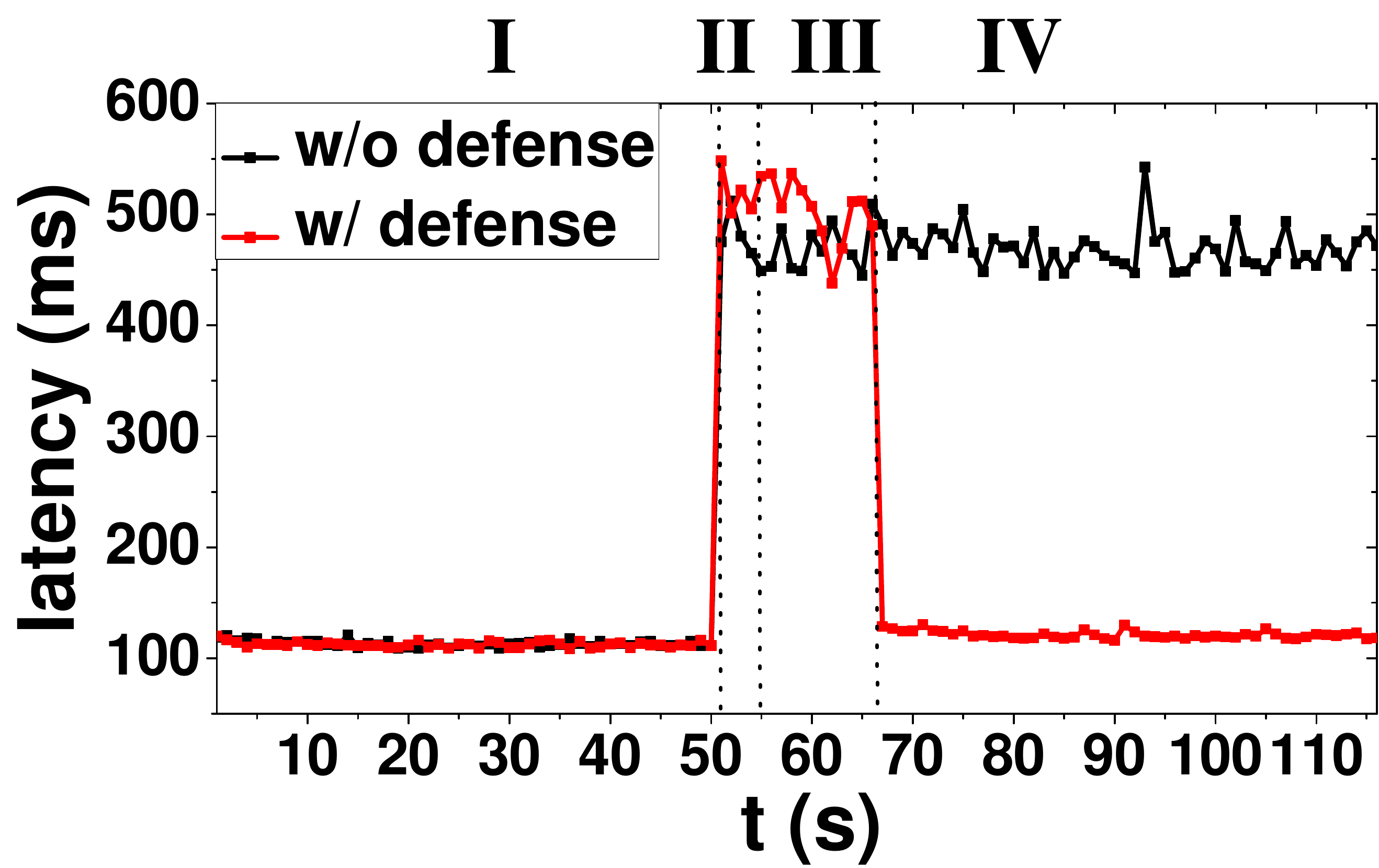}
     \label{fig:bus_detect_perf}}
    \caption{Request latency of Magento Application}
    \label{fig:vm_detect_perf}
\end{figure}

\bheading{Latency increase and mitigation.} 
We chose a latency-critical application, the Magento E-commerce application as the 
target victim. One Apache web server was selected as the \protectedVM, co-locating 
with an attacker and 7 benign VMs running linux utilities. Figure
\ref{fig:vm_detect_perf} shows the response latency with and without our defense. 
The detection phase does not affect the \protectedVM's performance 
(stage \RNum{1}), since the PMU collects monitored samples without interrupting 
the VM's execution. In stage \RNum{2}, the attack occurs and the defense system 
detects the \protectedVM's performance is degraded. In stage \RNum{3}, attacker VM 
identification is done. After throttling down the attacker VM in stage \RNum{4}, 
the \protectedVM's performance is not affected by the \attackname. The 
latency during the attack in Phase \RNum{2} increases significantly, but returns 
to normal after mitigation in Phase \RNum{4}.

\begin{figure}[ht]
\centerline{\mbox{\includegraphics[width=\linewidth]{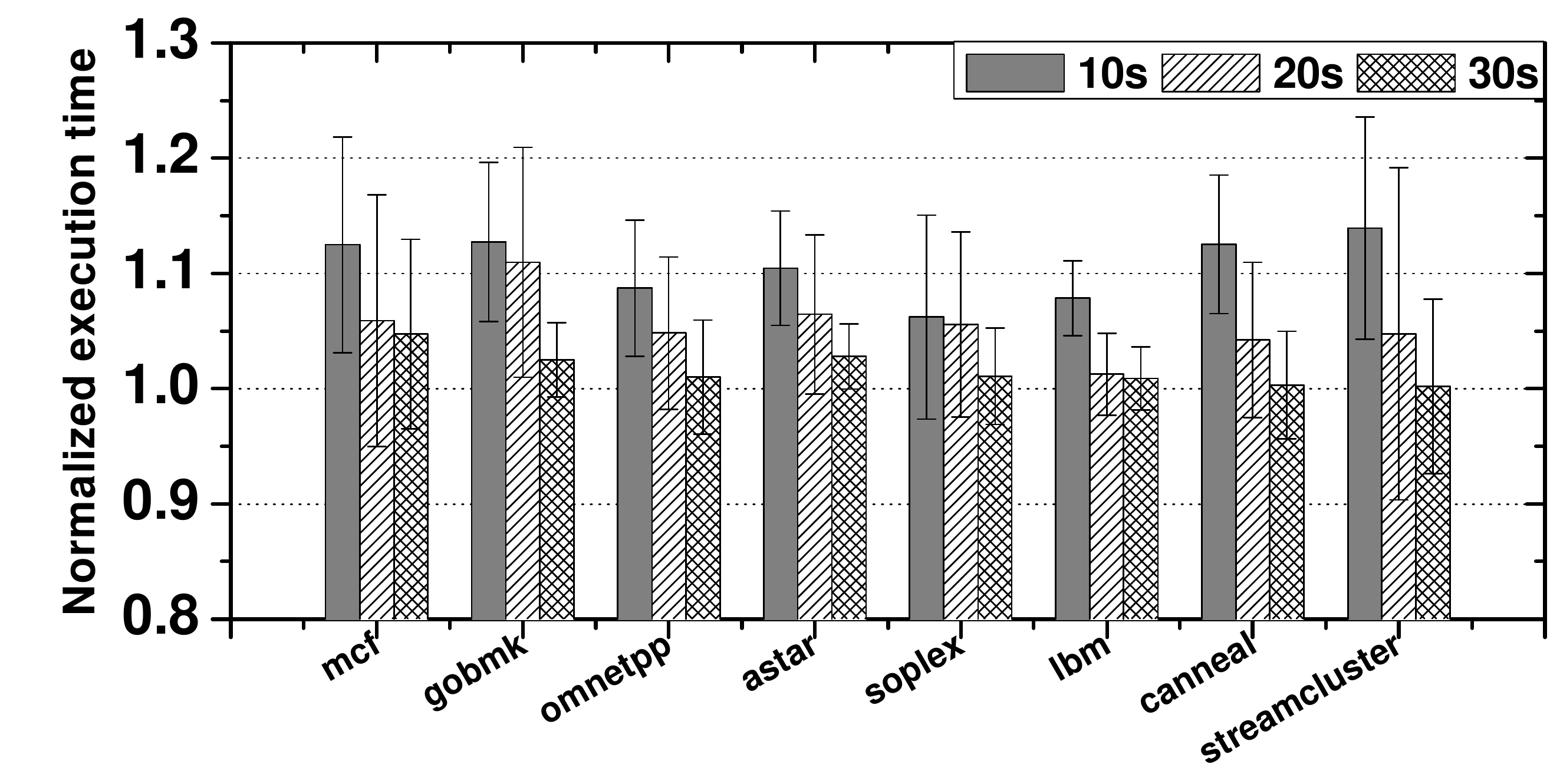}}}
\caption{Performance overhead of co-located VMs due to monitoring.}
\label{fig:coVM_perf}
\end{figure}

We also evaluated the performance overhead of co-located VMs due to \emph{execution
throttling} in the detection step. We launched one VM running one of the eight 
SPEC2006 or PARSEC benchmarks. Then we periodically throttle down this VM every 
10s, 20s or 30s. Each time throttling lasted for 1s (the same value for $W_R$ and 
$W_M$ used earlier). The normalized performance of this VM is shown in Figure 
\ref{fig:coVM_perf}. We can see that when the server throttles this VM every 10s, 
the performance penalty can be around 10\%. However, when the frequency is set to 
be 30s (our implementation choice), this penalty is smaller than 5\%.

\section{Related Work}
\label{sec:related}
\subsection{Resource Contention Attacks}
\label{sec:attack_past}

\bheading{Cloud DoS attacks.}
\cite{Li:10} proposed a DoS attack which can deplete the victim's network 
bandwidth from its subnet. \cite{BeSh:12} proposed a network-initiated 
DoS attack which causes contention in the shared Network Interface Controller. 
\cite{HuLe:13} proposed cascading performance attacks which exhaust hypervisor's 
I/O processing capability. \cite{AlWo:13} exploited VM migration 
to degrade the hypervisor's performance. Our work is different as we exploit 
failure of isolation in the hardware memory subsystem (which has not been 
addressed by cloud providers), and not attacks on networks or hypervisors.

\bheading{Cloud resource stealing attacks.}
\cite{VaKoFa:12} proposed the resource-freeing attack, where a malicious VM can
steal one type of resource from the co-located victim VM by increasing this VM's 
usage of other types of resources. \cite{ZhGoDe:11} designed a CPU resource attack 
where an attacker VM can exploit the boost mechanism in the Xen credit scheduler 
to obtain more CPU resource than paid for.  Our attacks do not steal extra cloud 
resources. Rather, we aim to induce the maximum performance degradation to the co-located victim VM targets.

\bheading{Hardware resource contention studies.}
\cite{GrGh:02} studied the effect of trace cache evictions
on the victim's execution with Hyper-Threading enabled in an Intel Pentium 4
Xeon processor. \cite{WoLe:07} explored frequently flushing shared
L2 caches on multicore platforms to slow down a victim program.  They studied
saturation and locking of buses that connect L1/L2 caches and the main
memory~\cite{WoLe:07}. \cite{MoMu:07} studied contention
attacks on the schedulers of memory controllers. However, due to advances in
computer hardware design, caches and DRAMs are larger and their management 
policies more sophisticated, so these prior attacks may not work in modern cloud settings.

\bheading{Timing channels in clouds.}
Prior studies showed that shared memory resources can be exploited by an attacker 
to extract crypto keys from the victim VM using cache side-channel attacks in 
cloud settings~\cite{ZhJuRe:12, ZhJuRe:14, YaLiGe:15}, or to transmit 
information, using cache operations \cite{RiTrSh:09, XuBaJa:11} or bus activities
\cite{WuXuWa:12} in covert channel communications between two VMs. Unlike 
side-channel attacks our \attackname aim to \emph{maximize} the effects of 
resource contention, while resource contention is an unintended side-effect of 
side-channel attacks. To maximize contention, we addressed various new challenges, 
\eg, finding which attacks cause greatest resource contention (exotic bus locking 
versus memory controller attacks), maximizing the frequency of resource depletion, 
and minimizing self-contention. To the best of our knowledge, we are 
the first to show that similar attack strategies (enhanced for resource 
contention) can be used as availability attacks as well as confidentiality 
attacks.

\subsection{Eliminating Resource Contention}
\label{sec:defense_past}

\bheading{VM performance monitoring.} Public clouds offer performance monitoring services
for customers' VMs and applications, \eg, Amazon CloudWatch \cite{amazon_cloudwatch}, Microsoft Azure Application Insights \cite{azure_insight}, Google Stackdriver \cite{google_stackdriver}, \etc. However, these services only monitor CPU usage, network traffic and disk bandwidth, but not low-level memory usage. To measure a VM's performance without contention for reference sampling, past work offer three ways: (1) collecting the VM's performance characteristics before it is deployed in the cloud \cite{DeKo:13, ZhLaMa:14}; (2) measuring the performance of other 
VMs which run similar tasks \cite{ZhTuHa:13, NoDeVa:13}; (3) measuring the 
\protectedVM while pausing all other co-located VMs \cite{GuSaTa:14, YaBrMa:13}. The drawback 
of (1) and (2) is that it only works for programs with predictable and stable 
performance characteristics, and does not support arbitrary programs running in 
the \protectedVM. The problem with (3) is the significant performance overhead 
inflicted on co-located VMs. In contrast, we use novel \emph{execution throttling} 
of the co-located VMs to collect the \protectedVM's baseline (reference) 
measurements with negligible performance overhead. While \emph{execution throttling} has been used to achieve resource fairness in prior work~\cite{ZhDwSh1:09, EbLeMu:10}; using it to collect Reference samples at runtime is, to our knowledge, novel.

\bheading{QoS-aware VM scheduling.} Prior research propose to predict interference 
between different applications (or VMs) by profiling their resource usage offline 
and then statically scheduling them to different servers if co-locating them will 
lead to excessive resource contention \cite{DeKo:13, ZhLaMa:14, YaBrMa:13}. The underlying assumption is that applications (or VMs), 
when deployed on the cloud servers, will not change their resource usage 
patterns. Unfortunately, these approaches fall short in defense against malicious 
applications, who can reduce their resource uses during the profiling stage, then 
run \attackname when deployed, thus evading these QoS scheduling mechanisms.

\bheading{Load-triggered VM migration.} Some studies propose to monitor the resource consumption of guest VMs or the entire server in real-time, and migrate VMs to different processor packages or servers when there is severe resource contention \cite{HuiBlZhFe:10, ZhBlFe:10, AhKiHa:12, WaIsSu:15}. By doing so these approaches can dynamically balance the workload among multiple packages or servers when some of them are overloaded, and achieve an optimal resource allocation. While they work well for performance optimization of a set of fully-loaded servers, they fail to detect carefully-crafted \attackname. First, the metrics in their methods cannot be used to detect the existence of \attackname. These works measure the LLC miss rate \cite{HuiBlZhFe:10, ZhBlFe:10} or memory bandwidth \cite{AhKiHa:12, WaIsSu:15} of guest VMs or the whole server. A high LLC miss rate or memory bandwidth indicates severe resource contention for the VMs or servers. However, a \aattackname does not need to cause high LLC miss rate or memory bandwidth in order to degrade a victim's performance. For instance, atomic locking attacks lock the bus temporarily but frequently, which could incur decreased LLC accesses and LLC misses in the victim VM. Our experiments show the victim VM's LLC miss rate does not change, and its memory bandwidth is even decreased. So such micro-architectural measurement can never reveal severe resource contention, or trigger VM migration in the above approaches. Second, these approaches aim to balance the system's performance. So they cannot guarantee to choose the victim or attacker VMs for migration when they achieve optimal workload placement. For instance, Adaptive LLC cleaning attacks can increase victim VMs' LLC miss rate. However, the above approaches have no means to figure out that the victim VM has the strongest desire for migration. It is possible that there exists another VM, which has even higher LLC miss rate than the victim VM, due to its internal execution behaviors, not the interaction and contention with the attacker. So the above approaches will migrate this VM instead of the victim VM, as this VM has the highest miss rate. Then the victim VM will still suffer LLC cleaning attacks.

\bheading{Performance isolation.} While cloud providers can offer single-tenant
machines to customers with high demand for security and performance, disallowing
resource sharing by VMs will lead to low resource utilization and thus is at odds 
with the cloud business model. Another option is to partition memory resources to
enforce performance isolation on shared resources (\eg, LLC~\cite{WaLe:07, 
SaKo:11, KiPeMa:12, CoMoBi:13, KiKiHu:14, intel_cat}, or DRAM
\cite{MoMu:07, MuSuMu:11, WaFeSu:14}). These works aim to achieve fairness 
between different domains and provide fair QoS. However, they cannot effectively 
defeat \attackname. For cache partitioning, software page coloring methods 
\cite{KiPeMa:12} can cause significant wastage of 
LLC space, while hardware cache partitioning mechanisms have insufficient 
partitions (\eg, Intel Cache Allocation Technology \cite{intel_cat} only provides 
four QoS partitions on the LLC). Furthermore, LLC cache partitioning methods 
cannot resolve atomic locking attacks.

To summarize, existing solutions fail to address \attackname because they assume
benign applications with non-malicious behaviors.  Also, they are often tailored
to only one type of attack so that they cannot be generalized to all memory DoS
attacks, unlike our proposed defense.

\section{Conclusions}
\label{sec:conclusion}
%This paper presents the first systematic exploration of insufficient isolation 
%in multiple layers of hardware memory resources of a computer system, 
%\eg, caches, memory scheduling resources and DRAMs, 
%that may lead to malicious 
%resource contention.
%which we call \attackname.

We presented \attackname, in which a malicious VM intentionally
induces memory resource contention to degrade the performance of co-located 
victim VMs. We proposed several advanced techniques to conduct such attacks,
and demonstrate the severity of the resulting performance
degradation. Our attacks work on modern memory systems in cloud servers, for which prior attacks on older memory systems are often ineffective.
We evaluated our attacks against two commonly used applications in
a public cloud, Amazon EC2, and show that the adversary can cause
significant performance degradation to not only co-located VMs, but to the entire
distributed application. %with one of the victim VMs as an integral component.

%We show that a malicious VM may cause significant performance degradation of the 
%victim VM, and hence the applications it supports, by causing 
%contention in shared hardware memory resources. We describe effective attack 
%techniques for LLC cleansing and atomic locking. We further explore the use of 
%\attackname against multi-tenant cloud servers to conduct low-cost DoS attacks in 
%public clouds. We evaluate our attacks against two commonly used applications in a 
%public cloud, Amazon EC2, showing the adversary can introduce huge performance 
%degradation (up to $38\times$ slowdown for an E-commerce application) to the victim 
%at very low cost. 

%Our study raises the research 
%question of how effective performance isolation can be achieved in multi-tenant 
%public clouds, when threats from misbehaving cloud tenants must be considered. 

We then designed a novel and generalizable method that can detect and mitigate
all known \attackname. Our approach collects the \protectedVM's reference and 
monitored behaviors at runtime using the Performance Monitor Unit. This is done by 
establishing a pseudo isolated collection environment by using the duty-cycle 
modulation feature to throttle the co-resident VMs for collecting Reference samples. Statistical tests are performed 
to detect differing performance probability distributions between Reference and Monitored samples, with desired confidence 
levels. Our evaluation shows this defense can detect and defeat \attackname with 
very low performance overhead.

%Both the attack and defense strategies and techniques we propose can be generalized 
%to other similar systems, and we hope that this paper will stimulate more work in 
%defending computer systems from other host-based DoS attacks.

%The novelty of our study lies not only in the systematic investigation 
%of  memory resource contention in
%modern computer architectures in spite of system virtualization, but also in
%proposing three new techniques with which an adversary may enhance
%the effects of resource contention and exacerbate the performance
%degradation. 

\bibliographystyle{abbrv}
\bibliography{ref}

\begin{thebibliography}{10}

\bibitem{ab}
Ab - the apache software foundation.
\newblock \url{http://httpd.apache.org/docs/2.2/programs/ab.html}.

\bibitem{amazon_cloudwatch}
Amazon {CloudWatch}.
\newblock \url{https://aws.amazon.com/cloudwatch/}.

\bibitem{amazon_vpc}
Amazon virtual private cloud.
\newblock \url{https://aws.amazon.com/vpc/}.

\bibitem{google_stackdriver}
Google {Stackdriver}.
\newblock \url{https://cloud.google.com/stackdriver/}.

\bibitem{intel_cat}
Improving real-time performance by utilizing cache allocation technology.
\newblock
  \url{http://www.intel.com/content/www/us/en/communications/cache-allocation-technology-white-paper.html}.

\bibitem{intel_manual}
Intel 64 and ia-32 architectures software developer's manual, volume 3: System
  programming guide.
\newblock
  \url{http://www.intel.com/content/www/us/en/processors/architectures-software-developer-manuals.html}.

\bibitem{magento}
Magento: ecommerce software and ecommerce platform.
\newblock \url{http://www.magento.com/}.

\bibitem{memtier}
memtier benchmark.
\newblock \url{https://github.com/RedisLabs/memtier_benchmark}.

\bibitem{azure_insight}
Microsoft {Azure Application Insights}.
\newblock
  \url{https://azure.microsoft.com/en-us/services/application-insights/}.

\bibitem{SPEC2006}
Spec cpu 2006.
\newblock \url{https://www.spec.org/cpu2006/}.

\bibitem{sysbench}
Sysbench: a system performance benchmark.
\newblock \url{https://launchpad.net/sysbench/}.

\bibitem{httperf}
Welcome to the httperf homepage.
\newblock \url{http://www.hpl.hp.com/research/linux/httperf/}.

\bibitem{AhKiHa:12}
J.~Ahn, C.~Kim, J.~Han, Y.-R. Choi, and J.~Huh.
\newblock Dynamic virtual machine scheduling in clouds for architectural shared
  resources.
\newblock In {\em USENIX Conference on Hot Topics in Cloud Computing}, 2012.

\bibitem{AlWo:13}
S.~Alarifi and S.~D. Wolthusen.
\newblock Robust coordination of cloud-internal denial of service attacks.
\newblock In {\em Intl. Conf. on Cloud and Green Computing}, 2013.

\bibitem{BeSh:12}
H.~S. Bedi and S.~Shiva.
\newblock Securing cloud infrastructure against co-resident {D}o{S} attacks
  using game theoretic defense mechanisms.
\newblock In {\em Intl. Conf. on Advances in Computing, Communications and
  Informatics}, 2012.

\bibitem{Bi:11}
C.~Bienia.
\newblock {\em Benchmarking Modern Multiprocessors}.
\newblock PhD thesis, Princeton University, 2011.

\bibitem{HuiBlZhFe:10}
S.~Blagodurov, S.~Zhuravlev, A.~Fedorova, and A.~Kamali.
\newblock A case for numa-aware contention management on multicore systems.
\newblock In {\em ACM Intl. Conf. on Parallel architectures and compilation
  techniques}.

\bibitem{CoMoBi:13}
H.~Cook, M.~Moreto, S.~Bird, K.~Dao, D.~A. Patterson, and K.~Asanovic.
\newblock A hardware evaluation of cache partitioning to improve utilization
  and energy-efficiency while preserving responsiveness.
\newblock In {\em Intl. Symp. on Computer Architecture}, 2013.

\bibitem{DeKo:13}
C.~Delimitrou and C.~Kozyrakis.
\newblock Paragon: Qos-aware scheduling for heterogeneous datacenters.
\newblock In {\em Intl. Conf. on Architectural Support for Programming
  Languages and Operating Systems}, 2013.

\bibitem{EbLeMu:10}
E.~Ebrahimi, C.~J. Lee, O.~Mutlu, and Y.~N. Patt.
\newblock Fairness via source throttling: A configurable and high-performance
  fairness substrate for multi-core memory systems.
\newblock In {\em Architectural Support for Programming Languages and Operating
  Systems}, 2010.

\bibitem{GrGh:02}
D.~Grunwald and S.~Ghiasi.
\newblock Microarchitectural denial of service: Insuring microarchitectural
  fairness.
\newblock In {\em ACM/IEEE Intl. Symp. on Microarchitecture}, 2002.

\bibitem{GuSaTa:14}
A.~Gupta, J.~Sampson, and M.~B. Taylor.
\newblock Quality time: A simple online technique for quantifying multicore
  execution efficiency.
\newblock In {\em IEEE Intl. Symp. on Performance Analysis of Systems and
  Software}, 2014.

\bibitem{HuLe:13}
Q.~Huang and P.~P. Lee.
\newblock An experimental study of cascading performance interference in a
  virtualized environment.
\newblock {\em SIGMETRICS Perform. Eval. Rev.}, 2013.

\bibitem{JaSzPe:13}
P.~Jamkhedkar, J.~Szefer, D.~Perez-Botero, T.~Zhang, G.~Triolo, and R.~B. Lee.
\newblock A framework for realizing security on demand in cloud computing.
\newblock In {\em Conf. on Cloud Computing Technology and Science}, 2013.

\bibitem{KiKiHu:14}
D.~Kim, H.~Kim, and J.~Huh.
\newblock vcache: Providing a transparent view of the llc in virtualized
  environments.
\newblock {\em Computer Architecture Letters}, 2014.

\bibitem{KiPeMa:12}
T.~Kim, M.~Peinado, and G.~Mainar-Ruiz.
\newblock Stealthmem: System-level protection against cache-based side channel
  attacks in the cloud.
\newblock In {\em USENIX Security Symp.}, 2012.

\bibitem{YaLiGe:15}
F.~Liu, Y.~Yarom, Q.~Ge, G.~Heiser, and R.~B. Lee.
\newblock Last-level cache side-channel attacks are practical.
\newblock In {\em IEEE Symp. on Security and Privacy}, 2015.

\bibitem{Li:10}
H.~Liu.
\newblock A new form of {D}o{S} attack in a cloud and its avoidance mechanism.
\newblock In {\em ACM Workshop on Cloud Computing Security}, 2010.

\bibitem{LiCuXi:12}
L.~Liu, Z.~Cui, M.~Xing, Y.~Bao, M.~Chen, and C.~Wu.
\newblock A software memory partition approach for eliminating bank-level
  interference in multicore systems.
\newblock In {\em Intl. Conf. on Parallel Architectures and Compilation
  Techniques}, 2012.

\bibitem{Ma:51}
F.~J. Massey~Jr.
\newblock The kolmogorov-smirnov test for goodness of fit.
\newblock {\em Journal of the American statistical Association}, 1951.

\bibitem{stream_benchmark}
J.~D. McCalpin.
\newblock Stream: Sustainable memory bandwidth in high performance computers.
\newblock \url{http://www.cs.virginia.edu/stream/}.

\bibitem{MoMu:07}
T.~Moscibroda and O.~Mutlu.
\newblock Memory performance attacks: Denial of memory service in multi-core
  systems.
\newblock In {\em USENIX Security Symp.}, 2007.

\bibitem{MuSuMu:11}
S.~P. Muralidhara, L.~Subramanian, O.~Mutlu, M.~Kandemir, and T.~Moscibroda.
\newblock Reducing memory interference in multicore systems via
  application-aware memory channel partitioning.
\newblock In {\em ACM/IEEE Intl. Symp. on Microarchitecture}, 2011.

\bibitem{NoDeVa:13}
D.~Novakovi\'{c}, N.~Vasi\'{c}, S.~Novakovi\'{c}, D.~Kosti\'{c}, and
  R.~Bianchini.
\newblock Deepdive: Transparently identifying and managing performance
  interference in virtualized environments.
\newblock In {\em USENIX Conf. on Annual Technical Conference}, 2013.

\bibitem{PoCaGa:11}
N.~Poggi, D.~Carrera, R.~Gavalda, and E.~Ayguade.
\newblock Non-intrusive estimation of qos degradation impact on e-commerce user
  satisfaction.
\newblock In {\em IEEE Intl. Symp. on Network Computing and Applications},
  2011.

\bibitem{RiTrSh:09}
T.~Ristenpart, E.~Tromer, H.~Shacham, and S.~Savage.
\newblock Hey, you, get off of my cloud: Exploring information leakage in
  third-party compute clouds.
\newblock In {\em ACM Conf. on Computer and Communications Security}, 2009.

\bibitem{SaKo:11}
D.~Sanchez and C.~Kozyrakis.
\newblock Vantage: Scalable and efficient fine-grain cache partitioning.
\newblock In {\em AMC Intl. Symp. on Computer Architecture}, 2011.

\bibitem{VaKoFa:12}
V.~Varadarajan, T.~Kooburat, B.~Farley, T.~Ristenpart, and M.~M. Swift.
\newblock Resource-freeing attacks: Improve your cloud performance (at your
  neighbor's expense).
\newblock In {\em ACM Conf. on Computer and Communications Security}, 2012.

\bibitem{Varadarajan:2015:PVS}
V.~Varadarajan, Y.~Zhang, T.~Ristenpart, and M.~Swift.
\newblock A placement vulnerability study in multi-tenant public clouds.
\newblock In {\em USENIX Security Symp.}, 2015.

\bibitem{WaIsSu:15}
H.~Wang, C.~Isci, L.~Subramanian, J.~Choi, D.~Qian, and O.~Mutlu.
\newblock A-drm: Architecture-aware distributed resource management of
  virtualized clusters.
\newblock In {\em ACM Intl. Conference on Virtual Execution Environments},
  2015.

\bibitem{WaFeSu:14}
Y.~Wang, A.~Ferraiuolo, and G.~E. Suh.
\newblock Timing channel protection for a shared memory controller.
\newblock In {\em IEEE Intl. Symp. on High Performance Computer Architecture},
  2014.

\bibitem{WaLe:07}
Z.~Wang and R.~B. Lee.
\newblock New cache designs for thwarting software cache-based side channel
  attacks.
\newblock In {\em ACM Intl. Symp. on Computer Architecture}, 2007.

\bibitem{WoLe:07}
D.~H. Woo and H.-H.~S. Lee.
\newblock Analyzing performance vulnerability due to resource denial-of-service
  attack on chip multiprocessors.
\newblock In {\em Workshop on Chip Multiprocessor Memory Systems and
  Interconnects}, 2007.

\bibitem{WuXuWa:12}
Z.~Wu, Z.~Xu, and H.~Wang.
\newblock Whispers in the hyper-space: High-speed covert channel attacks in the
  cloud.
\newblock In {\em USENIX Security Symp.}, 2012.

\bibitem{XuBaJa:11}
Y.~Xu, M.~Bailey, F.~Jahanian, K.~Joshi, M.~Hiltunen, and R.~Schlichting.
\newblock An exploration of {L2} cache covert channels in virtualized
  environments.
\newblock In {\em ACM Workshop on Cloud computing security}, 2011.

\bibitem{Xu:2015:MSC}
Z.~Xu, H.~Wang, and Z.~Wu.
\newblock A measurement study on co-residence threat inside the cloud.
\newblock In {\em USENIX Security Symp.}, 2015.

\bibitem{YaBrMa:13}
H.~Yang, A.~Breslow, J.~Mars, and L.~Tang.
\newblock Bubble-flux: Precise online qos management for increased utilization
  in warehouse scale computers.
\newblock In {\em ACM Intl. Symp. on Computer Architecture}, 2013.

\bibitem{ZhLe:15}
T.~Zhang and R.~B. Lee.
\newblock Cloudmonatt: An architecture for security health monitoring and
  attestation of virtual machines in cloud computing.
\newblock In {\em ACM Intl. Symp. on Computer Architecture}, 2015.

\bibitem{ZhDwSh1:09}
X.~Zhang, S.~Dwarkadas, and K.~Shen.
\newblock Hardware execution throttling for multi-core resource management.
\newblock In {\em USENIX Annual Technical Conference}, 2009.

\bibitem{ZhTuHa:13}
X.~Zhang, E.~Tune, R.~Hagmann, R.~Jnagal, V.~Gokhale, and J.~Wilkes.
\newblock Cpi2: Cpu performance isolation for shared compute clusters.
\newblock In {\em ACM European Conf. on Computer Systems}, 2013.

\bibitem{ZhJuRe:12}
Y.~Zhang, A.~Juels, M.~K. Reiter, and T.~Ristenpart.
\newblock {Cross-VM} side channels and their use to extract private keys.
\newblock In {\em ACM Conf. on Computer and Communications Security}, 2012.

\bibitem{ZhJuRe:14}
Y.~Zhang, A.~Juels, M.~K. Reiter, and T.~Ristenpart.
\newblock Cross-tenant side-channel attacks in {PaaS} clouds.
\newblock In {\em ACM Conf. on Computer and Communications Security}, 2014.

\bibitem{ZhLaMa:14}
Y.~Zhang, M.~A. Laurenzano, J.~Mars, and L.~Tang.
\newblock Smite: Precise qos prediction on real-system smt processors to
  improve utilization in warehouse scale computers.
\newblock In {\em IEEE/ACM Intl. Symp. on Microarchitecture}, 2014.

\bibitem{ZhGoDe:11}
F.~Zhou, M.~Goel, P.~Desnoyers, and R.~Sundaram.
\newblock Scheduler vulnerabilities and coordinated attacks in cloud computing.
\newblock In {\em IEEE Intl. Symp. on Network Computing and Applications},
  2011.

\bibitem{ZhBlFe:10}
S.~Zhuravlev, S.~Blagodurov, and A.~Fedorova.
\newblock Addressing shared resource contention in multicore processors via
  scheduling.
\newblock In {\em Intl. Conf. on Architectural Support for Programming
  Languages and Operating Systems}, 2010.

\end{thebibliography}

\end{document}